\documentclass{hch}
\usepackage{graphicx}

\begin{document}
\title{FIELD THEORY FOR TRAPPED ATOMIC GASES}
\author{H.T.C. Stoof}
\address{Institute for Theoretical Physics, \\
         University of Utrecht, \\
         Princetonplein 5, \\
         3584 CC Utrecht, \\
         The Netherlands.}
\runningtitle{H.T.C. Stoof: Field theory for trapped atomic gases}
\maketitle

\begin{abstract}
In this course we give a selfcontained introduction to the quantum field theory 
for trapped atomic gases, using functional methods throughout. We consider both 
equilibrium and nonequilibrium phenomena. In the equilibrium case, we first 
derive the appropriate Hartree-Fock theory for the properties of the gas in the 
normal phase. We then turn our attention to the properties of the gas in the 
superfluid phase, and present a microscopic derivation of the Bogoliubov and 
Popov theories of Bose-Einstein condensation and the Bardeen-Cooper-Schrieffer 
theory of superconductivity. The former are applicable to trapped bosonic gases 
such as rubidium, lithium, sodium and hydrogen, and the latter in particular to 
the fermionic isotope of atomic lithium. In the nonequilibrium case, we discuss 
various topics for which a field-theoretical approach is especially suited, 
because they involve physics that is not contained in the Gross-Pitaevskii 
equation. Examples are quantum kinetic theory, the growth and collapse of a Bose 
condensate, the phase dynamics of bosonic and fermionic superfluids, and the 
collisionless collective modes of a Bose gas below the critical temperature. 
\end{abstract}

\section{Introduction}
An important trend in the condensed matter physics of the last two decades, has
been the use of advanced field-theoretical methods to discuss various
subtle and fundamental properties of interacting many-particle systems at low
temperatures. There are several reasons for this trend. The first reason is of
course, that a traditional topic in statistical and condensed matter physics is 
the study of phase transitions and critical phenomena, for which the universal 
properties are independent of the microscopic details of the system and can 
therefore be determined by a field theory describing only the large-scale 
properties of the system of interest. Since the latter are usually solely 
determined by symmetry considerations, this has led to the important concept of 
spontaneous symmetry breaking, which has turned out to be not only highly 
successful in condensed matter physics, but also in high-energy physics and in 
particular in the Standard Model of elementary particles \cite{A}.

A second reason is that soon after the development of the renormalization group
methods for critical phenomena \cite{1}, it was realized that the same methods 
can in fact be used to describe the large-scale properties of many-particle 
systems at any temperature and not only near the critical one. Moreover, 
application of the renormalization group ideas does not only lead to an 
understanding of the static behaviour but also of the dynamical properties 
\cite{2}. As a result quantum field theories can be used to conveniently 
determine the dynamics of many-particle systems close to equilibrium, i.e., for 
example the linear hydrodynamical equations of motion. In addition, it can even 
be used in highly nonequilibrium situations where in general also 
nonlinearities play an important role. This feature, that also the dynamics of 
the system can be captured by a quantum field theory, has for instance led in 
recent years to the study of so-called quantum phase transitions \cite{3}.

Finally, the importance of field-theoretical methods in condensed
matter physics is associated with the observation that also the effects of
imperfections, i.e., disorder, can be treated in this way \cite{4}. Apart from 
the technological importance of disorder, for example for superconducting 
magnets, disorder leads also to fundamentally new physics such as the
phenomenon of localization \cite{5}. In quantum Hall systems, a combination of 
disorder and interaction effects give rise to the realization of various 
peculiar quantum fluids with fractionally charged excitations \cite{6}. The 
application of field theories has proven to be highly successful in this case 
and has led to a theory of the quantum Hall effect in terms of edge states that 
form a chiral Luttinger liquid \cite{7}. In mesoscopic physics, the study of 
disorder in small electronic structures has resulted in the so-called random 
matrix theory \cite{8}, which has also been of much use in the study of the 
quantization of classically chaotic systems. 

In this course we aim to show that quantum field theory is also very convenient 
for obtaining a detailed understanding of the equilibrium and nonequilibrium 
properties of trapped atomic gases. After the first observations of 
Bose-Einstein condensation in 1995 \cite{JILA,Rice,MIT}, degenerate atomic 
gases have received again a great deal of attention and are presently the main 
subject of study of a large number of experimental and theoretical groups around 
the world. The reason for all this excitement is, first of all, that 
Bose-Einstein condensation has never before been observed experimentally in a 
clear-cut manner, even though this phenomenon was already predicted by Einstein 
in 1925 \cite{bec}. Second, it is of fundamental interest because it is the only 
phase transition that occurs also in the absense of interactions and is, 
therefore, the textbook example for the use of statistical-physics methods. 
Finally, the goal of achieving Bose-Einstein condensation turned out to be much 
more difficult than anticipated at first. Just before the break\-through in 
1995, it had even acquired the nature of a `quest for the holy grail', since the 
pioneering experiments were already performed in 1980 \cite{jook}. 

From a theoretical point of view, a quantitative understanding of the  
experiments with cold atomic gases requires that we take into account the 
following two effects. First, the gas is magnetically trapped in an, usually 
axially symmetric, harmonic oscillator potential. This is necessary because, in 
order to obtain the required record low temperatures of $1-100$~nK, the gas 
cannot be allowed to have any contact with material walls. Second, the atoms of 
the gas interact with each other, which in general dramatically affects the 
behaviour of the condensed gas, even at the relevant low densities of 
$10^{12}-10^{14}$~cm$^{-3}$. An accurate description of these degenerate gases 
thus requires the solution of a highly inhomogeneous many-body problem. The 
theoretical challenge posed by these new quantum systems lies therein that the 
density of the gas is sufficiently small that it should be possible to 
accurately solve this many-body problem from first principles and to compare 
the outcome of the theory directly with experiment, i.e., without any adjustable 
parameters. In our opinion, quantum field theory is the most simple way in which 
we are able to meet this challenge.

\section{Equilibrium field theory}
\label{EFT}
We start our development of the quantum field theory of trapped atomic gases by 
considering first the equilibrium properties of these gases. We consider both 
Bose and Fermi gases, and the ultimate aim of this section is to arrive, for 
both cases, at an accurate description of the normal and superfluid phases of 
the gas. Although mixtures of Bose and Fermi gases are also of current 
interest, we do not consider them explicitly here, because they can be treated 
by a straightforward generalization of the theory. In section \ref{NFT} we then 
turn to the nonequilibrium properties, which are perhaps the most interesting 
and certainly the least understood at present. The reason for organizing the 
course in this way, is that the development of the equilibrium theory gives us 
an opportunity to introduce all the necessary tools that are required for a 
treatment of the more complicated nonequilibrium case. In particular, we 
present in detail the way in which we can make use of functional methods. To 
make also a connection with the more familiar operator language, however, we 
first briefly summarize the outcome of the second quantization formalism.

\subsection{Second quantization}
\label{SQ}
The atoms of interest to us have internal degrees of freedom due to the 
electron and nuclear spins. In principle this is very important, because it 
gives the atom a magnetic moment, which is used to trap the atoms in a magnetic 
field minumum. During this course, however, we restrict ourselves to atomic 
gases that are a mixture of at most two hyperfine states. Without loss of 
generality, we can then suppose to have $N$ identical atoms with mass $m$ and 
effective spin ${\bf s}$ in an external potential $V^{\rm ex}({\bf x})$. As a 
result, the time-dependent Schr\"odinger equation we have to solve is
\begin{equation}
i\hbar \frac{\partial}{\partial t} |\Psi (t)\rangle =
    \hat{H} |\Psi (t)\rangle~,
\end{equation}
where the hamiltonian is
\begin{equation}
\hat{H} 
= \sum_{i=1}^N \left\{ \frac{\hat{\bf p}_i^2}{2m} 
    + V^{\rm ex}(\hat{\bf x}_i)
    - \gamma \hat{\bf s}_i \cdot {\bf B}^{\rm eff} \right\}
    + \frac{1}{2} \sum_{i \neq j=1}^N V(\hat{\bf x}_i - \hat{\bf x}_j)~,
\end{equation}
$[\hat{\bf x}_j,\hat{\bf p}_j]_- = i\hbar$, and all other commutators of the 
positions and momenta vanish. The first term in the right-hand side is the sum 
of the one-particle hamiltonians, which includes an effective Zeeman 
interaction 
that accounts for a possible difference in the hyperfine energies. The second 
term represents the interactions. For simplicity, we have assumed that the 
interaction $V(\hat{\bf x}_i - \hat{\bf x}_j)$ is independent of the hyperfine 
states of the atoms $i$ and $j$. This is in general not justified for realistic 
atomic gases, but is valid for the specific applications that we have in mind. 
Moreover, in section \ref{HFT} we also discuss the general case. Finally, we 
have also neglected possible three-body forces. This is a result of the fact 
that we are interested in dilute quantum gases, for which it is highly 
improbable for three atoms to simultaneously interact with each other. 

Without interactions the eigenstates are, of course, given by the states 
$\{ |{\bf n}_1,\alpha_1\rangle_1 \otimes 
    |{\bf n}_2,\alpha_2\rangle_2 \otimes \dots \otimes
    |{\bf n}_N,\alpha_N\rangle_N \}$, 
where the specific quantum state for each atom is exactly known. Here 
${\bf n} = (n_x,n_y,n_z)$ and the nonnegative integers $n_x$, $n_y$, and 
$n_z$ denote the three quantum numbers that are required to specify the 
one-particle eigenstates in the external potential. The wave functions and 
energies of these eigenstates are 
$\chi_{\bf n}({\bf x}) \equiv \langle {\bf x}|{\bf n}\rangle$ and 
$\epsilon_{\bf n}$, respectively, and are found from the 
time-independent Schr\"odinger equation
\begin{equation}
\left\{ - \frac{\hbar^2 \mbox{\boldmath $\nabla$}^2}{2m} 
    + V^{\rm ex}({\bf x})
    - \epsilon_{\bf n} \right\} \chi_{\bf n}({\bf x}) = 0~.
\end{equation}
In addition, the internal state $|\alpha\rangle$ is a shorthand notation for 
$|s,m_s\rangle$. The many-body wave function $|\Psi (t)\rangle$, however, has 
to be symmetric or antisymmetric under permutations for bosonic or fermionic 
atoms, respectively. Therefore it is more convenient to use a properly 
(anti)symmetrized version of the above basis, i.e., the states 
$|\{N_{{\bf n},\alpha}\}\rangle$ with the occupation numbers 
$N_{{\bf n},\alpha} = 0,1,2, \dots , \infty$ for bosons and 
$N_{{\bf n},\alpha} = 0,1$ for fermions. The Hilbert space of all these states, 
without the constraint $N = \sum_{{\bf n},\alpha} N_{{\bf n},\alpha}$, is known 
as the Fock space.

Clearly, the many-body wave function $|\Psi (t)\rangle$ can be expanded in this 
basis as
\begin{equation}
|\Psi (t)\rangle = {\sum_{\{N_{{\bf n},\alpha}\}}}~
  \Psi(\{N_{{\bf n},\alpha}\},t) |\{N_{{\bf n},\alpha}\}\rangle~,
\end{equation}
where $\Psi(\{N_{{\bf n},\alpha}\},t)$ is the amplitude for the gas to be in 
state $|\{N_{{\bf n},\alpha}\}\rangle$ at time $t$. In this basis the 
Schr\"odinger 
equation becomes
\begin{equation}
i\hbar \frac{\partial}{\partial t} \Psi(\{N_{{\bf n},\alpha}\},t)
  = {\sum_{\{N'_{{\bf n},\alpha}\}}}~
     \langle \{N_{{\bf n},\alpha}\}|
                    \hat{H}|\{N'_{{\bf n},\alpha}\}\rangle
                           \Psi(\{N'_{{\bf n},\alpha}\},t)~,
\end{equation}
which shows that we need the matrix elements of the hamiltonian between 
different states in the Fock space. To calculate these most easily we introduce 
so-called annihilation operators $\hat{\psi}_{{\bf n},\alpha}$ by 
\begin{equation}
\hat{\psi}_{{\bf n},\alpha} |\dots,N_{{\bf n},\alpha},\dots\rangle =
  \sqrt{N_{{\bf n},\alpha}}
                   |\dots,N_{{\bf n},\alpha}-1,\dots\rangle~,
\end{equation}
from which it follows that the creation operators 
$\hat{\psi}^{\dagger}_{{\bf n},\alpha}$ obey
\begin{equation}
\label{cr}
\hat{\psi}^{\dagger}_{{\bf n},\alpha}
                 |\dots,N_{{\bf n},\alpha},\dots\rangle =
  \sqrt{1 \pm N_{{\bf n},\alpha}}
                   |\dots,N_{{\bf n},\alpha}+1,\dots\rangle~.
\end{equation}
As a result, we see that the operator 
$\hat{\psi}^{\dagger}_{{\bf n},\alpha} \hat{\psi}_{{\bf n},\alpha}$ counts the 
number of atoms in the state $|{\bf n},\alpha\rangle$, i.e.,
\begin{equation}
\hat{\psi}^{\dagger}_{{\bf n},\alpha} \hat{\psi}_{{\bf n},\alpha}
      |\dots,N_{{\bf n},\alpha},\dots\rangle
      = N_{{\bf n},\alpha} |\dots,N_{{\bf n},\alpha},\dots\rangle~.
\end{equation}
We have also that 
$[\hat{\psi}_{{\bf n},\alpha},\hat{\psi}_{{\bf n}',\alpha'}]_{\mp} =
 [\hat{\psi}^{\dagger}_{{\bf n},\alpha},
                           \hat{\psi}^{\dagger}_{{\bf n}',\alpha'}]_{\mp}=0$ 
and most importantly that
\begin{equation}
\label{com}
[\hat{\psi}_{{\bf n},\alpha},\hat{\psi}^{\dagger}_{{\bf n}',\alpha'}]_{\mp}
  = \delta_{{\bf n},{\bf n}'} \delta_{\alpha,\alpha'}~.
\end{equation}
In equations (\ref{cr}) and (\ref{com}) the upper sign refers to bosons and the 
lower to fermions. This will be true throughout the course.

From these results we can now easily show, first of all, that the basis in the 
Fock space is given by 
\begin{equation}
|\{N_{{\bf n},\alpha}\}\rangle = \prod_{{\bf n},\alpha}
   \frac{(\hat{\psi}^{\dagger}_{{\bf n},\alpha})^{N_{{\bf n},\alpha}}}
        {\sqrt{N_{{\bf n},\alpha}!}} |0\rangle~,
\end{equation}
with $|0\rangle$ the vacuum state containing no atoms. Second,
the hamiltonian is 
\begin{eqnarray}
&& \hspace*{-0.17in} 
\hat{H} = \sum_{{\bf n},\alpha} \epsilon_{{\bf n},\alpha} 
          \hat{\psi}^{\dagger}_{{\bf n},\alpha} \hat{\psi}_{{\bf n},\alpha} \\
&&+ \frac{1}{2} \sum_{\alpha,\alpha'} 
                 \sum_{{\bf n},{\bf n}',{\bf m},{\bf m}'}
      V_{{\bf n},{\bf n}';{\bf m},{\bf m}'} 
         \hat{\psi}^{\dagger}_{{\bf n},\alpha}  
         \hat{\psi}^{\dagger}_{{\bf n}',\alpha'}
         \hat{\psi}_{{\bf m}',\alpha'} \hat{\psi}_{{\bf m},\alpha}~, \nonumber
\end{eqnarray}
with 
$\epsilon_{{\bf n},\alpha} = \epsilon_{\bf n}
      - \gamma m_s B^{\rm eff} \equiv \epsilon_{\bf n} + \epsilon_{\alpha}$
the one-particle energies. Furthermore,  
\begin{equation}
V_{{\bf n},{\bf n}';{\bf m},{\bf m}'} = 
   \int d{\bf x} \int d{\bf x}'~ 
      \chi^*_{\bf n}({\bf x})\chi^*_{{\bf n}'}({\bf x}')V({\bf x}-{\bf x'})
          \chi_{\bf m}({\bf x}) \chi_{{\bf m}'}({\bf x}')
\end{equation}
is the amplitude for a collision which scatters two atoms out of the states 
$|{\bf m},\alpha\rangle$ and $|{\bf m}',\alpha'\rangle$ into the states 
$|{\bf n},\alpha\rangle$ and $|{\bf n}',\alpha'\rangle$, as schematically shown 
in figure 1(a). 

\begin{figure}
\begin{center}
\includegraphics[height=0.30\hsize]{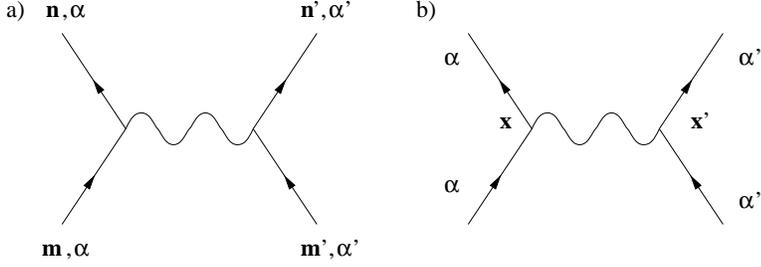}
\end{center}
\caption{Diagrammatic representation of the interaction terms in a) equation
         (2.11) and b) equation (2.13).}
\end{figure}

Introducing the field operators 
$\hat{\psi}_{\alpha}({\bf x}) = \sum_{{\bf n}} \hat{\psi}_{{\bf n},\alpha}
                                               \chi_{\bf n}({\bf x})$ and 
$\hat{\psi}^{\dagger}_{\alpha}({\bf x}) = \sum_{{\bf n}}   
    \hat{\psi}^{\dagger}_{{\bf n},\alpha} \chi^*_{\bf n}({\bf x})$, 
that annihilate and create atoms in the spin state $|\alpha\rangle$ at position 
${\bf x}$ respectively, we can rewrite this result into 
\begin{eqnarray}
\label{hamilton}
&& \hspace*{-0.17in}
\hat{H} = \sum_{\alpha} \int d{\bf x}~\hat{\psi}^{\dagger}_{\alpha}({\bf x})
\left\{ - \frac{\hbar^2 \mbox{\boldmath $\nabla$}^2}{2m} 
        + V^{\rm ex}({\bf x}) + \epsilon_{\alpha}
\right\} \hat{\psi}_{\alpha}({\bf x}) \\
&&+ \frac{1}{2}
\sum_{\alpha,\alpha'} \int d{\bf x} \int d{\bf x}'~
  \hat{\psi}^{\dagger}_{\alpha}({\bf x})
  \hat{\psi}^{\dagger}_{\alpha'}({\bf x}') V({\bf x}-{\bf x}')
  \hat{\psi}_{\alpha'}({\bf x}') \hat{\psi}_{\alpha}({\bf x})~. \nonumber
\end{eqnarray}
Note that, due to the (anti)commutation relations of the creation and 
annihilation operators and the orthogonality of the wave functions 
$\chi_{\bf n}({\bf x})$, the field operators obey 
$[\hat{\psi}_{\alpha}({\bf x}), 
            \hat{\psi}_{\alpha'}({\bf x}')]_{\mp}=
 [\hat{\psi}^{\dagger}_{\alpha}({\bf x}), 
            \hat{\psi}^{\dagger}_{\alpha'}({\bf x}')]_{\mp}= 0$ and           
$[\hat{\psi}_{\alpha}({\bf x}), 
            \hat{\psi}^{\dagger}_{\alpha'}({\bf x}')]_{\mp}=      
               \delta({\bf x}-{\bf x}')\delta_{\alpha,\alpha'}$.
Moreover, it is also important for the following to note that the number 
operator is
\begin{equation}
\label{number}
\hat{N} = \sum_{{\bf n},\alpha} 
            \hat{\psi}^{\dagger}_{{\bf n},\alpha} \hat{\psi}_{{\bf n},\alpha}
  = \sum_{\alpha} \int d{\bf x}~\hat{\psi}^{\dagger}_{\alpha}({\bf x})
                                \hat{\psi}_{\alpha}({\bf x})
\end{equation}
and, similarly, that the effective total spin operator is
\begin{equation}
\hat{{\bf S}} = \sum_{{\bf n},\alpha,\alpha'} 
           \hat{\psi}^{\dagger}_{{\bf n},\alpha}
                \langle \alpha|\hat{\bf s}|\alpha'\rangle    
                                            \hat{\psi}_{{\bf n},\alpha'}
= \sum_{\alpha,\alpha'} \int d{\bf x}~
               \hat{\psi}^{\dagger}_{\alpha}({\bf x})
                    \langle \alpha|\hat{\bf s}|\alpha'\rangle    
                          \hat{\psi}_{\alpha'}({\bf x})~.
\end{equation}
The density of atoms in the state $|\alpha\rangle$ is thus simply 
$\hat{n}_{\alpha}({\bf x}) = \hat{\psi}^{\dagger}_{\alpha}({\bf x})
                                               \hat{\psi}_{\alpha}({\bf x})$.

This completes our brief discussion of the second quantization formalism. In 
principle, we could now proceed to develop an operator formulation of the 
quantum field theory of interest to us. Since the experimentally most important 
observables can be expressed as appropriate products of the field operators, as 
we have just seen, this would essentially amount to the study of the 
(imaginary) 
time evolution of the Heisenberg operator
$\hat{\psi}({\bf x},\tau) = 
e^{(\hat{H}-\mu \hat{N})\tau/\hbar} \hat{\psi}_{\alpha}({\bf x})
                                        e^{-(\hat{H}-\mu \hat{N})\tau/\hbar}$
at a fixed chemical potential $\mu$ \cite{fetter1}. Put differently, the 
desired 
quantum field theory would be defined by the Heisenberg equation of motion
\begin{equation}
\hbar\frac{\partial}{\partial \tau} \hat{\psi}_{\alpha}({\bf x},\tau) =
                  [\hat{H}-\mu \hat{N}, \hat{\psi}_{\alpha}({\bf x},\tau)]_-~,
\end{equation}
and we would need to solve this equation in a sufficiently accurate 
approximation. As mentioned previously, however, we here want to develop 
Feynman's `path-integral' formulation of the problem, which will turn out to be 
much more convenient for our purposes. To do so in a manner that is the same 
for both bosonic and fermionic atomic gases, we first need to introduce some 
mathematical background. 

\subsection{Grassmann variables and coherent states}
\label{GV}
We have seen that in the case of fermions, we need to make use of anticommuting 
creation and annihilation operators. This automatically builds in the Pauli 
principle in the theory, because it implies that
$(\hat{\psi}^{\dagger}_{{\bf n},\alpha})^2|0\rangle =0$ and thus that the 
occupation numbers $N_{{\bf n},\alpha}$ are restricted to be either $0$ or $1$. 
For reasons that will become clear in a moment, it is in that case also 
convenient to introduce anticommuting complex numbers or Grassmann variables. 
The simplest example is to have two such Grassmann variables, say $\phi$ and 
$\phi^*$. The set $\{1,\phi,\phi^*,\phi^*\phi\}$, and linear combinations 
thereof with complex coefficients, form then a so-called Grassmann algebra.

By definition we have $[\phi,\phi]_+ = [\phi,\phi^*]_+ = [\phi^*,\phi^*]_+ = 0$ 
and thus in particular $\phi^2 = {\phi^*}^2 = 0$. Therefore, the above set is 
indeed complete. The complex conjugation in this algebra is defined by 
$(\phi)^*=\phi^*$, $(\phi^*)^*=\phi$, and $(\phi^*\phi)^*=(\phi)^*(\phi^*)^* = 
\phi^*\phi$. Moreover, we can also define an analytic function on this algebra 
by
\begin{equation}
A(\phi^*,\phi) = a_{11} + a_{12}\phi + a_{21}\phi^* 
                 + a_{22}\phi^*\phi~.
\end{equation}
As a result, it is natural to define also a differentiation by
\begin{equation}
\frac{\partial}{\partial\phi} 
     A(\phi^*,\phi) = a_{12} - a_{22}\phi^*~.
\end{equation} 
To be more precise this is in fact a left differentiation and the minus sign 
occurs, because we need to permute $\phi^*$ and $\phi$ before we can 
differentiate with respect to $\phi$. So, similarly, we have
\begin{equation}
\frac{\partial}{\partial\phi^*} 
     A(\phi^*,\phi) = a_{21} + a_{22}\phi
\end{equation}
and
\begin{equation}
\frac{\partial^2}{\partial\phi^*\partial\phi} A(\phi^*,\phi) =
- \frac{\partial^2}{\partial\phi\partial\phi^*} A(\phi^*,\phi) =
                                                - a_{22}~.
\end{equation}

Next we also need integrations over these Grassmann variables. Note that since 
$\phi^2=0$ we have only two possible integrals,
namely $\int d\phi~1$ and $\int d\phi~ \phi$. We define these by
\begin{equation}
\int d\phi~1 = 0 
\end{equation}
and
\begin{equation}
\int d\phi~\phi = 1~.
\end{equation}
This means that integration is equivalent to differentation. The main reason 
for 
the above definition is that we want the integration to obey the usual rules of 
partial integration. In particular, this implies that 
\begin{equation}
\int d\phi~ \frac{\partial F(\phi)}{\partial \phi} = 0~,
\end{equation}
for any function $F(\phi) = f_{1} + f_{2}\phi$. It is clear that this condition 
requires that $\int d\phi~1 = 0$. The result of $\int d\phi~\phi$ is then 
solely 
a question of normalization. It turns out that we are primarily interested in 
integrals of the form
\begin{equation}
\int d\phi^*d\phi~ A(\phi^*,\phi) =
 \int d\phi^*d\phi~ (a_{11} + a_{12}\phi + a_{21}\phi^* 
                      + a_{22}\phi^*\phi) = - a_{22}
\end{equation}
as we will see in section \ref{FI}.

Clearly, we can now also consider the Grassmann algebra based on the variables 
$\phi_n$ and $\phi^*_n$ with $n=1,2,\dots,\infty$. What we will need in the 
following are gaussian integrals over these variables. It is not difficult to 
show with the above definitions that
\begin{equation}
\int \prod_n d\phi^*_n d\phi_n~ 
\exp \left\{ - \sum_{n,n'} \phi_n^* A_{n,n'} \phi_{n'} \right\}
  = {\rm det} A = e^{{\rm Tr}[\ln A]}~.
\end{equation}
If the variables $\phi_n$ and $\phi^*_n$ were just ordinary complex numbers we 
would in contrast have the result
\begin{equation}
\int \prod_n \frac{d\phi^*_n d\phi_n}{2\pi i}~ 
\exp \left\{ - \sum_{n,n'} \phi_n^* A_{n,n'} \phi_{n'} \right\}
  = \frac{1}{{\rm det} A} = e^{- {\rm Tr}[\ln A]}~.
\end{equation}
These last two results will be used many times in the following.

One immediate use of these Grassmann variables is that we can now consider 
eigenstates of the annihilation operator $\hat{\psi}_{{\bf n},\alpha}$, also 
when we are dealing with fermions. These eigenstates are called coherent states. 
Consider the state
\begin{equation}
|\phi_{{\bf n},\alpha}\rangle \equiv 
  (1-\phi_{{\bf n},\alpha} \hat{\psi}^{\dagger}_{{\bf n},\alpha})|0\rangle
  = \exp \left\{ -\phi_{{\bf n},\alpha} 
                \hat{\psi}^{\dagger}_{{\bf n},\alpha} \right\} |0\rangle~,
\end{equation}
where $\phi_{{\bf n},\alpha}$ is a Grassmann variable that also anticommutes 
with the creation and annihilation operators in our Fock space. Clearly we have 
that
\begin{eqnarray}
&& \hspace*{-0.47in}
\hat{\psi}_{{\bf n},\alpha} |\phi_{{\bf n},\alpha}\rangle =
    \hat{\psi}_{{\bf n},\alpha} |0\rangle 
  + \phi_{{\bf n},\alpha} \hat{\psi}_{{\bf n},\alpha} 
                          \hat{\psi}^{\dagger}_{{\bf n},\alpha} |0\rangle \\
&& \hspace*{0.2in}
= \phi_{{\bf n},\alpha} |0\rangle = \phi_{{\bf n},\alpha}
    (1-\phi_{{\bf n},\alpha} \hat{\psi}^{\dagger}_{{\bf n},\alpha})|0\rangle
    = \phi_{{\bf n},\alpha} |\phi_{{\bf n},\alpha}\rangle~, \nonumber
\end{eqnarray}
so $|\phi_{{\bf n},\alpha}\rangle$ is indeed an eigenstate of 
$\hat{\psi}_{{\bf n},\alpha}$ with eigenvalue $\phi_{{\bf n},\alpha}$. In
general we can now make the states 
\begin{equation}
|\phi\rangle = \exp \left\{- \sum_{{\bf n},\alpha} 
                 \phi_{{\bf n},\alpha} 
                 \hat{\psi}^{\dagger}_{{\bf n},\alpha} \right\} |0\rangle
\end{equation}
that obey 
$\hat{\psi}_{{\bf n},\alpha}|\phi\rangle = \phi_{{\bf n},\alpha}|\phi\rangle$. 
Introducing the Grassmann-valued field 
$\phi_{\alpha}({\bf x}) = \sum_{{\bf n}} \phi_{{\bf n},\alpha}
                                                  \chi_{\bf n}({\bf x})$,
the latter two relations can be rewritten as 
\begin{equation}
|\phi\rangle = \exp \left\{ - \sum_{\alpha} \int d{\bf x}~ 
              \phi_{\alpha}({\bf x}) 
              \hat{\psi}^{\dagger}_{\alpha}({\bf x}) \right\} |0\rangle
\end{equation}
and $\hat{\psi}_{\alpha}({\bf x})|\phi\rangle = \phi_{\alpha}({\bf x})
                                                                |\phi\rangle$.
                                                                
It is important to note that these coherent states are not orthonormal. In
contrast, we find that 
\begin{eqnarray}
&& \hspace*{-0.4in}
\langle\phi|\phi'\rangle = \prod_{{\bf n},\alpha}
  \langle 0|(1-\hat{\psi}_{{\bf n},\alpha} \phi^*_{{\bf n},\alpha})
  (1-\phi'_{{\bf n},\alpha} \hat{\psi}^{\dagger}_{{\bf n},\alpha})|0\rangle \\
&&= \prod_{{\bf n},\alpha} (1 +
                 \phi^*_{{\bf n},\alpha} \phi'_{{\bf n},\alpha})
= \exp \left\{ \sum_{{\bf n},\alpha} \phi^*_{{\bf n},\alpha} 
                                \phi'_{{\bf n},\alpha} \right\}  \nonumber \\
&&= \exp \left\{ \sum_{\alpha} \int d{\bf x}~ 
       \phi^*_{\alpha}({\bf x}) \phi'_{\alpha}({\bf x}) \right\} 
  \equiv e^{(\phi|\phi')}~. \nonumber
\end{eqnarray}
Nevertheless they obey a closure relation, as can be seen explicitly from
\begin{eqnarray}
&& \hspace*{-0.2in}
\int \prod_{{\bf n},\alpha}
       d\phi^*_{{\bf n},\alpha} d\phi_{{\bf n},\alpha}
   \exp \left\{- \sum_{{\bf n},\alpha} \phi^*_{{\bf n},\alpha} 
                                \phi_{{\bf n},\alpha} \right\}
   |\phi\rangle\langle\phi| \\
&&= \prod_{{\bf n},\alpha} 
    \int d\phi^*_{{\bf n},\alpha} d\phi_{{\bf n},\alpha}~
    (1 - \phi^*_{{\bf n},\alpha} \phi_{{\bf n},\alpha})
(1-\phi_{{\bf n},\alpha} \hat{\psi}^{\dagger}_{{\bf n},\alpha})|0\rangle
\langle 0|(1-\hat{\psi}_{{\bf n},\alpha} \phi^*_{{\bf n},\alpha})
                                     \nonumber \\
&&= \prod_{{\bf n},\alpha} \left( |0\rangle\langle 0| +
                           |1\rangle\langle 1| \right) = \hat{1}~. \nonumber
\end{eqnarray}
We write this from now on simply as 
\begin{equation}
\int d[\phi^*]d[\phi]~e^{-(\phi|\phi)} |\phi\rangle\langle\phi| = \hat{1}~.
\end{equation}

The interesting observation at this point is that essentially the same 
formulaes also hold for bosons \cite{mandel,NO}. We have only a minus sign 
difference in
\begin{equation}
|\phi\rangle = \exp \left\{ \sum_{\alpha} \int d{\bf x}~ 
              \phi_{\alpha}({\bf x}) 
              \hat{\psi}^{\dagger}_{\alpha}({\bf x}) \right\} |0\rangle~,
\end{equation}
but then it can still be easily shown that
$\hat{\psi}_{\alpha}({\bf x})|\phi\rangle =
                      \phi_{\alpha}({\bf x})|\phi\rangle$. A convenient way to 
do so is, for instance, by noting that the commutation relation 
$[\hat{\psi}_{\alpha}({\bf x}),\hat{\psi}^{\dagger}_{\alpha'}({\bf x}')]_- 
                          = \delta({\bf x}-{\bf x}') \delta_{\alpha,\alpha'}$ 
implies that $\hat{\psi}_{\alpha}({\bf x})$ acts as 
$\delta/\delta\hat{\psi}^{\dagger}_{\alpha}({\bf x})$ on these states.  
Furthermore, the same expressions hold for the overlap 
$\langle\phi|\phi'\rangle$ and the closure relation, if we define the 
integration measure by 
$\int d[\phi^*]d[\phi] \equiv \int \prod_{{\bf n},\alpha} 
                    d\phi^*_{{\bf n},\alpha} d\phi_{{\bf n},\alpha}/(2\pi i)$
in this case. 

Summarizing, we have thus found for bosons and fermions that
\begin{equation}
|\phi\rangle = \exp \left\{ \pm \sum_{\alpha} \int d{\bf x}~ 
              \phi_{\alpha}({\bf x}) 
              \hat{\psi}^{\dagger}_{\alpha}({\bf x}) \right\} |0\rangle~,
\end{equation}
\begin{equation}
\langle\phi|\phi'\rangle = e^{(\phi|\phi')}~,
\end{equation}
and 
\begin{equation}
\int d[\phi^*]d[\phi]~e^{-(\phi|\phi)} |\phi\rangle\langle\phi| = \hat{1}~.
\end{equation}
The last ingredient we need is that in terms of these coherent states the trace 
of an operator $\hat{O}$ over the Fock space can be expressed as
\begin{equation}
\label{trace}
{\rm Tr}[\hat{O}] = \int d[\phi^*]d[\phi]~e^{-(\phi|\phi)}
                 \langle\pm\phi|\hat{O}|\phi\rangle~.
\end{equation}
The minus sign in the fermionic case is easily understood from the fact that we 
then have $\langle\{N_{{\bf n},\alpha}\}|\phi\rangle 
           \langle\phi|\{N_{{\bf n},\alpha}\}\rangle =
           \langle-\phi|\{N_{{\bf n},\alpha}\}\rangle
           \langle\{N_{{\bf n},\alpha}\}|\phi\rangle$, 
due to the anticommuting nature of the Grassmann
variables. After this mathematical interlude we can now return to physics and 
to our goal of arriving at a unified treatment of bosonic and fermionic atomic 
gases.

\subsection{Functional integrals}
\label{FI}
We are interested in determining the equilibrium properties of a trapped atomic 
gas at some temperature $T$. From statistical physics we know that all these 
properties can be obtained from the grand-canonical partition function 
\begin{equation}
Z = {\rm Tr} \left[ e^{-\beta(\hat{H}-\mu \hat{N})} \right]~,
\end{equation}
where $\beta = 1/k_BT$ and $\mu$ the chemical potential. We thus need to 
evaluate this quantity. As mentioned previously, we want to do so by making use 
of quantum field theory and, in addition, of Feynman's path-integral approach 
to 
quantum mechanics. We thus want to write the partition function as a functional 
integral over time-dependent fields $\phi_{\alpha}({\bf x},\tau)$, just like 
the 
partition function of a single particle in an external potential can be written 
as a functional integral over all time-dependent paths ${\bf x}(\tau)$. This 
can 
indeed be achieved with the coherent states that we have introduced in the 
previous section. 

We start with using our formula (\ref{trace}) for the trace of an operator,
\begin{equation}
Z = \int d[\phi^*]d[\phi]~e^{-(\phi|\phi)}
            \langle\pm\phi| e^{-\beta(\hat{H}-\mu \hat{N})} |\phi\rangle~,
\end{equation}
and observe that we are thus faced with the task of calculating the matrix 
elements $\langle\phi_M| e^{-\beta(\hat{H}-\mu \hat{N})} |\phi_0\rangle$ with 
$\phi_0({\bf x})=\phi({\bf x})$ and $\phi_M^*({\bf x})=\pm\phi^*({\bf x})$. 
This 
is difficult in general but can be simplified in the following way. We first 
realize that the operator $e^{-\beta(\hat{H}-\mu \hat{N})}$ is identical to the 
quantum mechanical evolution operator 
$U(t,0) = e^{-i(\hat{H}-\mu \hat{N})t/\hbar}$ evaluated at $t=-i\hbar\beta$. 
Put 
differently, we thus want to calculate the matrix elements of the 
imaginary-time 
evolution operator $U(-i\tau,0)$ for $\tau = \hbar\beta$. To do so, we next 
split the time interval $[0,\hbar\beta]$ into $M$ pieces,
with $\tau_m=m\hbar\beta/M$ and $m=0,1,\dots,M$. So $\Delta\tau =
\hbar\beta/M$. The procedure is summarized in figure 2. 

\begin{figure}
\begin{center}
\includegraphics[height=0.23\hsize]{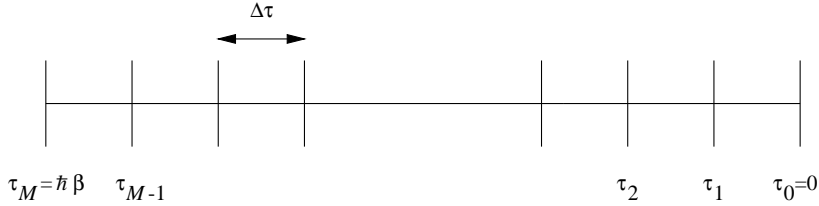}
\end{center}
\caption{Illustration of the slicing of the imaginary time interval  
         $[0,\hbar\beta]$ needed for Feynman's path-integral formulation
         of the partition function.}
\end{figure}

At each intermediate time $\tau_m$ we then apply a closure relation of the 
coherent states. This gives 
\begin{eqnarray}
&& \hspace*{-0.2in}
\langle\phi_M| e^{-\beta(\hat{H}-\mu \hat{N})} |\phi_0\rangle \\
&&= \int \left( \prod_{m=1}^{M-1} d[\phi^*_m]d[\phi_m]~
     e^{-(\phi_m|\phi_m)} \right) \prod_{m=1}^M \langle\phi_m|
     e^{-\Delta\tau(\hat{H}-\mu \hat{N})/\hbar} |\phi_{m-1}\rangle~. \nonumber
\end{eqnarray}
Now we can use that in the limit $M \rightarrow \infty$ we only need to know 
the 
latter matrix elements up to order $\Delta\tau$, because terms of order 
$(\Delta\tau)^2$ lead to corrections of order $M(\Delta\tau)^2 \propto 1/M$, 
which vanish in that limit. Hence
\begin{eqnarray}
\label{me}
&& \hspace*{-0.21in} 
\langle\phi_m| e^{-\Delta\tau(\hat{H}-\mu \hat{N})/\hbar} |\phi_{m-1}\rangle
\simeq \langle\phi_m| 1 -\Delta\tau(\hat{H}-\mu \hat{N})/\hbar 
                       |\phi_{m-1}\rangle \hspace*{0.4in} \\
&&\hspace*{3.55cm} \equiv \langle\phi_m|\phi_{m-1}\rangle 
   ( 1 - \Delta\tau H[\phi^*_m,\phi_{m-1}]/\hbar )~, \nonumber
\end{eqnarray}
with the grand-canonical hamiltonian functional resulting from equations 
(\ref{hamilton}) and (\ref{number}) equal to 
\begin{eqnarray}
&& \hspace*{-0.2in} 
H[\phi^*,\phi] = \sum_{\alpha} \int d{\bf x}~
 \phi^*_{\alpha}({\bf x})
  \left\{ - \frac{\hbar^2 \mbox{\boldmath $\nabla$}^2}{2m}         
                         + V^{\rm ex}({\bf x}) + \epsilon_{\alpha} - \mu
  \right\} \phi_{\alpha}({\bf x}) \hspace*{0.4in} \\
&&\hspace*{0.85cm} + \frac{1}{2}
  \sum_{\alpha,\alpha'} \int d{\bf x} \int d{\bf x}'~
           \phi^*_{\alpha}({\bf x})
           \phi^*_{\alpha'}({\bf x}') V({\bf x}-{\bf x}')
           \phi_{\alpha'}({\bf x}') \phi_{\alpha}({\bf x})~, \nonumber
\end{eqnarray}
since $\psi_{\alpha}({\bf x})|\phi\rangle =
         \phi_{\alpha}({\bf x})|\phi\rangle$ and 
$\langle\phi|\psi^{\dagger}_{\alpha}({\bf x}) = 
         \langle\phi|\phi^*_{\alpha}({\bf x})$.

Thus, neglecting terms of order $(\Delta\tau)^2$, we can again exponentiate the 
right-hand side of equation (\ref{me}), which leads in first instance to 
\begin{equation}
\langle\phi_m| e^{-\Delta\tau(\hat{H}-\mu \hat{N})/\hbar} |\phi_{m-1}\rangle
= e^{(\phi_m|\phi_{m-1})-\Delta\tau H[\phi^*_m,\phi_{m-1}]/\hbar}~,
\end{equation}
and therefore for the desired matrix element of the imaginary-time evolution 
operator to
\begin{eqnarray}
&& \hspace*{-0.4in} 
\langle\phi_M| e^{-\beta(\hat{H}-\mu \hat{N})} |\phi_0\rangle  
= \int \left( \prod_{m=1}^{M-1} d[\phi^*_m]d[\phi_m]~
     e^{-(\phi_m|\phi_m)} \right) \\
&& \hspace*{0.83in} \times
  \exp \left\{ \sum_{m=1}^{M} ((\phi_m|\phi_{m-1})-
                        \Delta\tau H[\phi^*_m,\phi_{m-1}]/\hbar)
      \right\}~. \nonumber   
\end{eqnarray}
This can then be manipulated into the suggestive form
\begin{eqnarray}
&& \hspace*{-0.2in} 
\langle\phi_M| e^{-\beta(\hat{H}-\mu \hat{N})} |\phi_0\rangle 
= e^{(\phi_M|\phi_M)} \int \left( \prod_{m=1}^{M-1} d[\phi^*_m]d[\phi_m]~  
                           \right) \\
&& \times \exp \left\{ - \frac{1}{\hbar} \sum_{m=1}^{M} \Delta\tau
\left( \hbar \frac{(\phi_m|\phi_{m})-(\phi_m|\phi_{m-1})}
                  {\Delta\tau} + H[\phi^*_m,\phi_{m-1}] \right)
      \right\}~. \nonumber 
\end{eqnarray} 

Taking now the continuum limit $M \rightarrow \infty$ and putting 
$\phi_m \equiv \phi(\tau_m)$, we find that
\begin{eqnarray}
\label{path1}
&& \hspace*{-0.5in}
\langle\phi_M| e^{-\beta(\hat{H}-\mu \hat{N})} |\phi_0\rangle \\
&& = e^{(\phi(\hbar\beta)|\phi(\hbar\beta))}
 \int_{\phi(0) = \phi_0}^{\phi^*(\hbar\beta) = \phi^*_M} d[\phi^*]d[\phi]~
                     e^{-S[\phi^*,\phi]/\hbar}~, \nonumber
\end{eqnarray}
where the so-called Euclidean action is 
\begin{eqnarray}
&& \hspace*{-0.2in}
S[\phi^*,\phi] \\
&&= \int_0^{\hbar\beta} d\tau~
  \left\{ \sum_{\alpha} \int d{\bf x}~
     \phi^*_{\alpha}({\bf x},\tau) 
               \hbar \frac{\partial}{\partial \tau}
                                   \phi_{\alpha}({\bf x},\tau)
    + H[\phi^*(\tau),\phi(\tau)] \right\}~. \nonumber
\end{eqnarray}
This is essentially the desired functional integral over the complex fields
$\phi_{\alpha}({\bf x},\tau)$ with the boundary conditions 
$\phi_{\alpha}({\bf x},0) = \phi_{0;\alpha}({\bf x})$ and
$\phi^*_{\alpha}({\bf x},\hbar\beta) =
                                  \phi^*_{M;\alpha}({\bf x})$. It is precisely 
the field theory analogue of the Feynman path integral. To obtain the partition 
function we only need to put $\phi_0({\bf x})$ equal to 
$\pm \phi_M({\bf x})$ and perform a last integration over 
$\phi_M({\bf x})$ and $\phi_M^*({\bf x})$. It then finally becomes
\begin{equation}
\label{path2}
Z = \int d[\phi^*]d[\phi]~ e^{-S[\phi^*,\phi]/\hbar} 
\end{equation}
with the boundary conditions 
$\phi_{\alpha}({\bf x},\hbar\beta) = \pm \phi_{\alpha}({\bf x},0)$,
i.e., the fields are periodic in $[0,\hbar\beta]$ for bosons and antiperiodic 
for fermions. Note that in equation (\ref{path1}) we have used the same 
notation 
for the integration measure as in equation (\ref{path2}), although there is in 
principle one more integration in the expression for the partition function. 
The 
difference is in the continuum limit accounted for in the boundary conditions, 
which are in practice usually left implicite. Having arrived at an exact 
identity between the partition function and a functional integral, we are now 
going to familiarize ourselves with this identity, and with how to perform 
functional integrals in general, by considering the ideal quantum gases.

\subsection{Ideal quantum gases}
Since the partition functions $Z_0$ of the ideal quantum gases are known 
exactly, they are ideal test cases for our field-theoretical methods. Moreover, 
a thorough knowledge of the ideal quantum gases is also an important first step 
in understanding experiments with trapped atomic gases, because the effects of 
the interatomic interaction can essentially be included perturbatively. How 
this 
perturbation theory is performed is discussed in detail in section \ref{IFD}, 
but before we can do that we need to understand the noninteracting gases first.
In that case, we have 
\begin{eqnarray}
&& \hspace*{-0.35in}
S_0[\phi^*,\phi] =  \sum_{\alpha} 
  \int_0^{\hbar\beta} d\tau \int d{\bf x}~ \\
&& \hspace*{0.2in} \times~ \phi^*_{\alpha}({\bf x},\tau) 
        \left\{ \hbar \frac{\partial}{\partial \tau} \right. 
        \left.  - \frac{\hbar^2 \mbox{\boldmath $\nabla$}^2}{2m} 
                + V^{\rm ex}({\bf x})
                + \epsilon_{\alpha} - \mu
        \right\} \phi_{\alpha}({\bf x},\tau) \nonumber
\end{eqnarray}
and the partition function is a gaussian integral, which explains why we were 
interested in gaussian integrals in section \ref{GV}. It will be illustrative 
to evaluate this partition function in three different ways.

\subsubsection{Semiclassical method}
In the first way, we perform the evaluation of the trace involved in the 
definition of the partition function at the end of the calculation, and start by 
considering the matrix element  
$\langle\pm\phi| e^{-\beta(\hat{H}-\mu \hat{N})} |\phi\rangle$ as the 
functional integral
\begin{eqnarray}
&& \hspace*{-0.5in}
\int_{\phi(0) = \phi}^{\phi^*(\hbar\beta) = \pm \phi^*} 
      \left( \prod_{{\bf n},\alpha}
          d[\phi^*_{{\bf n},\alpha}]d[\phi_{{\bf n},\alpha}] \right)~
  \exp \left\{ \sum_{{\bf n},\alpha}
       \phi^*_{{\bf n},\alpha}(\hbar\beta)
                              \phi_{{\bf n},\alpha}(\hbar\beta)
      \right\} 
       \nonumber \\ 
&& \times \exp \left\{- \frac{1}{\hbar} \int_0^{\hbar\beta} d\tau 
 \sum_{{\bf n},\alpha}
   \phi^*_{{\bf n},\alpha}(\tau)
        \left( \hbar \frac{\partial}{\partial \tau}
                 + \epsilon_{{\bf n},\alpha} - \mu
        \right) \phi_{{\bf n},\alpha}(\tau)
       \right\}~,                                \nonumber
\end{eqnarray}
which is the product for each ${\bf n}$ and $\alpha$ of the
path integral 
\begin{eqnarray}
\int d[\phi^*]d[\phi]~
  \exp \left\{ \phi^*(\hbar\beta) \phi(\hbar\beta)
             - \frac{1}{\hbar} \int_0^{\hbar\beta} d\tau~
   \phi^*(\tau)
        \left( \hbar \frac{\partial}{\partial \tau}
                 + \epsilon - \mu \right) \phi(\tau)
       \right\}                                \nonumber
\end{eqnarray}
with the boundary conditions $\phi(0) = \phi$ and 
$\phi(\hbar\beta) = \pm \phi^*$. It clearly corresponds to the matrix element 
$\langle\pm \phi| e^{-\beta(\epsilon-\mu)\hat{\psi}^{\dagger}\hat{\psi}}  
                                                                |\phi\rangle$ 
for one particular value of ${\bf n}$ and $\alpha$. We calculate this matrix 
element by performing a shift in the integration variables, i.e., 
$\phi(\tau) = \phi_{\rm cl}(\tau) + \xi(\tau)$ and 
$\phi^*(\tau) = \phi^*_{\rm cl}(\tau) + \xi^*(\tau)$, where
$\phi_{\rm cl}(\tau)$ obeys the `classical' equations of motion
\begin{equation}
\left. \frac{\delta S_0[\phi^*,\phi]}{\delta \phi^*(\tau)} 
 \right|_{\phi=\phi_{\rm cl}} = \left( \hbar \frac{\partial}{\partial \tau}
                 + \epsilon - \mu \right) \phi_{\rm cl}(\tau) =0
\end{equation}
and similarly for $\phi^*_{\rm cl}(\tau)$. The solutions with the correct 
boundary 
solutions are 
$\phi_{\rm cl}(\tau)= \phi e^{-(\epsilon-\mu)\tau/\hbar}$ and
$\phi^*_{\rm cl}(\tau)= 
          \pm \phi^* e^{(\epsilon-\mu)(\tau-\hbar\beta)/\hbar}$,
leading to the path integral
\begin{eqnarray}
&& \hspace*{-0.2in}
\exp \left\{ \pm e^{-\beta(\epsilon-\mu)} \phi^* \phi \right\} \nonumber \\ 
&& \times \int d[\xi^*]d[\xi]~
  \exp \left\{ - \frac{1}{\hbar} \int_0^{\hbar\beta} d\tau~
   \xi^*(\tau) \left( \hbar \frac{\partial}{\partial \tau}
                 + \epsilon - \mu \right) \xi(\tau)
       \right\}                                \nonumber
\end{eqnarray}
with the boundary conditions $\xi^*(\hbar\beta)=\xi(0)=0$. This means that the 
last path integral is just equal to
$\langle 0| e^{-\beta(\epsilon-\mu)\hat{\psi}^{\dagger}\hat{\psi}}|0\rangle = 
1$
and that the desired result is just the prefactor. 

Substituting this for each value of ${\bf n}$ and $\alpha$, we apparently have 
\begin{equation}
\langle\pm\phi| e^{-\beta(\hat{H}-\mu \hat{N})} |\phi\rangle =
\exp \left\{ \pm \sum_{{\bf n},\alpha} 
      e^{-\beta(\epsilon_{{\bf n},\alpha}-\mu)}
             \phi^*_{{\bf n},\alpha} \phi_{{\bf n},\alpha} \right\}
\end{equation}
and the partition function becomes the product
\begin{eqnarray}
&& \hspace*{-0.2in}
Z_0 = \prod_{{\bf n},\alpha} \int 
  \frac{d\phi^*_{{\bf k},\alpha}d\phi_{{\bf k},\alpha}}
       {(2\pi i)^{(1\pm 1)/2}}~
  \exp \left\{ - (1 \mp e^{-\beta(\epsilon_{{\bf n},\alpha}-\mu)})
             \phi^*_{{\bf n},\alpha} \phi_{{\bf n},\alpha} \right\} \\ 
&&= \prod_{{\bf n},\alpha} 
     (1 \mp e^{-\beta(\epsilon_{{\bf n},\alpha}-\mu)})^{\mp 1}
 = \exp \left\{ \mp \sum_{{\bf n},\alpha} 
    \ln (1 \mp e^{-\beta(\epsilon_{{\bf n},\alpha}-\mu)})
      \right\}~. \nonumber
\end{eqnarray}
This is the correct result, because from the usual thermodynamic identity
$\langle \hat{N} \rangle = \partial \ln Z_0/\partial(\beta\mu)$ we find 
\begin{equation}
\langle \hat{N} \rangle = \sum_{{\bf n},\alpha}
                 \frac{1}{e^{\beta(\epsilon_{{\bf n},\alpha}-\mu)} \mp 1}~,
\end{equation}
as desired. 

\subsubsection{Matsubara expansion}
The second way is easier and more common in practice. We immediately start with
\begin{equation}
Z_0 = \int d[\phi^*]d[\phi]~ e^{-S_0[\phi^*,\phi]/\hbar} 
\end{equation}
and incorporate the boundary conditions by expanding the fields
as
\begin{equation}
\phi_{\alpha}({\bf x},\tau) =
  \sum_{{\bf n},n} \phi_{{\bf n},n,\alpha} \chi_{\bf n}({\bf x})
    \frac{e^{-i\omega_n\tau}}
         {\sqrt{\hbar\beta}}
\end{equation}
where $\omega_n = \pi(2n)/\hbar\beta$ for bosons and 
$\omega_n = \pi(2n+1)/\hbar\beta$ for fermions. These are known as the even and 
odd Matsubara frequencies, respectively. Using this expansion we have
\begin{eqnarray}
&& \hspace*{-0.2in}
Z_0 = \int \prod_{{\bf n},n,\alpha}  
  \frac{d\phi^*_{{\bf n},n,\alpha}d\phi_{{\bf n},n,\alpha}}
       {(2\pi i)^{(1\pm 1)/2}} \frac{1}{(\hbar\beta)^{\pm 1}}~ \\
&&\times \exp \left\{ - \frac{1}{\hbar} \sum_{{\bf n},n,\alpha}
         \phi^*_{{\bf n},n,\alpha}
            (-i\hbar\omega_n + \epsilon_{{\bf n},\alpha} - \mu)
              \phi_{{\bf n},n,\alpha}
      \right\}~, \nonumber
\end{eqnarray}
if we also take account of the jacobian involved in the change of integration 
variables. Note that the difference between the jacobians in the bosonic and 
fermionic case, is a consequence of the fact that for Grassmann variables we 
have that $\int d\phi f_2\phi = f_2 = \int d(f_2\phi) f_2 (f_2\phi)$ instead of 
the result $\int d(f_2\phi) (1/f_2) (f_2\phi)$ that we expect on the basis of 
ordinary complex integration. Note also that these are all again gaussian 
integrals, so we find in first instance
\begin{eqnarray}
&& \hspace*{-0.2in}
Z_0 = \prod_{{\bf n},n,\alpha}
        (\beta(-i\hbar\omega_n 
               + \epsilon_{{\bf n},\alpha} - \mu))^{\mp 1} \\
&&= \exp \left\{ \mp \sum_{{\bf n},n,\alpha} 
         \ln( \beta(-i\hbar\omega_n 
               + \epsilon_{{\bf n},\alpha} - \mu) ) \right\}~. \nonumber
\end{eqnarray}

To evaluate the sum over Matsubara frequencies we need to add a convergence 
factor $e^{i\omega_n\eta}$ and finally take the limit $\eta \downarrow 0$. The 
precise reason for this particular procedure cannot be fully understood at this 
point but is explained in section \ref{IFD}. However, doing so we indeed find 
that
\begin{equation}
\lim_{\eta \downarrow 0} \sum_n \ln( \beta(-i\hbar\omega_n 
               + \epsilon - \mu) ) e^{i\omega_n\eta}
  = \ln(1 \mp e^{-\beta(\epsilon-\mu)})~.
\end{equation}
To see that this is correct, we differentiate the latter equation with respect 
to $\beta\mu$. This gives us
\begin{equation}
\lim_{\eta \downarrow 0} \frac{1}{\hbar\beta} \sum_n 
    \frac{e^{i\omega_n\eta}} 
         {i\omega_n - (\epsilon - \mu)/\hbar} = 
     \mp \frac{1}{e^{\beta(\epsilon-\mu)} \mp 1}~,                  
\end{equation}
which can be proved by contour integration in the following way. The function
$\hbar\beta/(e^{\hbar\beta z} \mp 1)$ has poles at the even and odd Matsubara 
frequencies with residu $\pm 1$. Hence, by Cauchy's theorem the left-hand side 
is equal to
\begin{eqnarray}
\lim_{\eta \downarrow 0} \frac{1}{2\pi i} \int_C dz~ 
   \frac{e^{\eta z}}{z-(\epsilon-\mu)/\hbar}~
   \frac{\pm 1}{e^{\hbar\beta z} \mp 1}
                                          \nonumber
\end{eqnarray} 
with $C$ a contour that fully encloses the imaginary axis in the direction 
shown in figure 3. Adding to the contour $C$ the contour $C'$, which gives no 
contribution to the integral, we can again apply Cauchy's theorem to obtain the 
desired result. Why does the integration over the contour $C'$ vanish? The 
reason for this is that the integrant behaves as 
$\pm e^{-(\hbar\beta-\eta){\rm Re}(z)}/|z|$ for ${\rm Re}(z) \rightarrow 
                                                                      \infty$ 
and as $-e^{\eta{\rm Re}(z)}/|z|$ for ${\rm Re}(z) \rightarrow -\infty$. The 
integrant thus always vanishes much faster than $1/|z|$ on the contour $C'$ for 
any $0<\eta<\hbar\beta$.

\begin{figure}
\begin{center}
\includegraphics[height=0.5\hsize]{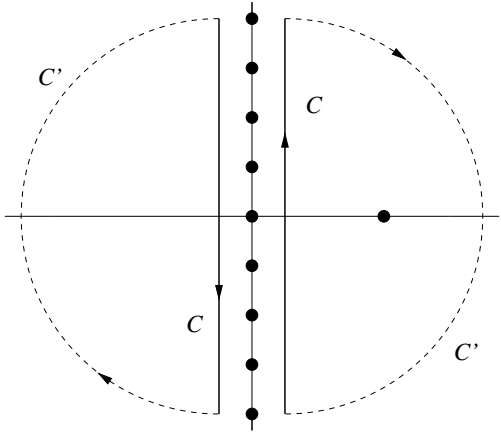}
\end{center}
\caption{Illustration of the contour integration that is required to 
         perform the summation over the Matsubara frequencies. The black
         dots indicate the position of the poles in the integrant.}
\end{figure}

\subsubsection{Green's function method}
The third and last method is simplest and, therefore, most common in practice. 
We first rewrite the partition function as
\begin{eqnarray}
&& \hspace*{-0.2in}
Z_0 = \int d[\phi^*]d[\phi]~ 
 \exp \left\{ 
      \sum_{\alpha} \int_0^{\hbar\beta} d\tau \int d{\bf x} \right. \\
&& \times \left.
      \sum_{\alpha'} \int_0^{\hbar\beta} d\tau' \int d{\bf x}'~
       \phi^*_{\alpha}({\bf x},\tau) 
          G^{-1}_{\alpha,\alpha'}({\bf x},\tau;{\bf x}',\tau')
        \phi_{\alpha'}({\bf x}',\tau')
      \right\} \nonumber
\end{eqnarray}
and see $G^{-1}$ as a `matrix' both in spin space as in coordinate space. We 
then know that this gaussian integral is just
\begin{equation}
Z_0 = [{\rm det}(-G^{-1})]^{\mp 1} = \exp \left\{ \mp {\rm Tr}[\ln (-G^{-1})]
                                          \right\}~.
\end{equation}
Clearly, we have from the action $S_0[\phi^*,\phi]$ that
\begin{eqnarray}
&& \hspace*{-0.2in}
G^{-1}_{\alpha,\alpha'}({\bf x},\tau;{\bf x}',\tau') \\
&&= - \frac{1}{\hbar} 
     \left\{ \hbar \frac{\partial}{\partial \tau}
                 - \frac{\hbar^2 \mbox{\boldmath $\nabla$}^2}{2m} 
                 + V^{\rm ex}({\bf x}) + \epsilon_{\alpha} - \mu
     \right\} \delta({\bf x}-{\bf x}') \delta(\tau-\tau')
              \delta_{\alpha,\alpha'} \nonumber
\end{eqnarray}
or equivalently that
\begin{eqnarray}
\left\{ \hbar \frac{\partial}{\partial \tau}
                 - \frac{\hbar^2 \mbox{\boldmath $\nabla$}^2}{2m} 
                 + V^{\rm ex}({\bf x}) + \epsilon_{\alpha} - \mu   \right\}
G_{\alpha,\alpha'}({\bf x},\tau;{\bf x}',\tau') 
                                                 \hspace*{0.8in} \\
= -\hbar \delta({\bf x}-{\bf x}') \delta(\tau-\tau')
       \delta_{\alpha,\alpha'}~, \nonumber
\end{eqnarray}
which means that 
$G_{\alpha,\alpha'}({\bf x},\tau;{\bf x}',\tau')$ is a Green's function. We 
come 
back to its physical meaning shortly. For now we only need to see that the last 
equation is solved by 
\begin{eqnarray}
\label{green}
&& \hspace*{-0.2in}
G_{\alpha,\alpha'}({\bf x},\tau;{\bf x}',\tau') \\
&&= \delta_{\alpha,\alpha'} 
  \sum_{{\bf n},n} 
       \frac{-\hbar}
            {-i\hbar\omega_n + \epsilon_{{\bf n},\alpha} - \mu}
            \chi_{\bf n}({\bf x}) \chi^*_{\bf n}({\bf x}')
    \frac{e^{-i\omega_n (\tau-\tau')}}{\hbar\beta} \nonumber
\end{eqnarray}
thus again
\begin{eqnarray}
&& \hspace*{-0.2in}
Z_0 = \exp \left\{ \mp \sum_{{\bf n},n,\alpha} 
         \ln( \beta(-i\hbar\omega_n 
               + \epsilon_{{\bf n},\alpha} - \mu) ) \right\} \\
&&= \exp \left\{ \mp \sum_{{\bf n},\alpha} 
         \ln(1 \mp e^{-\beta(\epsilon_{{\bf n},\alpha}-\mu)})
      \right\}~. \nonumber
\end{eqnarray}

In principle, we have slightly cheated in the last step of this derivation, 
because equation (\ref{green}) shows that 
\begin{equation}
G_{{\bf n},n,\alpha;{\bf n}',n',\alpha'}
  = \delta_{{\bf n},{\bf n}'} \delta_{n,n'} \delta_{\alpha,\alpha'}
       \frac{-\hbar}{-i\hbar\omega_n + \epsilon_{{\bf n},\alpha} - \mu}~.
\end{equation}
The problem with this last way of calculating the functional integral is that 
it 
does not correctly account for the jacobian involved in the change of variables 
from imaginary time $\tau$ to the Matsubara frequencies $\omega_n$. However, 
this problem can be avoided by calculating never a single determinant, but 
always the ratio of two determinants. In this manner the effect of the jacobian 
is exactly cancelled. We will see some examples of how this works in sections 
\ref{HFT} and \ref{SC}.

We now return to the physical meaning of the above Green's function. To see its 
meaning, we first consider the time-ordered expectation value 
\begin{eqnarray}
&& \hspace*{-0.2in}
\langle T[\hat{\psi}_{\alpha}({\bf x},\tau)
          \hat{\psi}^{\dagger}_{\alpha'}({\bf x}',\tau')] \rangle  \\
&&\equiv \theta(\tau-\tau') 
  \langle \hat{\psi}_{\alpha}({\bf x},\tau)
          \hat{\psi}^{\dagger}_{\alpha'}({\bf x}',\tau') \rangle   
\pm \theta(\tau'-\tau)
  \langle \hat{\psi}^{\dagger}_{\alpha'}({\bf x}',\tau')
          \hat{\psi}_{\alpha}({\bf x},\tau) \rangle~. \nonumber
\end{eqnarray}
Here the expectation value is taken in the grand-canonical ensemble and 
$\hat{\psi}_{\alpha}({\bf x},\tau)$ is the imaginary time 
Heisenberg operator, which is defined by
$e^{(\hat{H}-\mu \hat{N})\tau/\hbar} \hat{\psi}_{\alpha}({\bf x})
                                        e^{-(\hat{H}-\mu \hat{N})\tau/\hbar}$
and therefore obeys the Heisenberg equation of motion
$\hbar\partial_{\tau} \hat{\psi}_{\alpha}({\bf x},\tau) =
                  [\hat{H}-\mu \hat{N}, \hat{\psi}_{\alpha}({\bf x},\tau)]_-$.
For the noninteracting case it reads
\begin{equation}
\hbar \frac{\partial}{\partial\tau} \hat{\psi}_{\alpha}({\bf x},\tau) 
 = \left( \frac{\hbar^2\mbox{\boldmath $\nabla$}^2}{2m} 
          - V^{\rm ex}({\bf x}) - \epsilon_{\alpha} + \mu
   \right) \hat{\psi}_{\alpha}({\bf x},\tau)~.
\end{equation}
As a result
\begin{eqnarray}
&& \hspace*{-0.2in}
\hbar \frac{\partial}{\partial\tau}
\langle T[\hat{\psi}_{\alpha}({\bf x},\tau)
          \hat{\psi}^{\dagger}_{\alpha'}({\bf x}',\tau')] \rangle
= \hbar \delta(\tau-\tau') 
\langle [\hat{\psi}_{\alpha}({\bf x},\tau),
       \hat{\psi}^{\dagger}_{\alpha'}({\bf x}',\tau')]_{\mp} \rangle 
                                                            \hspace*{0.4in} \\
&&\hspace*{1.8cm} +
    \left( \frac{\hbar^2\mbox{\boldmath $\nabla$}^2}{2m} - V^{\rm ex}({\bf x})
           - \epsilon_{\alpha} + \mu
    \right)
\langle T[\hat{\psi}_{\alpha}({\bf x},\tau)
          \hat{\psi}^{\dagger}_{\alpha'}({\bf x}',\tau')] \rangle~. \nonumber
\end{eqnarray}
Substituting the equal-time relations that we derived in section \ref{SQ}, 
i.e.,
$[\hat{\psi}_{\alpha}({\bf x},\tau),
       \hat{\psi}^{\dagger}_{\alpha'}({\bf x}',\tau)]_{\mp} =
                            \delta({\bf x}-{\bf x}') \delta_{\alpha,\alpha'}$,
thus suggests that
\begin{equation}
G_{\alpha,\alpha'}({\bf x},\tau;{\bf x}',\tau') =
- \langle T[\hat{\psi}_{\alpha}({\bf x},\tau)
          \hat{\psi}^{\dagger}_{\alpha'}({\bf x}',\tau')] \rangle~.
\end{equation}

We can actually also prove this important relation, that bridges the gap between 
the functional formulation of quantum field theory used here and the more 
familiar operator formalism. First of all, it is clear from the slicing 
procedure used in our derivation of the functional integral in section \ref{FI}, 
that it automatically leads to time-ordered expectation values. We should, 
therefore, only be able to prove that
\begin{eqnarray}
&& \hspace*{-0.47in}
- G_{\alpha,\alpha'}({\bf x},\tau;{\bf x}',\tau')
   = \langle \phi_{\alpha}({\bf x},\tau)
                  \phi^*_{\alpha'}({\bf x}',\tau') \rangle \\
&& \hspace*{0.7in} \equiv \frac{1}{Z_0} \int d[\phi^*]d[\phi]~
   \phi_{\alpha}({\bf x},\tau) \phi^*_{\alpha'}({\bf x}',\tau')
   e^{-S_0[\phi^*,\phi]/\hbar}~. \nonumber
\end{eqnarray}
This is most easily achieved in the following way. We introduce a partition 
function in the presence of external currents $J_{\alpha}({\bf x},\tau)$ and 
$J^*_{\alpha}({\bf x},\tau)$, where in the fermionic case these currents are 
also Grassmann variables. The partition function is taken to be
\begin{eqnarray}
&& \hspace*{-0.4in}
Z_0[J,J^*] = \int d[\phi^*]d[\phi]~
 \exp \left\{ - \frac{1}{\hbar} S_0[\phi^*,\phi] 
                                  \raisebox{0.25in}{} \right. \\
&&\hspace*{0.15in}
  + \left. \sum_{\alpha} \int_0^{\hbar\beta} d\tau \int d{\bf x}~
       ( \phi^*_{\alpha}({\bf x},\tau) J_{\alpha}({\bf x},\tau)
       + J^*_{\alpha}({\bf x},\tau) \phi_{\alpha}({\bf x},\tau) )
     \right\}~, \nonumber
\end{eqnarray}
since then we must simply prove that
\begin{equation}
- G_{\alpha,\alpha'}({\bf x},\tau;{\bf x}',\tau') =
  \frac{\pm 1}{Z_0} 
  \left.  \frac{\delta^2 Z_0}
               {\delta J^*_{\alpha}({\bf x},\tau)
                \delta J_{\alpha'}({\bf x}',\tau')} 
  \right|_{J,J^*=0}~.
\end{equation}

Using a short-hand notation we have 
\begin{equation}
Z_0[J,J^*] = \int d[\phi^*]d[\phi]~
 \exp \left\{ (\phi|G^{-1}|\phi) + (\phi|J) + (J|\phi) \right\}~.
\end{equation}
The terms in the exponent can be rewritten as $(\phi+JG|G^{-1}|\phi+GJ) - 
(J|G|J)$, which is usually called completing the square. Performing a shift in 
the integration variables, we then easily see that
\begin{eqnarray}
&& \hspace*{-0.5in}
Z_0[J,J^*] = Z_0[0,0] e^{-(J|G|J)}  
 = Z_0[0,0] \exp \left\{
  - \sum_{\alpha} \int_0^{\hbar\beta} d\tau \int d{\bf x} 
                 \right. \hspace*{0.4in} \\  
&& \hspace*{0.08in} \times \left.  
    \sum_{\alpha'} \int_0^{\hbar\beta} d\tau'  \int d{\bf x}'~
    J^*_{\alpha}({\bf x},\tau)   
      G_{\alpha,\alpha'}({\bf x},\tau;{\bf x}',\tau')
    J_{\alpha'}({\bf x}',\tau')
   \right\}~, \nonumber
\end{eqnarray}
which after differentiation indeed leads to the desired result. We can now in 
fact calculate the expectation value of the time-ordered product of any number 
of operators. With the above expression for $Z_0[J,J^*]$ one can easily prove 
that this results in the sum of all possible products of time-ordered 
expectation values of two operators. For instance
\begin{eqnarray}
&& \hspace*{-0.5in} 
\langle \phi^*_{\alpha}({\bf x},\tau) \phi^*_{\alpha'}({\bf x}',\tau') 
        \phi_{\alpha''}({\bf x}'',\tau'') \phi_{\alpha'''}({\bf x}''',\tau''')  
\rangle \\
&&=
\langle \phi^*_{\alpha}({\bf x},\tau) \phi_{\alpha'''}({\bf x}''',\tau''') 
\rangle \langle \phi^*_{\alpha'}({\bf x}',\tau') 
                \phi_{\alpha''}({\bf x}'',\tau'') \rangle  \nonumber \\
&&\pm \langle \phi^*_{\alpha}({\bf x},\tau) \phi_{\alpha''}({\bf x}'',\tau'') 
\rangle \langle \phi^*_{\alpha'}({\bf x}',\tau') 
                \phi_{\alpha'''}({\bf x}''',\tau''') \rangle~. \nonumber
\end{eqnarray}
This is the famous Wick's theorem, which plays a crucial role in the next 
section where we start to discuss the profound effects that interactions can 
have on the results obtained thusfar.

\subsection{Interactions and Feynmann diagrams}
\label{IFD}
The Green's function, or one-particle propagator,  
$G_{\alpha,\alpha'}({\bf x},\tau;{\bf x}',\tau')$ is one of the most important 
quantities that we want to determine theoretically, because it gives us the 
possibility to calculate in principle the expectation value of any one-particle 
observable. It also gives us the elementary excitations of our system of 
interest. For example, the average density of the spin state $|\alpha\rangle$ 
is in an ideal gas given by
\begin{eqnarray}
&& \hspace*{-0.2in}
\langle \hat{\psi}^{\dagger}_{\alpha}({\bf x},\tau)
          \hat{\psi}_{\alpha}({\bf x},\tau) \rangle
= \mp G_{\alpha,\alpha}({\bf x},\tau;{\bf x},\tau^+) \\
&&= \lim_{\eta \downarrow 0} 
      \sum_{{\bf n},n} \frac{\pm e^{i\omega_n\eta}}{\beta(-i\hbar\omega_n 
                            + \epsilon_{{\bf n},\alpha} - \mu)}
    |\chi_{\bf n}({\bf x})|^2                          
= \sum_{{\bf n}} 
              \frac{1}{e^{\beta(\epsilon_{{\bf n},\alpha} - \mu)}
                  \mp 1}|\chi_{\bf n}({\bf x})|^2~, \nonumber
\end{eqnarray}
where we introduced the notation $\tau^+$ for the limit 
$\eta \downarrow 0$ of $\tau +\eta$. Note that this procedure is necessary due 
to the time-ordening involved in the definition of the Green's function and the 
fact that the field operators do not commute at equal times. 
It also gives a natural explanation for our previous use of the convergence 
factor $e^{i\omega_n\eta}$. Moreover, from the argument of the Bose or Fermi 
distribution function, we see that the elementary excitations have an energy of 
$\epsilon_{{\bf n},\alpha} - \mu$. If we replace in the Fourier transform of 
the Green's function, i.e., in 
\begin{equation}
G_{\alpha,\alpha'}({\bf x},{\bf x}';i\omega_n) =
- \hbar \delta_{\alpha,\alpha'} \sum_{\bf n}  
     \frac{1}{-i\hbar\omega_n + \epsilon_{{\bf n},\alpha} - \mu}
          \chi_{\bf n}({\bf x}) \chi^*_{\bf n}({\bf x}')~,
\end{equation}
$i\omega_n$ by $\omega$ we also see that 
$G_{\alpha,\alpha'}({\bf x},{\bf x}';\omega)$ has a pole at 
$\hbar\omega = \epsilon_{{\bf n},\alpha} - \mu$. This is a general result: 
Poles 
in $G_{\alpha,\alpha'}({\bf x},{\bf x}';\omega)$ correspond to the energies of 
the elementary excitations. These energies in general can also have a negative 
imaginary components, which then correspond to the lifetime of the excitation. 
The question that arises, therefore, is how to determine also the Green's 
function for an interacting system.

This can be done in perturbation theory, as follows. We want to calculate the 
expectation value
\begin{eqnarray}
\label{expec}
&& \hspace*{-0.5in}
- \langle \phi_{\alpha}({\bf x},\tau)
          \phi^*_{\alpha'}({\bf x}',\tau') \rangle  \\
&&= - \frac{1}{Z} \int d[\phi^*]d[\phi]~
   \phi_{\alpha}({\bf x},\tau) \phi^*_{\alpha'}({\bf x}',\tau')
   e^{-S[\phi^*,\phi]/\hbar} \nonumber
\end{eqnarray}
with the action $S[\phi^*,\phi]=S_0[\phi^*,\phi]+S_{\rm int}[\phi^*,\phi]$. We 
now expand both the numerator and the denominator in powers of $S_{\rm int}$. 
Up to first order we find for the partition function
\begin{eqnarray}
\label{z1}
&& \hspace*{-0.15in}
Z = \int d[\phi^*]d[\phi]~e^{-S_0[\phi^*,\phi]/\hbar}
  \left( 1 - \frac{S_{\rm int}[\phi^*,\phi]}{\hbar} \right) \\
&&\equiv Z_0 \left( 1 - \frac{1}{\hbar} 
                     \langle S_{\rm int}[\phi^*,\phi] \rangle_0
        \right)~. \nonumber
\end{eqnarray}
Using Wick's theorem, we thus have 
\begin{eqnarray}
\label{z2}
&& \hspace*{-0.5in}
- \frac{1}{\hbar} \langle S_{\rm int}[\phi^*,\phi] \rangle_0 
= \frac{1}{2} \sum_{\alpha,\alpha'} \int_0^{\hbar\beta} d\tau
  \int d{\bf x} \int d{\bf x}'~
G_{0;\alpha,\alpha}({\bf x},\tau;{\bf x},\tau^+) \\
&&\hspace*{1.5in} \times~ \frac{- V({\bf x}-{\bf x}')}{\hbar}
     G_{0;\alpha',\alpha'}({\bf x}',\tau;{\bf x}',\tau^+) \nonumber \\
&&\hspace*{0.52in} \pm \frac{1}{2} \sum_{\alpha} \int_0^{\hbar\beta} d\tau
  \int d{\bf x} \int d{\bf x}'~
G_{0;\alpha,\alpha}({\bf x}',\tau;{\bf x},\tau^+) \nonumber \\
&&\hspace*{1.5in} \times~ \frac{- V({\bf x}-{\bf x}')}{\hbar}
     G_{0;\alpha,\alpha}({\bf x},\tau;{\bf x}',\tau^+)~. \nonumber
\end{eqnarray}

\begin{figure}
\begin{center}
\includegraphics[height=0.16\hsize]{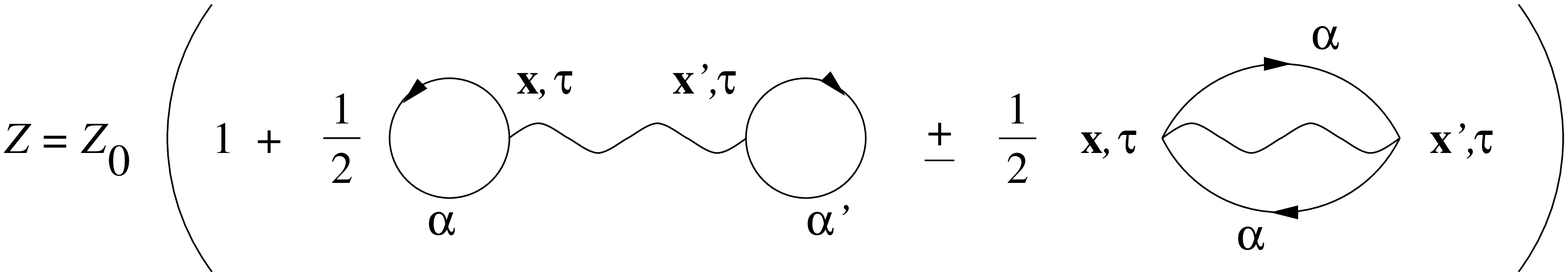}
\end{center}
\caption{Diagrammatic representation of the partition function up to first 
         order in the interaction.}
\end{figure}

\begin{figure}
\begin{center}
\includegraphics[height=0.34\hsize]{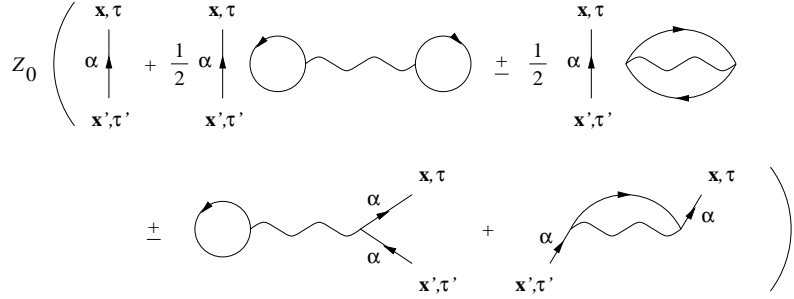}
\end{center}
\caption{Diagrammatic representation of the numerator of equation 
         (\ref{expec}) up to first order in the interaction.}
\end{figure}

To understand the general structure of the perturbation expansion, it is very 
convenient to represent equations (\ref{z1}) and (\ref{z2}) in terms of Feynman 
diagrams. The final result is shown in figure 4, where a wiggly line 
corresponds to the factor $-V({\bf x}-{\bf x}')/\hbar$ and a thin arrowed line 
pointing from 
$({\bf x}',\tau')$ to $({\bf x},\tau)$ represents the noninteracting Green's 
function $G_{0;\alpha,\alpha}({\bf x},\tau;{\bf x}',\tau')$. In figure 4 we 
have, for clarity sake, also explicitly indicated the various coordinates and 
spin degrees of freedom that we have to either integrate or sum over, 
respectively. It is, however, much more common in practice to suppress in  
Feynman diagrams those degrees of freedom that have to be integrated or summed 
over, and denote only the degrees of freedom on which the quantity of interest 
depends explicitly. For example, for the numerator of equation (\ref{expec}) we 
obtain
\begin{eqnarray}
&& \hspace*{-0.2in}
- \int d[\phi^*]d[\phi]~
   \phi_{\alpha}({\bf x},\tau) \phi^*_{\alpha'}({\bf x}',\tau')
  e^{-S_0[\phi^*,\phi]/\hbar}
    \left( 1 - \frac{S_{\rm int}[\phi^*,\phi]}{\hbar} \right) \hspace*{0.4in} 
\\
&&= Z_0 \left( - \langle \phi_{\alpha}({\bf x},\tau) 
                       \phi^*_{\alpha'}({\bf x}',\tau') \rangle_0
+ \frac{1}{\hbar} \langle \phi_{\alpha}({\bf x},\tau) 
           \phi^*_{\alpha'}({\bf x}',\tau') S_{\rm int}[\phi^*,\phi] \rangle_0
        \right), \nonumber
\end{eqnarray}
whose diagrammatic equivalent is then given in figure 5. Up to first order the 
interacting Green's function, which is represented by a thick arrowed line, 
obeys the equation shown in figure 6. Note that the so-called disconnected 
diagrams have exactly cancelled. This is a general feature that happens for any 
expectation value, i.e., the disconnected diagrams that occur in the numerator 
are always exactly cancelled by the denominator \cite{amit,zinn}. 

\begin{figure}[b]
\begin{center}
\includegraphics[height=0.16\hsize]{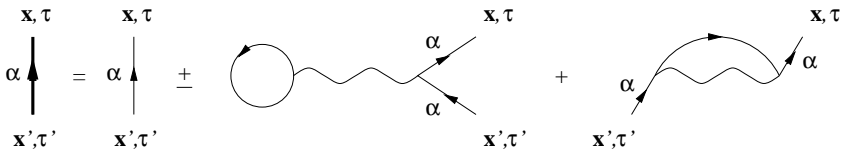}
\end{center}
\caption{Interacting Green's function up to first order in the interaction.}
\end{figure}

\begin{figure}
\begin{center}
\includegraphics[height=0.10\hsize]{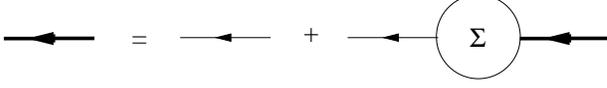}
\end{center}
\caption{Exact Dyson equation for the interacting Green's function.}
\end{figure}

One can also show that the general structure of the perturbation expansion of 
the Green's function is as in figure 7, in which only one-particle irreducible 
diagrams contribute to the selfenergy $\hbar\Sigma$. These one-particle 
irreducible diagrams can be distinguished from the fact that they do not become 
disconnected if we cut a single thin arrowed line. This is Dyson's equation
\cite{B}. It is particularly insightful for a homogeneous gas or for a trapped 
gas in the weak-coupling limit, since then the one-particle states 
$\chi_{\bf n}({\bf x})$ are not affected by the interactions and the exact 
one-particle propagator can be written as
\begin{equation}
G_{\alpha,\alpha'}({\bf x},\tau;{\bf x}',\tau') = 
  \sum_{{\bf n},n} G_{\alpha,\alpha'}({\bf n},i\omega_n)
                   \chi_{\bf n}({\bf x}) \chi^*_{\bf n}({\bf x}')
    \frac{e^{-i\omega_n (\tau-\tau')}}{\hbar\beta}~.
\end{equation}
Note that for realistic trapped atomic gases this weak-coupling limit is 
essentially always realized in the normal phase of the gas. It is, therefore, 
also of some experimental interest to consider that limit in more detail first. 

One then finds that the Dyson equation in first instance becomes
\begin{eqnarray}
&& \hspace*{-0.44in}
G_{\alpha,\alpha'}({\bf n},i\omega_n) =
  G_{0;\alpha,\alpha'}({\bf n},i\omega_n) \\
&& \hspace*{0.4in} + \sum_{\alpha'',\alpha'''}    
       G_{0;\alpha,\alpha''}({\bf n},i\omega_n)
         \Sigma_{\alpha'',\alpha'''}({\bf n},i\omega_n)
       G_{\alpha''',\alpha'}({\bf n},i\omega_n)~. \nonumber
\end{eqnarray}
Because 
$G_{0;\alpha,\alpha'}({\bf n},i\omega_n) 
                    = G_{0;\alpha}({\bf n},i\omega_n) \delta_{\alpha,\alpha'}$ 
and because we have assumed the interactions to be spin independent, we can 
easily convince ourselves that in perturbation theory it is always true that 
the 
selfenergy obeys
$\Sigma_{\alpha'',\alpha'''}({\bf n},i\omega_n) 
           = \Sigma_{\alpha''}({\bf n},i\omega_n) \delta_{\alpha'',\alpha'''}$ 
and as a result also that up to all orders in the interaction
$G_{\alpha,\alpha'}({\bf n},i\omega_n) 
                      = G_{\alpha}({\bf n},i\omega_n) \delta_{\alpha,\alpha'}$. 
As we will see in section \ref{HFT}, however, nonperturbative effect can change 
this result due to the phenomena of spontaneous broken symmetry. Ignoring the 
latter for the moment, we find for each spin state the uncoupled equation
\begin{equation}
G_{\alpha}({\bf n},i\omega_n) =
  G_{0;\alpha}({\bf n},i\omega_n) +    
       G_{0;\alpha}({\bf n},i\omega_n)
         \Sigma_{\alpha}({\bf n},i\omega_n)
       G_{\alpha}({\bf n},i\omega_n)~,
\end{equation}
or
\begin{eqnarray}
&& \hspace*{-0.73in}
\frac{1}{G_{\alpha}({\bf n},i\omega_n)} =
   \frac{1}{G_{0;\alpha}({\bf n},i\omega_n)} -   
      \Sigma_{\alpha}({\bf n},i\omega_n) \\
&&= - \frac{1}{\hbar}(-i\hbar\omega_n + \epsilon_{{\bf n},\alpha}
                    - \mu) - \Sigma_{\alpha}({\bf n},i\omega_n)~. \nonumber
\end{eqnarray}
Hence
\begin{equation}
G_{\alpha}({\bf n},i\omega_n) =
 \frac{-\hbar}{-i\hbar\omega_n + \epsilon_{{\bf n},\alpha}
               + \hbar\Sigma_{\alpha}({\bf n},i\omega_n) - \mu}~,
\end{equation}
which shows that the poles in the Green's function are indeed shifted by the 
interactions. 

\begin{figure}
\begin{center}
\includegraphics[height=0.12\hsize]{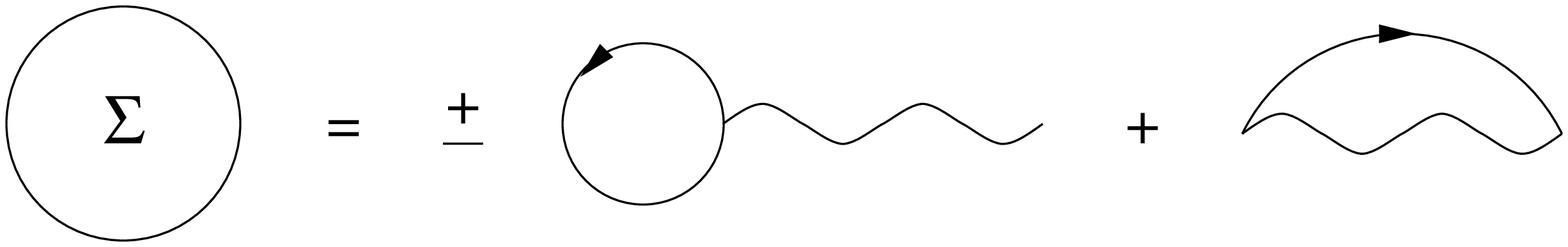}
\end{center}
\caption{Selfenergy up to first order in the interaction.}
\end{figure}

In our lowest order calculation we have found for the selfenergy the 
diagrammatic result shown in figure 8. This is the most simple approximation 
we can think of. In the weak-coupling limit it is mathematically equal to
\begin{eqnarray}
\label{self}
&& \hspace*{-0.78in}
\hbar\Sigma_{\alpha}({\bf n},i\omega_n) = 
  \sum_{{\bf n}',\alpha'} V_{{\bf n},{\bf n}';{\bf n},{\bf n}'} 
    \frac{1}{e^{\beta(\epsilon_{{\bf n}',\alpha'}-\mu)} \mp 1}  \\
&&\hspace*{1.16in}
  \mp \sum_{{\bf n}'} V_{{\bf n}',{\bf n};{\bf n},{\bf n}'} 
    \frac{1}{e^{\beta(\epsilon_{{\bf n}',\alpha}-\mu)} \mp 1} \nonumber \\
&&= \sum_{{\bf n}',\alpha'} 
  (V_{{\bf n},{\bf n}';{\bf n},{\bf n}'} 
           \pm V_{{\bf n}',{\bf n};{\bf n},{\bf n}'} \delta_{\alpha,\alpha'})
    \frac{1}{e^{\beta(\epsilon_{{\bf n}',\alpha'}-\mu)} \mp 1}~. \nonumber 
\end{eqnarray}
The first term is known as the direct or Hartree, and the second term as the 
exchange or Fock contribution to the selfenergy. From equation (\ref{self}) we 
conclude that the matrix elements of the interaction enter only in the 
combination 
$V_{{\bf n},{\bf n}';{\bf n},{\bf n}'} 
           \pm V_{{\bf n}',{\bf n};{\bf n},{\bf n}'} \delta_{\alpha,\alpha'}$. 
Clearly, this is a refection of the Pauli principle, which forces the effective 
interaction between two fermionic atoms in the same state 
$|{\bf n},\alpha\rangle$ to vanish. 

\begin{figure}[b]
\begin{center}
\includegraphics[height=0.12\hsize]{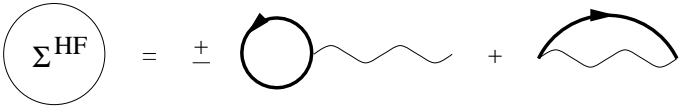}
\end{center}
\caption{Hartree-Fock approximation for the selfenergy.}
\end{figure}

To make the theory selfconsistent we should use in the expression for the 
selfenergy not the noninteracting propegators, but precisely those that follow 
from the Dyson equation. This is the Hartree-Fock approximation, which is 
diagrammatically given in figure 9. It leads in the weak-coupling limit to the 
new dispersion relation 
$\hbar\omega_{{\bf n},\alpha} = \epsilon_{{\bf n},\alpha} +
                                 \hbar\Sigma^{HF}_{\alpha}({\bf n},0) - \mu$ 
for the `dressed' particles or quasiparticles of the gas, where the selfenergy 
is essentially also found from equation (\ref{self}). We only have to replace 
in the right-hand side $\epsilon_{{\bf n}',\alpha'}-\mu$ by 
$\epsilon_{{\bf n'},\alpha'} + \hbar\Sigma^{HF}_{\alpha'}({\bf n}',0) - \mu$. 
In this manner we have thus obtained an approximation to the interacting Green's 
function that is nonperturbative in the interaction and effectively sums an 
infinite number of Feynman diagrams. Of course, the Hartree-Fock approximation 
can also be used in the strong-coupling limit. In that case it diagrammatically 
still corresponds to the solution of the Dyson equation of figure 7, with a 
selfenergy as shown in figure 9. However, we are then no longer allowed to 
assume that the exact Green's function is diagonal in the eigenstates 
$\chi_{\bf n}({\bf x})$ of the external trapping potential. To see more 
explicitly what this means physically, it is convenient to rederive the 
Hartree-Fock theory without making use of perturbation theory.  

\subsection{Hartree-Fock theory for an atomic Fermi gas}
\label{HFT}
Because the Hartree-Fock theory is very useful in a large number of 
circumstances, and because we want to illustrate a very useful technique which 
is nowadays often used in the literature, we are now going to reproduce the 
above results without making use of a diagramatic expansion. We consider for 
simplicity a fermionic mixture with an equal number of atoms in two hyperfine 
states and start by splitting our spin-independent interaction 
$V({\bf x}-{\bf x}')$ into two spin-dependent parts such that one contributes 
only to the Hartree diagram and the other only to the Fock diagram. Denoting a 
spin-dependent interaction by 
$V_{\alpha',\beta';\alpha,\beta}({\bf x}-{\bf x}')$
we thus want that
\begin{equation} 
V({\bf x}-{\bf x}') 
    \delta_{\alpha,\alpha'} \delta_{\beta,\beta'}
 = V^H_{\alpha',\beta';\alpha,\beta}({\bf x}-{\bf x}') +
   V^F_{\alpha',\beta';\alpha,\beta}({\bf x}-{\bf x}')
\end{equation}
with
$\sum_{\beta} V^H_{\beta,\alpha;\alpha,\beta}({\bf x}-{\bf x}') =
\sum_{\beta} V^F_{\alpha,\beta;\alpha,\beta}({\bf x}-{\bf x}') = 0$. 
Using operators in spin space, a possible solution to these equations is 
\begin{equation}
\hat{V}^H = \frac{2}{3}(2-\hat{P}_{12}) V({\bf x}-{\bf x}')
\end{equation}
and
\begin{equation}
\hat{V}^F = \frac{1}{3}(2\hat{P}_{12}-1) V({\bf x}-{\bf x}')~,
\end{equation} 
where 
$\hat{P}_{12} = (1+\hat{\mbox{\boldmath $\sigma$}}_1 
                \cdot \hat{\mbox{\boldmath $\sigma$}}_2)/2$ 
is the exchange operator of the spins of atoms $1$ and $2$, and 
$\mbox{\boldmath $\sigma$}_{\alpha,\alpha'}$ are the usual Pauli matrices. 
Clearly, we now have 
\begin{eqnarray}
&& \hspace*{-0.2in}
S_{\rm int}[\phi^*,\phi] =
\frac{1}{2}
  \sum_{\alpha,\alpha';\beta,\beta'} \int_0^{\hbar\beta} d\tau 
    \int d{\bf x} \int d{\bf x}'~ \\
&&\times \left\{\phi^*_{\alpha'}({\bf x},\tau)
           \phi_{\alpha}({\bf x},\tau) 
           V^H_{\alpha',\beta';\alpha,\beta}({\bf x}-{\bf x}')
           \phi^*_{\beta'}({\bf x}',\tau)
           \phi_{\beta}({\bf x}',\tau) \raisebox{0.15in}{} \right. \nonumber \\
&&\hspace*{0.5in} - \left. \raisebox{0.15in}{} \phi^*_{\alpha'}({\bf x},\tau)
           \phi_{\beta}({\bf x}',\tau) 
           V^F_{\alpha',\beta';\alpha,\beta}({\bf x}-{\bf x}')
           \phi^*_{\beta'}({\bf x}',\tau)
           \phi_{\alpha}({\bf x},\tau) \right\}~, \nonumber
\end{eqnarray}
which we write as
\begin{equation}
S_{\rm int}[\phi^*,\phi] \equiv 
 \frac{1}{2} (\phi^*\phi|V^H|\phi^*\phi)
  - \frac{1}{2} (\phi^*\phi||V^F||\phi^*\phi)~.
\end{equation}
We next apply a so-called Hubbard-Stratonovich transformation 
\cite{stratonovich,hubbard} to both the Hartree and the Fock parts of the 
interaction. First the Hartree part.

We note that $e^{-S^H_{\rm int}[\phi^*,\phi]}$ can be written as a functional 
integral over the four real fields contained in 
$\kappa_{\alpha,\alpha'}({\bf x},\tau) \equiv \kappa_0({\bf x},\tau) 
                                                   \delta_{\alpha,\alpha'}
 + \mbox{\boldmath $\kappa$}({\bf x},\tau) \cdot 
                                   \mbox{\boldmath $\sigma$}_{\alpha,\alpha'}$
as
\begin{eqnarray}
&& \hspace*{-0.2in}
\exp \left\{ - \frac{1}{2\hbar} (\phi^*\phi|V^H|\phi^*\phi)
    \right\} \\
&&= \int d[\kappa]~ \exp \left\{ \frac{1}{2\hbar} (\kappa|{V^H}^{-1}|\kappa)
         - \frac{1}{2\hbar} (\kappa|\phi^*\phi)
         - \frac{1}{2\hbar} (\phi^*\phi|\kappa) \right\}~. \nonumber
\end{eqnarray}
If we put $S^F_{\rm int}[\phi^*,\phi] = 0$ for a moment, we can substitute this 
equality in the partition function to obtain 
$Z = \int d[\kappa]d[\phi^*]d[\phi] e^{-S[\kappa,\phi^*,\phi]/\hbar}$
with the action
\begin{eqnarray}
&& \hspace*{-0.2in}
S[\kappa,\phi^*,\phi] = - \frac{1}{2} (\kappa|{V^H}^{-1}|\kappa) +
\sum_{\alpha,\alpha'} \int_0^{\hbar\beta} d\tau \int d{\bf x}~
    \phi^*_{\alpha}({\bf x},\tau) 
    \left\{ \left( \hbar \frac{\partial}{\partial \tau}  \right. \right. 
                                                         \hspace*{0.4in} \\
&& \hspace*{1.0cm} \left. \left.
            - \frac{\hbar^2 \mbox{\boldmath $\nabla$}^2}{2m} 
            + V^{\rm ex}({\bf x}) 
            + \epsilon_{\alpha} - \mu \right)
              \delta_{\alpha,\alpha'} 
                    + \kappa_{\alpha,\alpha'}({\bf x},\tau)
        \right\} \phi_{\alpha'}({\bf x},\tau)~. \nonumber
\end{eqnarray}
We see that in this manner the action for the fermions has become quadratic 
with a selfenergy 
$\hbar\Sigma_{\alpha,\alpha'}({\bf x},\tau;{\bf x}',\tau')
  = \kappa_{\alpha,\alpha'}({\bf x},\tau) \delta({\bf x}-{\bf x}')
                                        \delta(\tau-\tau')$.
Hence, we can now integrate out the fermion fields to obtain 
$Z = \int d[\kappa] e^{-S^{\rm eff}[\kappa]/\hbar}$ and
\begin{equation}
\label{haction}
S^{\rm eff}[\kappa] = - \frac{1}{2} (\kappa|{V^H}^{-1}|\kappa) 
                                -\hbar {\rm Tr}[\ln (-G^{-1})]~, 
\end{equation}
with $G^{-1} = G_0^{-1} - \Sigma$ also a functional of 
$\kappa_{\alpha,\alpha'}({\bf x},\tau)$. Up to now we have not made any 
approximations and have only performed an exact rewriting of the partition 
function. However, the action $S^{\rm eff}[\kappa]$ contains all powers of the 
fields $\kappa_{\alpha,\alpha'}({\bf x},\tau)$ and is thus quite complicated. 
It may therefore appear that we have not made any progress.

This is not the case, because the partition function will be largest for 
configurations that minimize the action $S^{\rm eff}[\kappa]$. To make use of 
this observation, we expand the action around its minimum, i.e., we put 
$\kappa_{\alpha,\alpha'}({\bf x},\tau) =   
   \langle \kappa_{\alpha,\alpha'}({\bf x}) \rangle +    
                      \delta\kappa_{\alpha,\alpha'}({\bf x},\tau)$
and require that 
$\delta S^{\rm eff}[\kappa]/\delta\kappa_{\alpha,\alpha'}({\bf x},\tau) 
                                          |_{\kappa=\langle\kappa\rangle} = 0$.
If we neglect the fluctuations, we find that
$Z \simeq e^{-S^{\rm eff}[\langle\kappa\rangle]/\hbar}$. This will turn out to 
be the Hartree approximation, as we will show now. First of all we have
\begin{equation}
G^{-1} = G_0^{-1} - \langle\kappa\rangle/\hbar - \delta\kappa/\hbar \equiv
  {G^H}^{-1} - \delta\kappa/\hbar 
  = {G^H}^{-1}(1 - G^H \delta\kappa/\hbar)~.
\end{equation}
Substituting this in equation (\ref{haction}) gives us for the terms linear in 
the fluctuations
\begin{eqnarray}
- \hbar {\rm Tr}[-G^H\delta\kappa/\hbar] 
              - (\delta\kappa|{V^H}^{-1}|\langle\kappa\rangle)~. \nonumber
\end{eqnarray}
If $\langle\kappa_{\alpha,\alpha'}({\bf x})\rangle$ is indeed a minimum of the 
action $S^{\rm eff}[\kappa]$, these linear terms have to vanish, which implies 
that
\begin{equation}
\langle\kappa_{\alpha',\alpha}({\bf x})\rangle = \sum_{\beta,\beta'}
  \int d{\bf x}~
      V^H_{\alpha',\beta';\alpha,\beta}({\bf x}-{\bf x}')
      G^H_{\beta,\beta'}({\bf x}',\tau;{\bf x}',\tau^+)~.
\end{equation}
As promised, this is precisely the most general mathematical expression for the 
Hartree contribution to the selfenergy in figure 9. Taking again 
$\langle\kappa_{\alpha',\alpha}({\bf x})\rangle = 
                                    \kappa({\bf x}) \delta_{\alpha',\alpha}$,
we also have that the Hartree approximation to the one-particle propagator 
obeys 
$G^H_{\beta,\beta'}({\bf x}',\tau;{\bf x}',\tau^+) = 
                    n({\bf x}') \delta_{\beta,\beta'}/2$,
with $n({\bf x}')$ the average total atomic density, and we obtain the usual 
Hartree selfenergy
\begin{equation}
\kappa({\bf x}) = \sum_{\beta} \int d{\bf x}'~
   V^H_{\alpha,\beta;\alpha,\beta}({\bf x}-{\bf x}') 
              \frac{n({\bf x}')}{2}
      = \int d{\bf x}'~ V({\bf x}-{\bf x}') n({\bf x}')~.
\end{equation}

Now we again want to include the Fock part of the interaction and treat this 
also by a Hubbard-Stratonovich transformation. This requires introducing four 
real fields that depend on two spatial coordinates. They are denoted by 
$\lambda_{\alpha,\alpha'}({\bf x},{\bf x}',\tau)$. Schematically we then use
\begin{eqnarray}
&& \hspace*{-0.2in}
\exp \left\{ \frac{1}{2\hbar} (\phi^*\phi||V^F||\phi^*\phi)
    \right\} \\
&&= \int d[\lambda]~ 
     \exp \left\{ - \frac{1}{2\hbar} (\lambda||{V^F}^{-1}||\lambda)
         + \frac{1}{2\hbar} (\lambda||\phi^*\phi)
         + \frac{1}{2\hbar} (\phi^*\phi||\lambda) \right\}~, \nonumber
\end{eqnarray}
which leads to the total selfenergy
\begin{equation}
\hbar\Sigma_{\alpha,\alpha'}({\bf x},\tau;{\bf x}',\tau')
  = [\kappa_{\alpha,\alpha'}({\bf x},\tau) \delta({\bf x}-{\bf x}')
     - \lambda_{\alpha,\alpha'}({\bf x},{\bf x}',\tau)]
                                        \delta(\tau-\tau')
\end{equation}
and after integration over the fermion fields to the effective action
\begin{equation}
S^{\rm eff}[\kappa,\lambda] =  - \frac{1}{2} (\kappa|{V^H}^{-1}|\kappa)
                   + \frac{1}{2} (\lambda||{V^F}^{-1}||\lambda)
                   - \hbar {\rm Tr}[\ln (-G^{-1})]~. 
\end{equation}
Requiring now that also  
$\delta S^{\rm eff}[\kappa,\lambda]/
   \delta\lambda_{\alpha,\alpha'}({\bf x},{\bf x}',\tau) 
                                     |_{\lambda=\langle\lambda\rangle} = 0$,
then leads to the expected Fock selfenergy 
\begin{equation}
\langle\lambda_{\alpha',\alpha}({\bf x}',{\bf x}) \rangle=
  \sum_{\beta,\beta'}
  \int d{\bf x}'~
      V^F_{\beta',\alpha';\alpha,\beta}({\bf x}-{\bf x}')
      G^{HF}_{\beta',\beta}({\bf x}',\tau;{\bf x},\tau^+)~.
\end{equation}
In the symmetric case it simply becomes 
$\langle\lambda_{\alpha',\alpha}({\bf x}',{\bf x})\rangle 
    = \lambda ({\bf x}',{\bf x}) \delta_{\alpha',\alpha}$ with
\begin{equation}
\lambda ({\bf x}',{\bf x}) 
            =  V({\bf x}-{\bf x}') n({\bf x}',{\bf x})
\end{equation}
and $n({\bf x}',{\bf x}) = \sum_{\beta}
          G^{HF}_{\beta,\beta}({\bf x}',\tau;{\bf x},\tau^+)/2 =
      \sum_{\beta} \langle \psi^{\dagger}_{\beta}({\bf x},\tau) 
                           \psi_{\beta}({\bf x}',\tau) \rangle/2$
exactly the Hartree-Fock approximation to the off-diagonal part of the 
one-particle density matrix. 

Note that to actually perform the Hartree-Fock calculation in the latter case, 
we need to be able to determine the Green's function 
$G^{HF}_{\alpha,\alpha'}({\bf x},\tau;{\bf x}',\tau')$. The easiest way to do 
so, is by realizing that it is the Green's function of the operator in the 
fermionic piece of the action $S[\kappa,\lambda,\phi^*,\phi]$ obtained after 
the 
Hubbard-Stratonovich transformations. If we diagonalize this operator by 
solving 
the eigenvalue problem
\begin{eqnarray}
&& \hspace*{-1.1in}
\left\{ - \frac{\hbar^2 \mbox{\boldmath $\nabla$}^2}{2m} + V^{\rm ex}({\bf x}) 
    + \kappa({\bf x}) - \epsilon'_{\bf n} \right\} \chi'_{\bf n}({\bf x}) \\
&& \hspace*{0.1in}
   - \int d{\bf x}'~ \lambda({\bf x},{\bf x}') \chi'_{\bf n}({\bf x}') = 0~,
    \nonumber
\end{eqnarray}    
the desired one-particle propagator aquires the ideal gas form
\begin{eqnarray}
&& \hspace*{-0.2in} 
G^{HF}_{\alpha,\alpha'}({\bf x},\tau;{\bf x}',\tau') \\
&&= \delta_{\alpha,\alpha'}
  \sum_{{\bf n},n} 
     \frac{-\hbar}{-i\hbar\omega_n + \epsilon'_{{\bf n},\alpha} - \mu}
                   \chi'_{\bf n}({\bf x}) \chi'^*_{\bf n}({\bf x}')
    \frac{e^{-i\omega_n (\tau-\tau')}}{\hbar\beta}~, \nonumber
\end{eqnarray}
with new one-particle energies 
$\epsilon'_{{\bf n},\alpha} = \epsilon'_{{\bf n}} + \epsilon_{\alpha}$ and
eigenstates $\chi'_{\bf n}({\bf x})$ that incorporate the average effect of the 
interactions of an atom with all the other atoms in the gas. Clearly, in the 
cases that the eigenstates are not affected by these so-called mean-field 
effects, we recover the weak-coupling results of the previous section. 

Although we have thus precisely reproduced our diagrammatic result, there are 
two important advantages in using the Hubbard-Stratonovich transformation.
First, it is in principle exact, and allows us to also calculate corrections to 
the Hartree-Fock approximation. For example, if we expand 
$S^{\rm eff}[\kappa,\lambda]$ up to quadratic order in $\delta\kappa$ and 
$\delta\lambda$, and neglect all higher orders, we find the so-called 
Generalized Random Phase Approximation. The latter approach actually gives us 
also the opportunity to study the density fluctuations and therefore the 
collective excitations of the gas. We return to this important topic in section 
\ref{modes}. Second, it allows for a beautiful way to describe phase 
transitions. As mentioned previously, in perturbation theory we always find for 
a homogeneous gas that 
$G_{\alpha,\alpha'}({\bf x},\tau;{\bf x},\tau^+) 
           = n \delta_{\alpha,\alpha'}/2$
due to the same feature of 
$G_{0;\alpha,\alpha'}({\bf x},\tau;{\bf x},\tau^+)$. From a fundamental point 
of 
view this is a result of the translational invariance of the gas and of the 
rotational symmetry in spin space. However, we can imagine that in principle we 
can also find selfconsistent solutions that do not have this property. We then 
have a spontaneous breaking of symmetry and therefore a phase transition in our 
system. For example, if below a certain temperature
$G_{\alpha,\alpha'}({\bf x},\tau;{\bf x},\tau^+) = 
                                    n({\bf x}) \delta_{\alpha,\alpha'}/2$,
we are dealing with a transition to a charge density wave or (Wigner) crystal. 
If on the other hand 
$G_{\alpha,\alpha'}({\bf x},\tau;{\bf x},\tau^+) = 
        n \delta_{\alpha,\alpha'}/2 
               + {\bf m} \cdot \mbox{\boldmath $\sigma$}_{\alpha,\alpha'}$
the gas is in a ferromagnetic phase. For a spin-density wave we even have that 
$G_{\alpha,\alpha'}({\bf x},\tau;{\bf x},\tau^+) = 
   n \delta_{\alpha,\alpha'}/2 
      + {\bf m}({\bf x}) \cdot \mbox{\boldmath $\sigma$}_{\alpha,\alpha'}$.
In all these cases the Hubbard-Stratonovich approach used above leads in a 
natural way to the appropriate Landau theory of the phase transition. Since the 
Landau theory is also very useful for the understanding of superfluidity in 
atomic gases, which is clearly the primary goal of the present course, we use 
the next section to give a short introduction into this subject.

\subsection{Landau theory of phase transitions}
In this summary of the Landau theory of phase transitions, we restrict 
ourselves 
in first instance to the homogeneous case because this makes the discussion 
more 
transparent. However, at the end of the section we also briefly mention how the 
inhomogeneity enters the theory. We have seen, for example in the Hartree-Fock 
theory discussed above, that the Green's function 
$G_{\alpha,\alpha'}({\bf x},\tau;{\bf x},\tau^+) = 
  n \delta_{\alpha,\alpha'}/2 
    + \langle {\bf s} \rangle \cdot \mbox{\boldmath $\sigma$}_{\alpha,\alpha'}$
can have a nonzero value of the average spin density 
$\langle {\bf s} \rangle = \langle \hat{\psi}^{\dagger}_{\alpha}({\bf x},\tau)
             \mbox{\boldmath $\sigma$}_{\alpha,\alpha'} 
               \hat{\psi}_{\alpha'}({\bf x},\tau) \rangle/2$, which is usually 
called the magnetization ${\bf m}$ in this context. This signals a phase 
transition to a ferromagnetic phase and the magnetization is called the order 
parameter of this transition. In the previous section we have also seen how we 
can, in principle exactly, obtain an expression for the partition function as a 
functional integral over the field 
${\bf s}({\bf x},\tau)$, i.e.,
\begin{equation}
Z = \int d[{\bf s}]~e^{-S^{\rm eff}[{\bf s}]/\hbar}~.
\end{equation}
We just have to integrate out the fields $\kappa_0$ and 
$\lambda_{\alpha,\alpha'}$. For space and time independent values of the 
magnetization ${\bf s}$ the action $S^{\rm eff}[{\bf s}]$ is 
$\hbar\beta V f_L({\bf s})$,
where $f_L({\bf s})$ is the Landau free-energy density and $V$ is the total 
volume of system. Because of the symmetry of $S[\phi^*,\phi]$ under rotations 
of 
the spin, this free-energy density must also be rotationally invariant and can 
therefore only depend on the magnitude of ${\bf s}$, which we denote by $s$. If 
a phase transition occurs the behaviour of $f_L(s)$ can essentially fall only 
into two categories. 

At temperatures very high compared to the critical temperature $T_c$, the 
system is fully disorded and the free energy $f_L(s)$ must have a single minimum 
at $s=0$ to make sure that the order parameter
$\langle s \rangle$ is zero. Bringing the temperature closer to $T_c$, 
however, the free energy can develop a second local minimum. As long as the 
free energy in this second local minimum is higher than the minimum at $s=0$, 
the equilibrium value of $\langle s \rangle$ will still be zero and no 
phase transition has occured. Lowering the temperature further, the value of the 
free energy in the second minimum decrease until, precisely at the critical 
temperature $T_c$, it is equal to the free energy at $s=0$. For temperatures 
below this critical one the second minimum has actually become the global 
minimum of the free energy, which implies that $\langle s \rangle \neq 0$ 
and we are in the ordered phase. In this scenario the order parameter has 
always 
a discontinuity at the critical temperature. As a result, this corresponds to a 
discontinuous, or first-order, phase transition. The behaviour of the free 
energy and the order parameter is illustarted in figure 10. This should be 
compared with the behaviour of the free energy and the order parameter for a 
continuous, i.e., second or higher order, phase transition, which is quite 
different and depicted in figure 11. Now the Landau free energy $f_L(s)$ has 
always a single minimum, which for temperatures above $T_c$ is at $s=0$ but at 
temperatures below the critical temperature shifts to a nonzero value of $s$. 
In particular, the order parameter thus shows no discontinuity at $T_c$.

\begin{figure}
\begin{center}
\includegraphics[height=0.30\hsize]{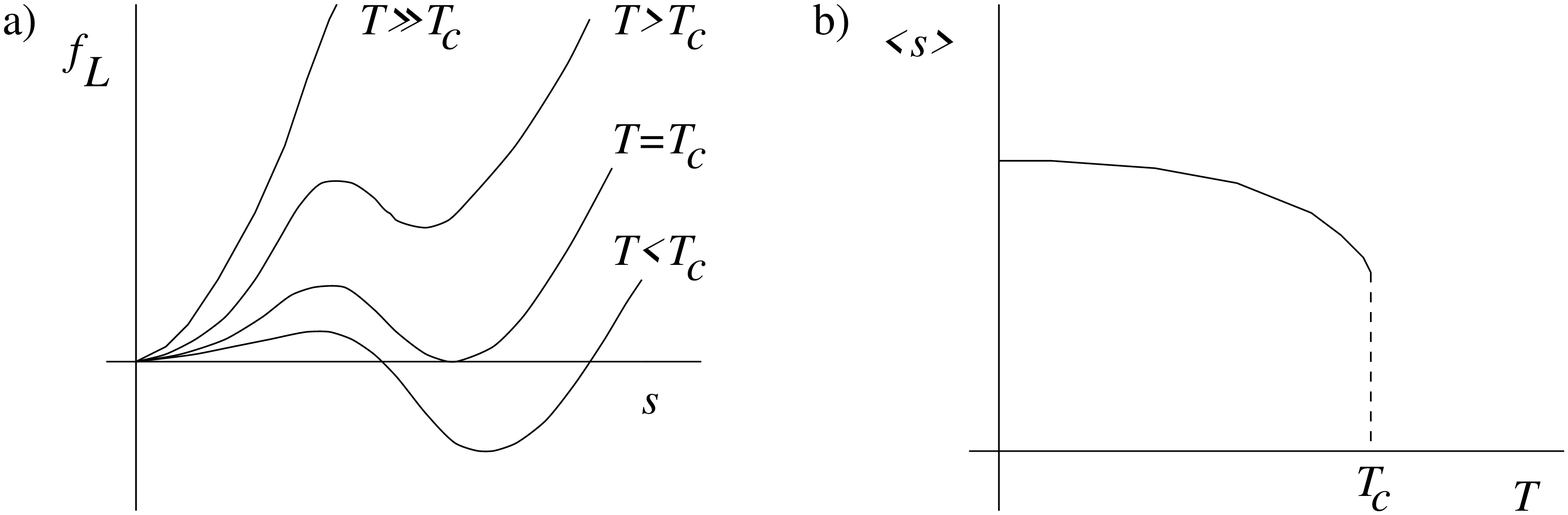}
\end{center}
\caption{Qualitative behaviour of a) the Landau free energy and b) the 
         order parameter for a discontinuous phase transition.}
\end{figure}

In the case of a second-order phase transition, we conclude that near the 
critical temperature $\langle s \rangle$ is very small. As a result we can then 
expand the Landau free-energy density in powers of $s$. Because the free energy 
must also be symmetric under $s \rightarrow -s$, we have
\begin{equation}
f_L({\bf s}) = \alpha(T) |{\bf s}|^2 
               + \frac{\beta}{2} |{\bf s}|^4~,
\end{equation}
with $\beta >0$ and $\alpha(T) = \alpha_0(T/T_c-1)$. Thus if $T>T_c$ we have 
$\alpha(T)>0$ and 
$\langle s \rangle = 0$. But for $T<T_c$, $\alpha(T)$ becomes negative and we 
have
\begin{equation}
\langle s \rangle = 
  \sqrt{\frac{\alpha_0}{\beta} \left( 1 - \frac{T}{T_c} \right)}~.
\end{equation}
Note that the free-energy density in this minimum is 
$-\alpha_0^2(1-T/T_c)^2/2\beta <0$ and has a discontinuity in its second 
derivative with respect to the temperature. Historically, this is the reason 
why 
the corresponding phase transition was named to be of second order.

\begin{figure}
\begin{center}
\includegraphics[height=0.30\hsize]{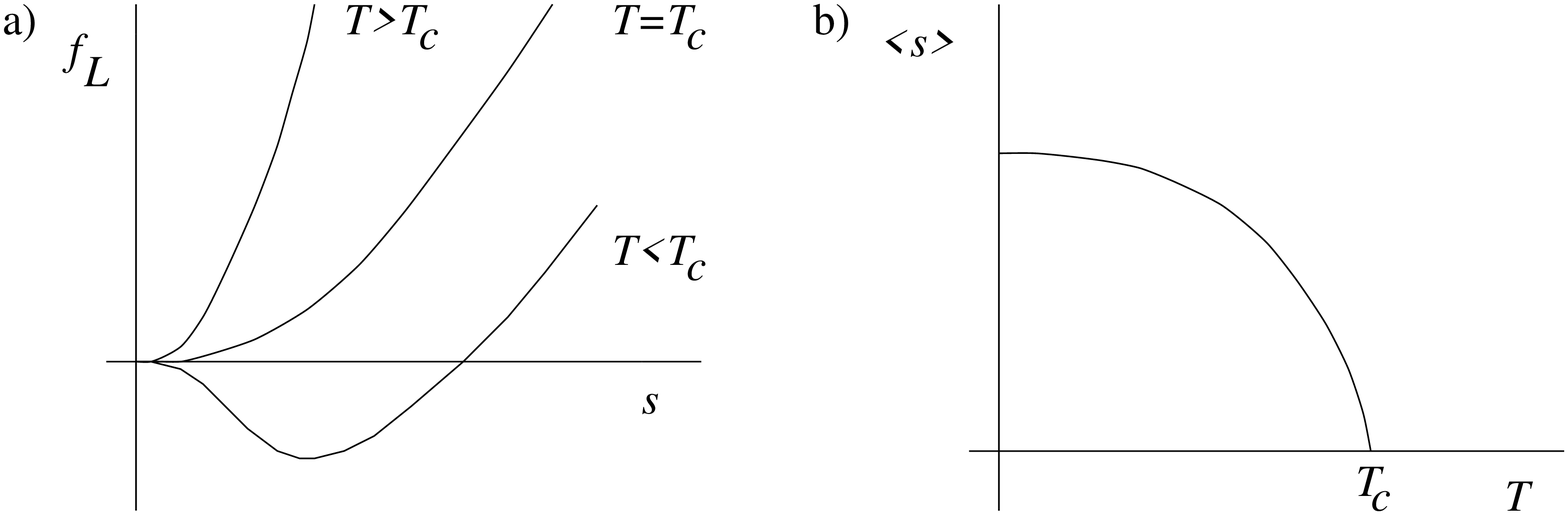}
\end{center}
\caption{Qualitative behaviour of a) the Landau free energy and b) the 
         order parameter for a continuous phase transition.}
\end{figure}

In fact, the Landau theory of second order phase transitions is slightly more 
involved, because it also considers slow spatial fluctuations in ${\bf s}$. 
Since a uniform rotation of ${\bf s}$ costs no energy, we must have that the 
free energy is
\begin{equation}
\label{LFE}
F_L[{\bf s}] = \int d{\bf x}~
  \left( \gamma (\mbox{\boldmath $\nabla$} {\bf s}({\bf x}))^2 
               + \alpha(T) |{\bf s}({\bf x})|^2 
               + \frac{\beta}{2} |{\bf s}({\bf x})|^4 \right)
\end{equation}
and $Z \simeq \int d[{\bf s}]~e^{-\beta F_L[{\bf s}]}$. Landau theory now 
esentially corresponds to minimizing $F_L[{\bf s}]$ and taking only quadratic 
fluctuations into account. Taking also higher order fluctuations into account 
turns out to be very difficult. It requires Renormalization Group methods, 
which 
we are not going to treat here and that can be found in the literature 
\cite{amit}. The effects of these fluctuation corrections to Landau theory are 
generally known as critical phenomena. Interestingly, they do not occur in 
first-order phase transitions. More importantly for our purposes is, however, 
that for atomic gases the fluctuations are only important in a small 
temperature interval around the critical temperature. For many applications it 
is, therefore, possible to neglect them. What usually cannot be negected is the 
effect of the inhomogeneity of the gas. In the context of Landau theory that 
implies that the coefficients $\alpha(T)$, $\beta$ and $\gamma$ in equation 
(\ref{LFE}) become also dependend on the spatial position in the trapping 
potential. We will see several examples of this feature in the following.   

\subsection{Superfluidity and superconductivity}
We finally want to consider two important second-order phase transitions, that 
are purely due to quantum effects and can be conveniently treated with the 
methods that we have developed sofar. Moreover, they occur very often in 
nature, 
for example in metals, in liquid helium and recently of course also in atomic 
gases of rubidium \cite{JILA}, lithium \cite{Rice}, sodium \cite{MIT} and 
hydrogen \cite{MITH}.

\subsubsection{Superfluidity}
\label{SF}
Let us first consider the last case, which is associated with a gas of 
spin-less 
bosons. For the low temperatures of interest the action is 
\begin{eqnarray}
&& \hspace*{-0.2in}
S[\phi^*,\phi] \\
&&= \int_0^{\hbar\beta} d\tau \int d{\bf x}~
 \phi^*({\bf x},\tau)
  \left\{ \hbar \frac{\partial}{\partial\tau} 
      - \frac{\hbar^2 \mbox{\boldmath $\nabla$}^2}{2m} 
      + V^{\rm ex}({\bf x}) - \mu
  \right\} \phi({\bf x},\tau)                                \nonumber \\
&&+ \frac{1}{2}
  \int_0^{\hbar\beta} d\tau \int d{\bf x} ~ V_{{\bf 0}}
           \phi^*({\bf x},\tau) \phi^*({\bf x},\tau)
             \phi({\bf x},\tau) \phi({\bf x},\tau)~, \nonumber
\end{eqnarray}
where $V({\bf x}-{\bf x}')$ is taken to be equal to 
$V_{{\bf 0}} \delta({\bf x}-{\bf x}')$. The justification for this 
simplificaties is, roughly speaking, that the thermal de Broglie wavelength 
$\Lambda_{\rm th} = (2\pi\hbar^2/mk_BT)^{1/2}$ of the atoms is much larger than 
the range of the interatomic interaction. In this system the phase transition 
of 
interest is Bose-Einstein condensation. The associated order parameter is
$\langle \phi({\bf x},\tau) \rangle$, since for time independent 
$\phi({\bf x},\tau)$ the above action has presicely the form of a Landau theory 
with a `free-energy' of
\begin{eqnarray}
&& \hspace*{-0.2in}
F_L[\phi^*,\phi] \\
&&= \int d{\bf x}~ 
    \left( \frac{\hbar^2}{2m} |\mbox{\boldmath $\nabla$}\phi({\bf x})|^2
           + (V^{\rm ex}({\bf x})-\mu) |\phi({\bf x})|^2 
           + \frac{V_{{\bf 0}}}{2} |\phi({\bf x})|^4
    \right)~. \nonumber
\end{eqnarray}
We conclude that in lowest order the critical temperature is determined by 
$\mu(T_c)=\epsilon_{\bf 0}$, because then the configurations 
$\phi({\bf x}) \propto \chi_{\bf 0}({\bf x})$ precisely make a vanishing  
contribution to the quadratic part of the `free energy'. This condition makes 
sense, because it is exactly the condition that we have a Bose-Einstein 
condensation in the ideal case: In the ideal Bose gas the number of particles 
in 
the one-particle ground state is 
$N_{{\bf 0}} = 1/(e^{\beta(\epsilon_{\bf 0}-\mu)}-1)$, which indeed diverges 
for 
$\mu=\epsilon_{\bf 0}$. 

To determine the corrections to this result we now explicitly substitute
$\phi({\bf x},\tau) = \phi_{\bf 0}({\bf x}) + \phi'({\bf x},\tau)$ into our 
functional integral. It is for lateron important to realize that to 
consistently 
define the fluctuations $\phi'({\bf x},\tau)$ in this manner, we also have to 
require that
\begin{equation}
\label{ort}
\int d{\bf x}~\phi_{\bf 0}^*({\bf x}) \phi'({\bf x},\tau) +
    \int d{\bf x}~\phi_{\bf 0}({\bf x}) {\phi'}^*({\bf x},\tau) = 0~.
\end{equation}
The physical reason behind this condition is that $\phi'({\bf x},\tau)$ should 
contain all the configurations that are orthogonal to $\phi_{\bf 0}({\bf x})$. 
In principle, therefore, it should also contain configurations that in effect 
only multiply $\phi_{\bf 0}({\bf x})$ by a global phase. Such fluctuations lead 
to the phenomena of phase `diffusion' and are discussed in section \ref{PD}. In 
full detail we find after the above substitution that
\begin{equation}
S[{\phi'}^*,\phi'] = \hbar\beta F_L[\phi_{\bf 0}^*,\phi_{\bf 0}]  
                   + S_0[{\phi'}^*,\phi'] + S_{\rm int}[{\phi'}^*,\phi']~,
\end{equation}
where the linear and quadratic terms are given by 
\begin{eqnarray}
\label{s0}
&& \hspace*{-0.2in}
S_0[{\phi'}^*,\phi']  \\             
&& = \int_0^{\hbar\beta} d\tau \int d{\bf x}~
    {\phi'}^*({\bf x},\tau) 
     \left\{ - \frac{\hbar^2 \mbox{\boldmath $\nabla$}^2}{2m} 
             + V^{\rm ex}({\bf x}) - \mu + V_{{\bf 0}} |\phi_{\bf 0}({\bf 
x})|^2
     \right\} \phi_{\bf 0}({\bf x})                        \nonumber \\
&& + \int_0^{\hbar\beta} d\tau \int d{\bf x}~ \phi'({\bf x},\tau) 
     \left\{ - \frac{\hbar^2 \mbox{\boldmath $\nabla$}^2}{2m} 
             + V^{\rm ex}({\bf x}) - \mu + V_{{\bf 0}} |\phi_{\bf 0}({\bf 
x})|^2
     \right\} \phi_{\bf 0}^*({\bf x})                      \nonumber \\     
&& + \int_0^{\hbar\beta} d\tau \int d{\bf x}~ {\phi'}^*({\bf x},\tau)
                                                           \nonumber \\      
&& \hspace*{0.85in} \times \left\{ \hbar \frac{\partial}{\partial\tau} 
      - \frac{\hbar^2 \mbox{\boldmath $\nabla$}^2}{2m} 
              + V^{\rm ex}({\bf x}) - \mu 
              + 2V_{{\bf 0}}|\phi_{\bf 0}({\bf x})|^2
    \right\} \phi'({\bf x},\tau)                           \nonumber \\
&& + \frac{1}{2} \int_0^{\hbar\beta} d\tau \int d{\bf x}~
   V_{{\bf 0}} (\phi_{\bf 0}({\bf x}))^2 {\phi'}^*({\bf x},\tau)
                       {\phi'}^*({\bf x},\tau)         \nonumber \\
&& + \frac{1}{2} \int_0^{\hbar\beta} d\tau \int d{\bf x}~
   V_{{\bf 0}} (\phi_{\bf 0}^*({\bf x}))^2
                       \phi'({\bf x},\tau) \phi'({\bf x},\tau)~, \nonumber
\end{eqnarray}
and the cubic and quartic terms by
\begin{eqnarray}
&& \hspace*{-0.6in}
S_{\rm int}[{\phi'}^*,\phi'] = 
\int_0^{\hbar\beta} d\tau \int d{\bf x}~   
   V_{{\bf 0}} \phi_{\bf 0}({\bf x}) {\phi'}^*({\bf x},\tau) 
               {\phi'}^*({\bf x},\tau) \phi'({\bf x},\tau) \hspace{0.4in} \\
&& \hspace*{0.1in} + \int_0^{\hbar\beta} d\tau \int d{\bf x}~
   V_{{\bf 0}} \phi_{\bf 0}^*({\bf x}) {\phi'}^*({\bf x},\tau) 
               \phi'({\bf x},\tau)  
               \phi'({\bf x},\tau)                    \nonumber \\
&& \hspace*{0.1in} + \frac{1}{2} \int_0^{\hbar\beta} d\tau \int d{\bf x}~
   V_{{\bf 0}} {\phi'}^*({\bf x},\tau) {\phi'}^*({\bf x},\tau)
               \phi'({\bf x},\tau) \phi'({\bf x},\tau)~, \nonumber 
\end{eqnarray}
respectively.

In the Bogoliubov approximation we neglect the last three interaction terms
\cite{bog}. Furthermore, to make sure that 
$\langle \phi({\bf x},\tau) \rangle = \phi_{\bf 0}({\bf x})$ or
$\langle \phi'({\bf x},\tau) \rangle = 0$, we need to require that the terms 
linear in $\phi'$ and ${\phi'}^*$ drop out of the action 
$S_0[{\phi'}^*,\phi']$. Clearly, this implies that
\begin{equation}
\left( - \frac{\hbar^2 \mbox{\boldmath $\nabla$}^2}{2m} 
             + V^{\rm ex}({\bf x}) + V_{{\bf 0}} |\phi_{\bf 0}({\bf x})|^2
     \right) \phi_{\bf 0}({\bf x}) = \mu \phi_{\bf 0}({\bf x})~,
\end{equation} 
which is the same result as obtained from minimizing the Landau `free-energy' 
$F_L[\phi^*,\phi]$. In the context of trapped atomic gases, this equation is 
known as the Gross-Pitaevskii equation \cite{GP}. It determines the macroscopic 
wave function of the condensate. The reason for calling $\phi_{\bf 0}({\bf x})$ 
the macroscopic wave function follows from the fact that the total density of 
the 
gas now obeys
\begin{equation}
\label{dens}
n({\bf x}) = \langle \phi({\bf x},\tau) \phi^*({\bf x},\tau^+) \rangle
           = |\phi_{\bf 0}({\bf x})|^2 + 
              \langle \phi'({\bf x},\tau) {\phi'}^*({\bf x},\tau^+) \rangle~.
\end{equation}
The total number of condensate atoms thus equals 
$N_{\bf 0} = \int d{\bf x}~|\phi_{\bf 0}({\bf x})|^2$. As equation (\ref{dens}) 
shows, it is in general always smaller that the total number of atoms in the 
gas 
due to the effect of the fluctuations. Note that in our present formulation the 
average $\langle \phi'({\bf x},\tau) {\phi'}^*({\bf x},\tau^+) \rangle$ 
physically describes not only the depletion of the condensate due to the usual 
thermal fluctuations known from the ideal Bose gas, but also due to the 
interactions, i.e., purely due to quantum fluctuations.  

Assuming that we have solved for the Gross-Pitaevskii equation, the fluctuation 
corrections are in the Bogoliubov theory determined by a quadratic action of
the form
\begin{eqnarray}
&& \hspace*{-0.2in}
S_B[{\phi'}^*,\phi'] \\
&& = - \frac{\hbar}{2} \int_0^{\hbar\beta} d\tau \int d{\bf x}~ 
  \left[ {\phi'}^*({\bf x},\tau) , \phi'({\bf x},\tau) \right]
\cdot {\bf G}^{-1} \cdot
\left[ 
\begin{array}{c} \phi'({\bf x},\tau) \\ {\phi'}^*({\bf x},\tau) \end{array}
\right]~, \nonumber
\end{eqnarray}
where the associated Green's function ${\bf G}$ has now a matrix structure 
because not only the normal average
$\langle \phi'({\bf x},\tau) {\phi'}^*({\bf x}',\tau') \rangle$ but also
the anomalous average 
$\langle \phi'({\bf x},\tau) \phi'({\bf x}',\tau') \rangle$ is now unequal to 
zero. We thus have that
\begin{equation}
- {\bf G}({\bf x},\tau;{\bf x}',\tau') = 
  \left\langle 
     \left[ 
       \begin{array}{c} \phi'({\bf x},\tau) \\ {\phi'}^*({\bf x},\tau)  
       \end{array}
     \right] \cdot
     \left[ {\phi'}^*({\bf x}',\tau') , \phi'({\bf x}',\tau') \right]
  \right\rangle~.
\end{equation}
From equation (\ref{s0}) we in fact find that in the Bogoliubov approximation
\begin{eqnarray} 
\label{gb}
&& \hspace*{-0.4in}                                
{\bf G}^{-1}({\bf x},\tau;{\bf x}',\tau') = 
                   {\bf G}_0^{-1}({\bf x},\tau;{\bf x}',\tau') \\
&& \hspace*{0.6in} - \frac{1}{\hbar}
\left[ 
\begin{array}{cc}
2V_{{\bf 0}}|\phi_{\bf 0}({\bf x})|^2  & V_{{\bf 0}}(\phi_{\bf 0}({\bf x}))^2 
\\
V_{{\bf 0}}(\phi_{\bf 0}^*({\bf x}))^2 & 2V_{{\bf 0}}|\phi_{\bf 0}({\bf x})|^2 
\end{array} \right] \delta({\bf x}-{\bf x}') \delta(\tau-\tau')~. \nonumber
\end{eqnarray}
with the noninteracting Green's function ${\bf G}_0$ defined by
\begin{equation} 
{\bf G}_0^{-1}({\bf x},\tau;{\bf x}',\tau') =
\left[ 
\begin{array}{cc}
G_0^{-1}({\bf x},\tau;{\bf x}',\tau') & 0 \\
0 & G_0^{-1}({\bf x}',\tau';{\bf x},\tau)
\end{array}
\right]    
\end{equation}
and, of course, 
\begin{eqnarray}
&& \hspace*{-0.2in}
G_0^{-1}({\bf x},\tau;{\bf x}',\tau') \\
&&= - \frac{1}{\hbar} \left\{ \hbar\frac{\partial}{\partial\tau}
    - \frac{\hbar^2 \mbox{\boldmath $\nabla$}^2}{2m} 
    + V^{\rm ex}({\bf x}) - \mu  \right\}
 \delta({\bf x}-{\bf x}') \delta(\tau-\tau')~. \nonumber
\end{eqnarray}
This is clearly only the lowest order result for the Green's function, because 
a perturbative treatment of $S_{\rm int}[{\phi'}^*,\phi']$ leads, in the same 
way as in section \ref{IFD}, to higher order corrections. In general, the Dyson 
equation is, however, always of the form
\begin{equation}
\left[ \begin{array}{cc}
G_{11} & G_{12} \\ G_{21} & G_{22}
\end{array} \right]^{-1} =
\left[ \begin{array}{cc}
G_0^{-1} & 0 \\ 0 & G_0^{-1}
\end{array} \right] -
\left[ \begin{array}{cc}
\Sigma_{11} & \Sigma_{12} \\ \Sigma_{21} & \Sigma_{22}
\end{array} \right]~.
\end{equation}
The off-diagonal elements are again called anomalous, since they vanish in the 
normal phase of the gas. Diagrammatically the Dyson equations for $G_{11}$ and 
$G_{21}$ are shown in figure 12. 

\begin{figure}
\begin{center}
\includegraphics[height=0.23\hsize]{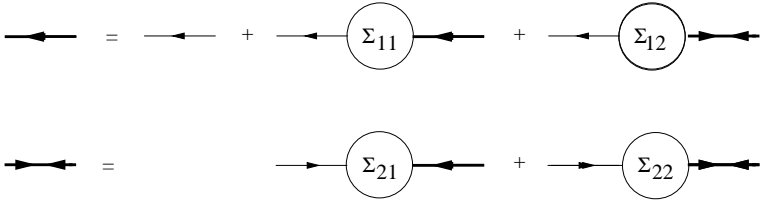}
\end{center}
\caption{Exact Dyson equation for the interacting normal and
         anomalous Green's functions.}
\end{figure}

The selfenergy in the Bogoliubov approximation is
\begin{eqnarray}
&& \hspace*{-0.2in}
\hbar\mbox{\boldmath $\Sigma$}({\bf x},\tau;{\bf x}',\tau') \\
&&= \left[ 
\begin{array}{cc}
2V_{{\bf 0}}|\phi_{\bf 0}({\bf x})|^2  &  V_{{\bf 0}}(\phi_{\bf 0}({\bf x}))^2 
\\
V_{{\bf 0}}(\phi_{\bf 0}^*({\bf x}))^2 & 2V_{{\bf 0}}|\phi_{\bf 0}({\bf x})|^2  
\end{array}
\right] \delta({\bf x}-{\bf x}') \delta(\tau-\tau')~.~~
\end{eqnarray}
Because we have completely negelected the cubic and quartic terms in the action 
to obtain this selfenergy, the Bogoliubov approach is only valid if the 
fluctuations are sufficiently small. Physically, this impies that the 
deplection 
of the condensate must be small. It can, therefore, not be applied to liquid 
helium, but is expected to be valid for a weakly-interacting atomic gas at such 
low temperature that it essentially only consists of a condensate. Under these 
conditions the Bogoliubov theory not only predicts the condensate density 
profile $n_{\bf 0}({\bf x}) = |\phi_{\bf 0}({\bf x})|^2$ but also the 
collective 
modes of the condensate. Both these predictions have been accurately verified 
in 
recent experiments \cite{coll1,coll2,hau}. Theoretically, the eigenfrequencies 
of the collective modes are again determined by the poles in the one-particle 
Green's function. To understand more clearly how these can be determined, we 
first consider a homogeneous Bose gas.

In a box with volume $V=L^3$ the one-particle states are most easily 
characterized by the wavevector ${\bf k} = (2\pi/L) {\bf n}$ and equal to 
$e^{i{\bf k}\cdot{\bf x}}/\sqrt{V}$. The one-particle energies are thus
$\epsilon_{\bf k} = \hbar^2 {\bf k}^2/2m$. Moreover, the Gross-Pitaevskii 
equation reduces to $\mu = V_{\bf 0} |\phi_{\bf 0}|^2$ in that case, because the 
Landau `free-energy' is minimized for a macroscopic wave function that is 
independent of the position in the box. Making use of this fact, equation 
(\ref{gb}) can immediately be solved by a Fourier transformation. The result is
\begin{eqnarray}
&& \hspace*{-0.2in}
\label{gbf}
-\hbar {\bf G}^{-1}({\bf k},i\omega_n) \\
&&= \left[  
\begin{array}{cc}
-i\hbar\omega_n + \epsilon_{{\bf k}} + V_{{\bf 0}}|\phi_{\bf 0}|^2 
                                           &    V_{{\bf 0}}\phi_{\bf 0}^2 \\
V_{{\bf 0}}{\phi_{\bf 0}^*}^2 
          & i\hbar\omega_n + \epsilon_{{\bf k}} + V_{{\bf 0}}|\phi_{\bf 0}|^2  
\end{array}
\right]~. \nonumber
\end{eqnarray}
Clearly there are poles in $G({\bf k},\omega)$ if the determinant of the 
right-hand side is zero or if
\begin{equation}
\hbar\omega = \hbar\omega_{{\bf k}} \equiv
 \sqrt{\epsilon_{{\bf k}}^2
           +2V_{{\bf 0}}|\phi_{\bf 0}|^2\epsilon_{{\bf k}}}
 = \sqrt{\epsilon_{{\bf k}}^2
           +2V_{{\bf 0}}n_{\bf 0}\epsilon_{{\bf k}}}~.
\end{equation}
This is the famous Bogoliubov dispersion of the collective excitations. 
Note that to finish the calculation we still have to obtain the condensate 
density $n_{\bf 0}=|\phi_{\bf 0}|^2$. This is determined by the total density 
of the gas, which obeys
\begin{equation}
n = |\phi_{\bf 0}|^2 - G_{11}({\bf x},\tau;{\bf x},\tau^+) = n_{\bf 0} + n'~,
\end{equation}
with $n'$ the density of the `above' condensate particles. By inverting the 
right-hand side of equation (\ref{gbf}), we find that the noncondensate density
is equal to
\begin{eqnarray}
\label{ncd}
&& \hspace*{-0.2in}
n' = \lim_{\eta \downarrow 0} \frac{\hbar}{V\hbar\beta} 
     \sum_{{\bf k} \neq {\bf 0},n} 
     e^{i\omega_n\eta}
     \frac{i\hbar\omega_n + \epsilon_{{\bf k}} + V_{{\bf 0}}n_{\bf 0}}
          {(\hbar\omega_n)^2 + (\hbar\omega_{\bf k})^2} \\
&&= \frac{1}{V} \sum_{{\bf k} \neq {\bf 0}} 
    \left( \frac{\epsilon_{\bf k} + V_{{\bf 0}}n_{\bf 0}}{\hbar\omega_{\bf k}}
           \frac{1}{e^{\beta\hbar\omega_{\bf k}}-1} 
         + \frac{\epsilon_{\bf k} + V_{{\bf 0}}n_{\bf 0} - \hbar\omega_{\bf k}}
                {2\hbar\omega_{\bf k}}
    \right)~. \nonumber
\end{eqnarray}
For a given density and temperature, the last two equations thus fully 
determine the condensate density. Note that equation (\ref{ncd}) explicitly 
shows that the condensate is indeed depleted both by thermal as well as quantum 
effects.

The generalization to the inhomogeneous case is straightforward. First we again 
have to solve the Gross-Pitaevskii equation at a fixed chemical potential. 
Given 
the condensate wave function, we can then calculate the collective mode 
frequencies by finding the poles of ${\bf G}$, or equivalently but more 
conveniently, the zero's of ${\bf G}^{-1}$. Clearly, the latter are located at 
$\hbar\omega = \hbar\omega_{\bf n}$, where $\hbar\omega_{\bf n}$ is found from 
the eigenvalue problem
\begin{eqnarray}
\left[ 
\begin{array}{cc}
\hat{K} + 2V_{{\bf 0}}|\phi_{\bf 0}({\bf x})|^2  &  
                                        V_{{\bf 0}}(\phi_{\bf 0}({\bf x}))^2 \\
V_{{\bf 0}}(\phi_{\bf 0}^*({\bf x}))^2 
                              & \hat{K} + 2V_{{\bf 0}}|\phi_{\bf 0}({\bf x})|^2  
\end{array}
\right] \cdot  
     \left[ 
       \begin{array}{c} u_{\bf n}({\bf x}) \\ v_{\bf n}({\bf x})
       \end{array}
     \right] \hspace*{0.4in} \\
= \hbar\omega_{\bf n} 
\left[
\begin{array}{cc}
1 & 0 \\
0 & - 1 
\end{array}
\right] \cdot
     \left[ 
       \begin{array}{c} u_{\bf n}({\bf x}) \\ v_{\bf n}({\bf x})
       \end{array}
     \right]~ \nonumber         
\end{eqnarray} 
and we introduced the operator 
$\hat{K} = - \hbar^2 \mbox{\boldmath $\nabla$}^2/2m 
                                                 + V^{\rm ex}({\bf x}) - \mu$.
This is de Bogoliubov-de Gennes equation that has recently been applied with 
great succes to the collective modes of a Bose condensed rubidium and sodium 
gas \cite{singh,keith2}. Note that a special solution with 
$\hbar\omega_{\bf 0} = 0$ is given by 
$[u_{\bf 0}({\bf x}),v_{\bf 0}({\bf x})] 
                        = [\phi_{\bf 0}({\bf x}),-\phi^*_{\bf 0}({\bf x})]$.
As a result we can, by making use of the fact that the left-hand side of the 
Bogoliubov-de Gennes equation involves a hermitian operator, easily proof that 
all the solutions obey 
\begin{equation}
\int d{\bf x}~ \phi^*_{\bf 0}({\bf x}) u_{\bf n}({\bf x}) 
+ \int d{\bf x}~ \phi_{\bf 0}({\bf x}) v_{\bf n}({\bf x}) = 0~,~~
\end{equation}
as required by the condition in equation (\ref{ort}). Moreover, we can similary 
show that the solutions with $\hbar\omega_{\bf n} > 0$ can always be normalized 
as \cite{J}
\begin{equation}
\int d{\bf x}~ (|u_{\bf n}({\bf x})|^2 - |v_{\bf n}({\bf x})|^2) = 1~.
\end{equation}

Physically, the zero frequency solution ${\bf n} = {\bf 0}$ describes the 
dynamics of the global phase of the condensate \cite{maciek}. Because of the 
so-called $U(1)$ symmetry of the action, i.e., its invariance under the 
simultaneous phase changes 
$\phi({\bf x},\tau) \rightarrow e^{i\vartheta} \phi({\bf x},\tau)$ and
$\phi^*({\bf x},\tau) \rightarrow e^{-i\vartheta} \phi^*({\bf x},\tau)$, this
solution is essentially of no importance to the thermodynamic properties of a 
macroscopic gas sample and is therefore usually neglected. Nevertheless, it has 
from a fundamental point of view some interesting consequences, as we will see 
in section \ref{PD}. Knowing all the eigenstates of ${\bf G}^{-1}$ we can then 
easily perform the inversion and finally again determine the density profile of 
the noncondensed atoms. Keeping the physical significance of the zero frequency 
mode in mind, we ultimately find,
\begin{equation}
\label{ncda}
n'({\bf x}) = \sum_{{\bf n} \neq {\bf 0}} 
  \left( (|u_{\bf n}({\bf x})|^2 + |v_{\bf n}({\bf x})|^2)
                          \frac{1}{e^{\beta\hbar\omega_{\bf n}}-1}
         + |v_{\bf n}({\bf x})|^2
  \right)~,
\end{equation}
which may be compared with equation (\ref{ncd}).

At temperatures near absolute zero, we have as a good approximation that 
$n'({\bf x})=0$ and the Bogoliubov theory applies. However at nonzero 
temperatures we thermally excite particles and $n'({\bf x})$ becomes nonzero. 
If 
we treat the effect of the noncondensate part of the gas in the Hartree-Fock 
approximation, we find that $S_{\rm int}[{\phi'}^*,\phi']$ on average adds  
\begin{eqnarray}
&& \hspace*{-0.75in}
S^{\rm HF}_{\rm int}[{\phi'}^*,\phi'] = 
2 \int_0^{\hbar\beta} d\tau \int d{\bf x}~  
    V_{{\bf 0}}  n'({\bf x}) {\phi'}^*({\bf x},\tau) \phi_{\bf 0}({\bf x}) \\
&&+ 2 \int_0^{\hbar\beta} d\tau \int d{\bf x}~ 
         V_{{\bf 0}} n'({\bf x}) \phi'({\bf x},\tau) \phi_{\bf 0}^*({\bf x}) 
                \nonumber \\
&&+ 2 \int_0^{\hbar\beta} d\tau \int d{\bf x}~ 
     V_{{\bf 0}}  n'({\bf x}) {\phi'}^*({\bf x},\tau) \phi'({\bf x},\tau) ~,
       \nonumber 
\end{eqnarray}
to the action $S_0[{\phi'}^*,\phi']$. In figure 13 we indicate how this can be 
understood diagramatically. Performing the same analysis as before, we conclude 
that the Gross-Pitaevskii equation is modified to
\begin{equation}
\left( - \frac{\hbar^2 \mbox{\boldmath $\nabla$}^2}{2m} 
             + V^{\rm ex}({\bf x}) + 2V_{{\bf 0}}  n'({\bf x})
             + V_{{\bf 0}} |\phi_{\bf 0}({\bf x})|^2
     \right) \phi_{\bf 0}({\bf x}) = \mu \phi_{\bf 0}({\bf x})~,
\end{equation} 
and the normal selfenergies are changed into 
$2V_{{\bf 0}}|\phi_{\bf 0}({\bf x})|^2 + 2V_{{\bf 0}} n'({\bf x}) 
                                                 = 2V_{{\bf 0}}n({\bf x})$. 
The Bogoliubov-de Gennes equation for the elementary excitations is, therefore, 
now given by
\begin{eqnarray}
\label{bdgt}
\left[ 
\begin{array}{cc}
\hat{K} + 2V_{{\bf 0}} n({\bf x})  &  V_{{\bf 0}}(\phi_{\bf 0}({\bf x}))^2 \\
V_{{\bf 0}}(\phi_{\bf 0}^*({\bf x}))^2 
                              & \hat{K} + 2V_{{\bf 0}} n({\bf x})  
\end{array}
\right] \cdot  
     \left[ 
       \begin{array}{c} u_{\bf n}({\bf x}) \\ v_{\bf n}({\bf x})
       \end{array}
     \right] \hspace*{0.4in} \\
= \hbar\omega_{\bf n} 
\left[
\begin{array}{cc}
1 & 0 \\
0 & - 1 
\end{array}
\right] \cdot
     \left[ 
       \begin{array}{c} u_{\bf n}({\bf x}) \\ v_{\bf n}({\bf x})
       \end{array}
     \right]~. \nonumber         
\end{eqnarray} 

These last two equations in combination with equation (\ref{ncda}) are known as 
the Popov theory in the recent literature \cite{popov}. It is much studied at 
present in the context of Bose-Einstein condensation in atomic gases, and has 
been applied with succes to the equilibrium density profile of the gas below the 
critical temperature \cite{K}. It has also been used to determine the 
collective mode frequencies of the gas at nonzero temperatures, however, with 
much less success \cite{L1,L2}. The reason for the failure of the Popov theory 
in this case is that the Bogoliubov-de Gennes equation in equation (\ref{bdgt}) 
describes physically only the motion of the condensate in the presence of a 
static noncondensed cloud and not the dynamics of the noncondensed cloud 
itself. How that can be achieved is discussed lateron in section \ref{modes}, 
when we have obtained a better understanding of the nonequilibrium theory. Now 
we briefly want to make a connection to the interaction parameter $V_{\bf 0}$ 
used in the Bogoliubov and Popov theories and the specific two-body scattering 
properties of the atomic gas of interest in a particular experiment.

\begin{figure}
\begin{center}
\includegraphics[height=0.45\hsize]{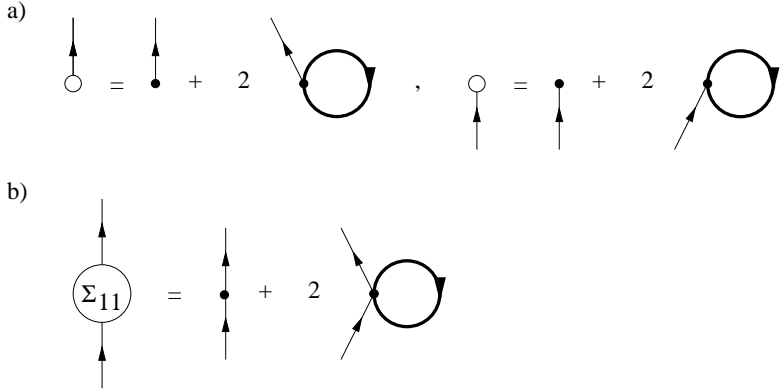}
\end{center}
\caption{Hartree-Fock corrections to a) the linear and b) the quadratic
         interaction terms of the Bogoliubov theory. This represents the
         Popov theory.}
\end{figure}
  
\subsubsection{Some atomic physics}
\label{AP}
In the previous section we mentioned that the Bogoliubov theory, and to a 
certain extent also the Popov theory, agree very well with experiment. It is 
clear, however, that to apply these theories to an actual experiment we need to 
know the interaction parameter $V_{\bf 0}$. In the case of atomic gases it is 
indeed possible to perform an {\it ab initio} calculation of this quantity, 
something which for instance cannot be done for liquid helium. The reason why 
$V_{\bf 0}$ can be determined for atomic gases, is that under the experimental 
conditions of interest the densities are always so low that we only need to 
consider all two-body processes taking place in the gas and we can neglect 
three-body and higher-body processes. This implies physically that we only have 
to calculate and add the quantum mechanical amplitudes for two atoms to scatter 
of each other an arbitrary number of times. Diagrammatically the procedure can 
essentially be summarized by the T-matrix equation in figure 14, because by 
iteration of this equation we easily see that we are indeed summing all two-body 
interacion processes.  

\begin{figure}
\begin{center}
\includegraphics[height=0.4\hsize]{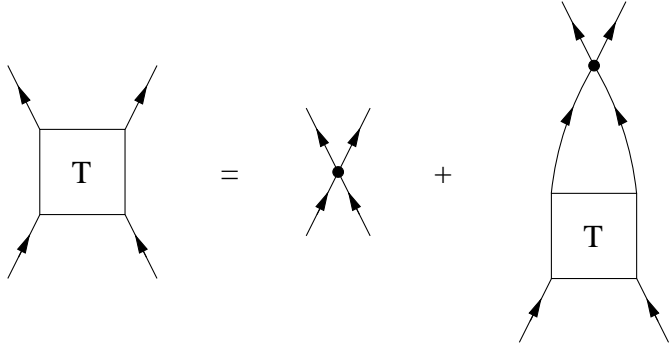}
\end{center}
\caption{T-matrix equation for the effective interatomic interaction.}
\end{figure} 

Denoting the total momentum of the two incoming particles by $\hbar {\bf K}$ 
and the sum of the two Matsubara frequencies by $\Omega_n$, the T-matrix 
equation  in figure 14 can, with our knowledge of how to perform the Matsubara 
sum in the right-hand side, easily be shown to be mathematically equivalent to
\begin{eqnarray}
\label{tmatrix}
&& \hspace*{-0.45in}
T({\bf K},i\Omega_n) = V_{\bf 0} + \frac{V_{\bf 0}}{V} \sum_{\bf k} 
   \frac{1}{i\hbar\Omega_n - \epsilon_{{\bf K}/2+{\bf k}}
            - \epsilon_{{\bf K}/2-{\bf k}} +2\mu} \\
&& \hspace*{0.22in} \times
   \left(1 + \frac{1}{e^{\beta(\epsilon_{{\bf K}/2+{\bf k}}-\mu)}-1}
           + \frac{1}{e^{\beta(\epsilon_{{\bf K}/2-{\bf k}}-\mu)}-1} \right)
   T({\bf K},i\Omega_n)~. \nonumber        
\end{eqnarray}
It is immediately solved by 
\begin{eqnarray}
&& \hspace*{-0.5in}
\frac{1}{T({\bf K},i\Omega_n)} = \frac{1}{V_{\bf 0}} - \frac{1}{V} \sum_{\bf k} 
   \frac{1}{i\hbar\Omega_n - \epsilon_{{\bf K}/2+{\bf k}}
            - \epsilon_{{\bf K}/2-{\bf k}} +2\mu} \\
&& \hspace*{0.55in} \times
   \left(1 + \frac{1}{e^{\beta(\epsilon_{{\bf K}/2+{\bf k}}-\mu)}-1}
           + \frac{1}{e^{\beta(\epsilon_{{\bf K}/2-{\bf k}}-\mu)}-1} 
   \right)~. \nonumber        
\end{eqnarray}  
Note that this is a somewhat formal result, because the momentum sum in the 
right-hand side has an ultra-violet divergence, which from a field-theoretical 
point of view requires a renormalization procedure.

The physical reason for the divergence is that we have used the approximate 
potential $V_{\bf 0} \delta({\bf x}-{\bf x}')$ instead of the actual 
interactomic potential $V({\bf x}-{\bf x}')$. The argument for using the 
$\delta$-function approximation was that the thermal de Broglie wavelength of 
the atoms is for the ultra-low temperatures of interest always much larger than 
the typical range of the interaction. We now see that this argument is not 
fully correct, because if we calculate corrections in perturbation theory, we 
have to deal with momentum sums which are not always restricted to momenta of 
order $\hbar/\Lambda_{\rm th}$ and are therefore sensitive to the precise 
details of the interaction potential. 

To cure this disease we note that if we put in equation (\ref{tmatrix}) the 
Bose occupation numbers equal to zero, we precisely get the T-matrix or 
Lippmann-Schwinger equation \cite{M} for two atoms interacting with the 
potential $V_{\bf 0} \delta({\bf x}-{\bf x}')$. Furthermore, if the atoms 
interact with the potential $V({\bf x}-{\bf x}')$ the solution of the 
corresponding Lippmann-Schwinger equation is known from elementary scattering 
theory to be equal to $4\pi a\hbar^2/m$ for small incoming momenta and 
energies, where $a$ is the $s$-wave scattering length. We thus conclude that we 
must interpret
\begin{eqnarray}
\frac{1}{V_{\bf 0}} - \frac{1}{V} \sum_{\bf k} 
   \frac{1}{i\hbar\Omega_n - \epsilon_{{\bf K}/2+{\bf k}}
            - \epsilon_{{\bf K}/2-{\bf k}} +2\mu}              \nonumber
\end{eqnarray}
as being equal to $m/4\pi a\hbar^2$ and, therefore, that the desired T-matrix 
in principle obeys
\begin{eqnarray}
\label{tmb}
&& \hspace*{-0.5in}
\frac{1}{T({\bf K},i\Omega_n)} 
 = \frac{m}{4\pi a\hbar^2} - \frac{1}{V} \sum_{\bf k} 
   \frac{1}{i\hbar\Omega_n - \epsilon_{{\bf K}/2+{\bf k}}
            - \epsilon_{{\bf K}/2-{\bf k}} +2\mu} \hspace{0.4in} \\
&& \hspace*{0.76in} \times
   \left(\frac{1}{e^{\beta(\epsilon_{{\bf K}/2+{\bf k}}-\mu)}-1}
           + \frac{1}{e^{\beta(\epsilon_{{\bf K}/2-{\bf k}}-\mu)}-1} 
   \right)~. \nonumber  
\end{eqnarray}

To obtain an accurate theory for a trapped atomic Bose gas that includes the 
effect of all two-body processes, we should now use instead of $V_{\bf 0}$ the 
above T-matrix in the Bogoliubov or Popov theories. Of course, in the normal 
phase the same is true for the Hartree-Fock theory. If the temperature is not 
too close to the critical temperature for Bose-Einstein condensation, it turns 
out that $T({\bf K},i\Omega_n) \simeq 4\pi a\hbar^2/m$ and we arrive at the 
conclusion that for an application to realistic atomic gases we must replace 
everywhere in section \ref{SF} the interaction parameter $V_{\bf 0}$ by 
$4\pi a\hbar^2/m$. For temperatures close to the critical temperature, this is 
however no longer true and the Bose distribution functions have a nonnegligible 
effect of the T-matrix \cite{N1,N2}. The theory that, by using `dressed' 
one-particle propagators in figure 14, selfconsistently includes these effects 
of the medium on the scattering properties of the atoms is known as the 
many-body T-matrix theory. It results in an effective interaction that, to avoid 
confusion, is denoted by $T^{(+)}({\bf K},\Omega)$ in section \ref{CF}.
As we will see then, it plays a crucial role if we want to arrive at a 
consistent discription of condensate formation in an atomic Bose gas.

\subsubsection{Superconductivity}
\label{SC}
Finally, we want to briefly discuss the Bardeen-Cooper-Schrieffer or BCS theory 
of superconductivity \cite{BCS}, which has received considerable attention 
recently in connection with ongoing experiments with the fermionic isotope of  
lithium. The reasons for this will become clear shortly. We are in that case 
dealing with effective spin $1/2$ fermions and the action is taken to be
\begin{eqnarray}
&& \hspace*{-0.2in}
S[\phi^*,\phi] \\
&& = \sum_{\alpha=\uparrow,\downarrow}
\int_0^{\hbar\beta} d\tau \int d{\bf x}~
 \phi^*_{\alpha}({\bf x},\tau)
  \left\{ \hbar \frac{\partial}{\partial\tau} 
      - \frac{\hbar^2 \mbox{\boldmath $\nabla$}^2}{2m} - \mu
  \right\} \phi_{\alpha}({\bf x},\tau) \nonumber \\
&&+ \int_0^{\hbar\beta} d\tau \int d{\bf x}~ V_{\bf 0}
           \phi^*_{\uparrow}({\bf x},\tau) 
           \phi^*_{\downarrow}({\bf x},\tau)
             \phi_{\downarrow}({\bf x},\tau)
             \phi_{\uparrow}({\bf x},\tau)~. \nonumber
\end{eqnarray}
Note that the use of a single chemical potential implies that we only consider 
the optimal case of an equal density in each hyperfine state. This situation is 
optimal in the sense that, for unequal densities in each hyperfine state, it is 
no longer possible to pair up all the atoms in the gas. As a result the critical 
temperature of the gas drops dramatically, and in general becomes experimentally 
inaccessible. The effect is to a large extent analogous to putting superfluid 
$^3$He in an homogeneous magnetic field. Furthermore, we in first instance 
consider the homogeneous case, because we want to illustrate in this section the 
local-density approximation to include the effect of the external trapping 
potential. Physically, this approximation treats the gas as consisting of a 
large number of independent gases that are in diffusive equilibrium with each 
other. Such an approach only works if the correlation length of the gas is much 
smaller than the typical length scale associated with changes of the external 
potential. Fortunately, this is almost always the case for realistic trapped 
atomic gases and for that reason the local-density approximation is often used 
in practice. 

The BCS theory is the theory of Bose-Einstein condensation of so-called Cooper 
pairs. This means that the order parameter is
$\langle \phi_{\downarrow}({\bf x},\tau)
             \phi_{\uparrow}({\bf x},\tau) \rangle$, in analogy with the order 
parameter $\langle \phi({\bf x},\tau) \rangle$ for the Bose case just 
discussed. 
Furthermore, it requires that the interaction parameter $V_{\bf 0}$ is 
negative, 
since otherwise the formation of pairs would not be energetically favorable. 
From now on we assume, therefore, that this is the case. The condensate of 
Cooper pairs can also be nicely treated with a Hubbard-Stratonovich 
transformation. We now introduce a complex field $\Delta({\bf x},\tau)$ and use
\begin{eqnarray}
&& \hspace*{-0.4in}
\exp \left\{ - \frac{1}{\hbar}
  \int_0^{\hbar\beta} d\tau \int d{\bf x}~ V_{\bf 0}
           \phi^*_{\uparrow}({\bf x},\tau) 
           \phi^*_{\downarrow}({\bf x},\tau)
           \phi_{\downarrow}({\bf x},\tau)
           \phi_{\uparrow}({\bf x},\tau)
     \right\} \\
&&= \int d[\Delta^*]d[\Delta]~
  \exp \left\{ \frac{1}{\hbar}
    \int_0^{\hbar\beta} d\tau \int d{\bf x}~ \left(
     \frac{|\Delta({\bf x},\tau)|^2}{V_{\bf 0}} 
   \right. \right.  \nonumber \\
&&+ \left. \left. \raisebox{0.2in}{} 
     \Delta^*({\bf x},\tau)
              \phi_{\downarrow}({\bf x},\tau)
              \phi_{\uparrow}({\bf x},\tau) 
   + \phi^*_{\uparrow}({\bf x},\tau) 
            \phi^*_{\downarrow}({\bf x},\tau)
            \Delta({\bf x},\tau)
   \right) \raisebox{0.25in}{} \right\}~. \nonumber
\end{eqnarray}
This leads to a partition function with the action
\begin{eqnarray}
&& \hspace*{-0.2in}
S[\Delta^*,\Delta,\phi^*,\phi] = - \int_0^{\hbar\beta} d\tau \int d{\bf x}~
     \frac{|\Delta({\bf x},\tau)|^2}{V_{\bf 0}} \\
&&- \frac{\hbar}{2}
  \int_0^{\hbar\beta} d\tau \int d{\bf x}~ 
  \left[ \phi^*_{\downarrow}({\bf x},\tau) ,
         \phi_{\uparrow}({\bf x},\tau) \right] \cdot {\bf G}^{-1}  \cdot
\left[ \begin{array}{c}
\phi_{\downarrow}({\bf x},\tau) \\
\phi^*_{\uparrow}({\bf x},\tau) \end{array} \right]~, \nonumber
\end{eqnarray}
where 
\begin{eqnarray}
&& \hspace*{-0.4in}                                
{\bf G}^{-1}({\bf x},\tau;{\bf x}',\tau') = 
                   {\bf G}_0^{-1}({\bf x},\tau;{\bf x}',\tau') \\
&& \hspace*{0.6in} - \frac{1}{\hbar}
\left[ 
\begin{array}{cc}
0  & \Delta({\bf x},\tau) \\
\Delta^*({\bf x},\tau) & 0
\end{array} \right] \delta({\bf x}-{\bf x}') \delta(\tau-\tau')~, \nonumber
\end{eqnarray}
and the noninteracting Green's function ${\bf G}_0$ is defined by
\begin{equation} 
{\bf G}_0^{-1}({\bf x},\tau;{\bf x}',\tau') =
\left[ 
\begin{array}{cc}
G_0^{-1}({\bf x},\tau;{\bf x}',\tau') & 0 \\
0 & -G_0^{-1}({\bf x}',\tau';{\bf x},\tau)
\end{array}
\right]    
\end{equation}
and 
\begin{eqnarray}
G_0^{-1}({\bf x},\tau;{\bf x}',\tau') 
= - \frac{1}{\hbar} \left\{ \hbar\frac{\partial}{\partial\tau}
    - \frac{\hbar^2 \mbox{\boldmath $\nabla$}^2}{2m} 
     - \mu  \right\}
 \delta({\bf x}-{\bf x}') \delta(\tau-\tau')~.~~
\end{eqnarray}

We thus see that the fermionic part has exactly the same matrix structure as in 
the case of a condensed Bose gas, only the selfenergy is now
\begin{equation}
\left[ \begin{array}{cc}
\Sigma_{11} & \Sigma_{12} \\ \Sigma_{21} & \Sigma_{22}
\end{array} \right] = \frac{1}{\hbar}
\left[ 
\begin{array}{cc}
0  &  \Delta({\bf x},\tau) \\
\Delta^*({\bf x},\tau) & 0  
\end{array}
\right] \delta({\bf x}-{\bf x}') \delta(\tau-\tau')~.~~
\end{equation}
If we again integrate out the fermion fields, we get the effective action
\begin{equation}
S^{\rm eff}[\Delta^*,\Delta] = 
        - \int_0^{\hbar\beta} d\tau \int d{\bf x}~
            \frac{|\Delta({\bf x},\tau)|^2}{V_{\bf 0}} 
        - \hbar {\rm Tr}[\ln (-{\bf G}^{-1})]~,
\end{equation}
which we can expand in powers of $\Delta$ by using 
${\bf G}^{-1}={\bf G}_0^{-1} - \mbox{\boldmath $\Sigma$} 
     = {\bf G}_0^{-1}(1 - {\bf G}_0 \mbox{\boldmath $\Sigma$})$ and therefore
\begin{equation}
- \hbar {\rm Tr}[\ln (-{\bf G}^{-1})] = - \hbar {\rm Tr}[\ln (-{\bf G}_0^{-1})]
  + \hbar \sum_{m=1}^{\infty} \frac{1}{m} 
                          {\rm Tr}[({\bf G}_0 \mbox{\boldmath $\Sigma$})^m]~.
\end{equation}
Explicite calculation \cite{hagen1} shows that for space and time independent 
$\Delta$ we obtain a `free-energy' density of the form of the Landau-theory of 
second-order phase transitions, i.e.,
\begin{equation}
f_L(|\Delta|) = \alpha(T) |\Delta|^2 
  + N(0) \frac{7\zeta(3)}{16(\pi T_c)^2} |\Delta|^4 ~,
\end{equation}
with as expected
\begin{eqnarray}
&& \hspace*{-0.35in}
\alpha(T) = -\frac{1}{V_{\bf 0}} + \frac{1}{V} \sum_{{\bf k}}
            \frac{1}{2(\epsilon_{{\bf k}}-\mu)}   
              \left( 1 - \frac{2}{e^{\beta(\epsilon_{{\bf k}}-\mu)}+1} 
              \right)  \hspace{0.4in} \\
&&\equiv N(0) \ln \left( \frac{T}{T_c} \right)
          \simeq N(0) \left( \frac{T}{T_c} - 1 \right)~.
\end{eqnarray}
Here $N(0)$ is the density of states of a single spin state at the Fermi energy 
$\epsilon_F \equiv \hbar^2k_F^2/2m$ and the critical temperature is given by
\begin{equation}
T_c = \frac{8e^{\gamma-2}}{\pi} \frac{\epsilon_F}{k_B}
       \exp \left\{- \frac{\pi}{2k_F|a|} \right\}~,~~
\end{equation}
if we use the same renormalization procedure as in section \ref{AP} to 
eliminate 
the interaction parameter $V_{\bf 0}$ in favor of the negative $s$-wave 
scattering length $a$. 

Below $T_c$ we thus have a nonzero average 
$\langle \Delta({\bf x},\tau) \rangle \equiv \Delta_0$. Using this in the 
Green's function for the fermions and neglecting fluctations, we find in 
momentum space that
\begin{equation}
-\hbar {\bf G}^{-1}({\bf k},i\omega_n) =
\left[ 
\begin{array}{cc}
-i\hbar\omega_n + \epsilon_{{\bf k}} -\mu & \Delta_0 \\
\Delta^*_0 &
-(i\hbar\omega_n + \epsilon_{{\bf k}} -\mu)  
\end{array}
\right]
\end{equation}
and therefore poles in ${\bf G}({\bf k},\omega)$ if
\begin{equation}
\hbar\omega_{{\bf k}} 
  = \sqrt{(\epsilon_{{\bf k}}-\mu)^2 + |\Delta_0|^2}~.
\end{equation}
This disperion relation has clearly a gap of magnitude $|\Delta_0|$ at the 
Fermi 
surface. As a consequence $|\Delta_0|$ is known as the BCS gap parameter. Note 
that the critical temperature below which the gap becomes nonzero depends 
exponentally on the parameter $\pi/2k_F|a|$. For typical atomic gases this is a 
quantity which is much larger than one, and the BCS transition is 
experimentally 
inaccesible. The only exception at the moment appears to be $^6$Li, with its 
anomalously large and negative triplet scattering length of $-2160~a_0$ 
\cite{O}. This explains the present interest in spin-polarized atomic lithium. 

In view of this exciting possibility, we have recently studied the equilibrium 
properties of atomic lithium is a harmonic oscillator potential 
$V^{\rm ex}({\bf x}) = m\omega^2{\bf x}^2/2$, with a trapping frequency of
$\omega/2\pi = 144$ Hz \cite{marianne2}. To incorporate the effect of the 
external potential we have, as mentioned above, applied the local-density 
approximation. The result is shown in figure 15. Note that the use of the 
local-density approximation implies that we perform the above outlined 
homogeneous calculation for each point in space with a chemical potential that 
is equal to
\begin{equation}
\mu({\bf x}) = \mu - \frac{1}{2} m\omega^2{\bf x}^2
               - \frac{4\pi a\hbar^2}{m} \frac{n({\bf x})}{2}~.~~
\end{equation} 
The third term in the right-hand side represents the mean-field effect of the 
Hartree contribution to the selfenergy of the fermions. Since our 
Hubbard-Stratonovich procedure is in principle exact, it is not immediately 
clear why such a term must be included in the theory. It can, however, be shown 
that it arises from the fluctuations of the BCS gap parameter. With this 
remark, we end our development of the equilibrium field theory of trapped atomic 
gases. We now turn our attention to the nonequilibrium field theory. 

\begin{figure}
\begin{center}
\includegraphics[height=0.93\hsize]{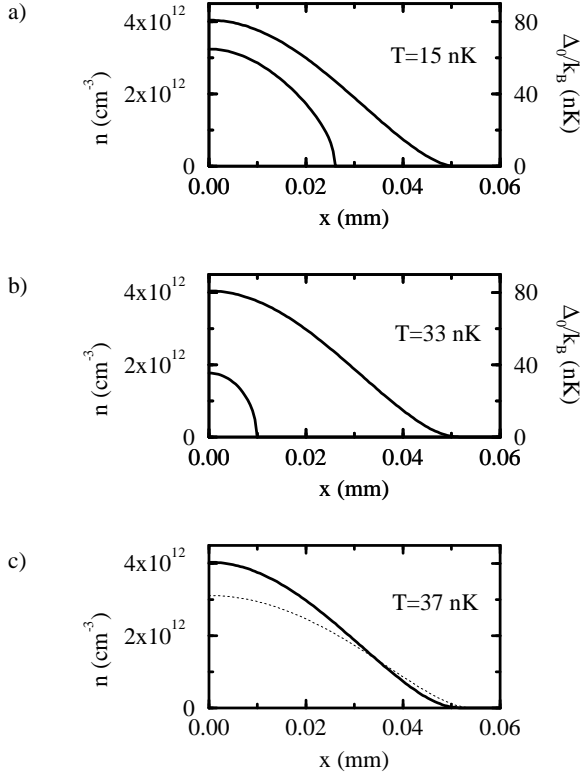}
\end{center}
\caption{Density distribution $n({\bf x})$ and energy gap $\Delta_0({\bf x})$ 
         for a $^6$Li atomic gas consisting
         of $2.865 \times 10^5$ atoms in each spin state a) at $T=15$ nK,
         b) at $T=33$ nK, slightly below $T_c$, and c) at $T=T_c=37$ nK.
         The left scale of each plot refers to the density and the right scale
         to the energy gap. The dotted line in c) shows the density 
         distribution for an ideal Fermi gas with the same number of 
         particles and at the same temperature.}
\end{figure} 
\section{Nonequilibrium field theory}
\label{NFT}
As we have already seen several times, the equlibrium field theory gives us 
also 
information on dynamical properties of the gas by means of the substitution 
$\omega_n \rightarrow -i\omega$ or equivalently $\tau \rightarrow it$. However, 
the Popov theory of Bose-Einstein condensation has shown us that we have to be 
very careful with this procedure, because we may not always end up with the 
correct physics needed for a discription of a particular experiment. It is for 
these cases that a truly nonequilibrium field theory is required and the 
development of such a theory is the topic of section \ref{QKT}. Before 
embarking 
on that, however, we first consider two important dynamical problems where the 
equilibrium theory does give us all the answers.

\subsection{Macroscopic quantum tunneling of a condensate}
\label{NSL}
In a Bose gas with effectively attractive interactions, i.e., with a negative 
scattering length $a$, a condensate will always have the tendency to collapse 
to a high density state due to the gain in energy that can be obtained in this 
way \cite{neg}. 
The most important question in this context is, therefore, if a condensate can 
exist sufficiently long to be experimentally observed. Neglecting the variation 
of the noncondensate density on the size of the condensate 
\cite{marianne1,tom}, we know from the Popov theory that the dynamics of the 
collapse is, apart from an unimportant shift in the chemical potential, 
determined by the Gross-Pitaevskii equation \cite{GP}
\begin{eqnarray}
\label{GPeq}
&& \hspace*{-0.2in}
i\hbar \frac{\partial}{\partial t} \langle \phi({\bf x},t) \rangle \\
&& = \left\{ - \frac{\hbar^2 \mbox{\boldmath $\nabla$}^2}{2m} 
             + V^{\rm ex}({\bf x}) - \mu   
             + \frac{4\pi a\hbar^2}{m}
                   |\langle \phi({\bf x},t) \rangle|^2 \right\}
                         \langle \phi({\bf x},t) \rangle~. \nonumber
\end{eqnarray}         
More precisely, this determines only the semiclassical dynamics. If we also 
want to study the quantum fluctuations, which is necessary if we are also 
interested in how the condensate tunnels through the macroscopic energy 
barrier, it is most convenient to calculate the grand canonical partition 
function of the condensate \cite{tunnel}. Quantizing the Gross-Pitaevskii 
equation we obtain for this partition function the functional integral
\begin{equation}
Z(\mu) = \int d[\phi^*]d[\phi]~ 
                    \exp \left\{ - \frac{1}{\hbar} S[\phi^*,\phi] \right\}~,
\end{equation}
over the complex field $\phi({\bf x},\tau)$ and with the Euclidian action
\begin{eqnarray}
&& \hspace*{-0.5in}
S[\phi^*,\phi] =                            
   \int_0^{\hbar\beta} d\tau \int d{\bf x}~ 
     \phi^*({\bf x},\tau) \left( \hbar \frac{\partial}{\partial \tau}
            + \frac{\hbar^2 \mbox{\boldmath $\nabla$}^2}{2m} \right. \\
&&+ \left. V^{\rm ex}({\bf x}) - \mu   
            + \frac{2\pi a\hbar^2}{m}
                   |\phi({\bf x},\tau)|^2 \right) \phi({\bf x},\tau)~. 
\nonumber
\end{eqnarray}
As always for Bose systems, the integration is only over fields that are 
periodic on the imaginary time axis.

Although it has recently been shown by Freire and Arovas that the tunneling 
process can also be studied in terms of the complex field $\phi({\bf x},\tau)$ 
\cite{arovas}, we believe that it leads to somewhat more physical insight if we 
use instead the fields $\rho({\bf x},\tau)$ and $\theta({\bf x},\tau)$ that 
correspond to the density and phase fluctuations of the condensate, 
respectively. They are introduced by performing the canonical variable 
transformation \cite{popov}
\begin{eqnarray}
\phi({\bf x},\tau) = \sqrt{\rho({\bf x},\tau)} e^{i \theta({\bf x},\tau)}
                                                             \nonumber 
\end{eqnarray}
in the functional integral for the partition function. As a result we find
\begin{equation}
\label{gr}
Z(\mu) = \int d[\rho]d[\theta]~ 
              \exp \left\{ - \frac{1}{\hbar} S[\rho,\theta;\mu] \right\}~,
\end{equation} 
with
\begin{eqnarray}
&& \hspace*{-0.56in}
S[\rho,\theta;\mu] =
   \int_0^{\hbar\beta} d\tau \int d{\bf x}~ 
   \left( i\hbar \rho({\bf x},\tau) 
                 \frac{\partial \theta({\bf x},\tau)}{\partial \tau} \right. \\
&&+ \frac{\hbar^2 \rho({\bf x},\tau)}{2m} 
                        (\mbox{\boldmath $\nabla$} \theta({\bf x},\tau))^2 
        + \frac{\hbar^2}{8m\rho({\bf x},\tau)} 
                        (\mbox{\boldmath $\nabla$} \rho({\bf x},\tau))^2 
                        \nonumber \\
&&+ V^{\rm ex}({\bf x})\rho({\bf x},\tau) - \mu\rho({\bf x},\tau) 
      + \left. \frac{2\pi a\hbar^2}{m} \rho^2({\bf x},\tau) \right)~. \nonumber
\end{eqnarray}
Next, we notice that this action is only quadratic in the phase fluctuations. 
The field $\theta({\bf x},\tau)$ can therefore be integrated over exactly, 
because it only involves the evaluation of a gaussian integral. 

Compared to ordinary gaussian integrals there is, however, one slight 
complication which is associated with the fact that $\theta({\bf x},\tau)$ are 
phase variables. This implies that the periodicity of the original field 
$\phi({\bf x},\tau)$ only constraints the phase field $\theta({\bf x},\tau)$ to 
be periodic up to a multiple of $2\pi$. To evaluate the grand canonical 
partition function in equation~(\ref{gr}) we must therefore first integrate 
over 
all fields $\theta({\bf x},\tau)$ that obey the boundary condition 
$\theta({\bf x},\hbar\beta) = \theta({\bf x},0) + 2\pi j$ and subsequently sum 
over all possible integers $j$. Because these different boundary conditions 
only affect the zero-momentum part of $\theta({\bf x},\tau)$ we first have to 
evaluate the sum
\begin{eqnarray}
\sum_{j} \int^{\theta_{\bf 0}(\hbar\beta)=\theta_{\bf 0}(0)+2\pi j}   
 d[\theta_{\bf 0}]~ 
   \exp \left\{ -i \int_0^{\hbar\beta} d\tau~ N_{\bf 0}(\tau)
                        \frac{\partial \theta_{\bf 0}(\tau)}{\partial \tau}
        \right\}~,                                             \nonumber
\end{eqnarray}
with $N_{\bf 0}(\tau) = \int d{\bf x}~ \rho({\bf x},\tau)$ the number of 
condensate 
particles. After performing a partial integration on the integral in the 
exponent, we can carry out the path integration over $\theta_{\bf 0}(\tau)$ to 
obtain
\begin{eqnarray}
\sum_{j} e^{2\pi i N_{\bf 0} j} \delta \left[ 
               \frac{\partial N_{\bf 0}(\tau)}{\partial \tau} \right]~. 
                                         \nonumber
\end{eqnarray}
As expected, the integration over the global phase of the condensate leads to 
the constraint of a constant number of condensate particles, i.e., 
$N_{\bf 0}(\tau)=N_{\bf 0}$. Moreover, we have $\sum_{j} e^{2\pi i N_{\bf 0} j} 
= \sum_{j} 
\delta(N_{\bf 0} - j)$, which restricts the number of condensate particles to 
an integer. Putting all these results together, we see that the integration over 
the zero-momentum part of $\rho({\bf x},\tau)$ is only a sum over the number of 
condensate particles and we have that
\begin{equation}
Z(\mu) = \sum_{N_{\bf 0}} e^{\beta\mu N_{\bf 0}} Z_{N_{\bf 0}}~.
\end{equation}
Here we introduced the canonical partition function of the condensate, which is 
apparently equal to the functional integral  
\begin{equation}
Z_{N_{\bf 0}} = \int d[\rho]d[\theta]~ 
                 \exp \left\{ - \frac{1}{\hbar} S[\rho,\theta;0] \right\}
\end{equation} 
over all the nonzero momentum components of the density and phase fields.

The integration over the nonzero momentum components of the phase field 
$\theta({\bf x},\tau)$ is easily performed, because it now involves an ordinary 
gaussian integral. Introducing the Green's function for the phase fluctuations 
$G({\bf x},{\bf x}';\rho)$ by
\begin{equation}
\label{geq}
\frac{\hbar}{m} \left( (\mbox{\boldmath $\nabla$} \rho) 
  \cdot \mbox{\boldmath $\nabla$}
      + \rho \mbox{\boldmath $\nabla$}^2 \right) G({\bf x},{\bf x}';\rho) =
                                 \delta({\bf x}-{\bf x}')~,
\end{equation}
we immediately obtain the desired effective action for the density field
\begin{eqnarray}
&& \hspace*{-0.4in}
S^{\rm eff}[\rho] = \int_0^{\hbar\beta} d\tau \int d{\bf x} \int d{\bf x}'~
    \left(-\frac{\hbar}{2} \frac{\partial \rho({\bf x},\tau)}
                                 {\partial \tau}
             G({\bf x},{\bf x}';\rho)
                            \frac{\partial \rho({\bf x}',\tau)}
                                 {\partial \tau}
    \right)  \\
     &&+ \int_0^{\hbar\beta} d\tau \int d{\bf x}~
             \left(
               \frac{\hbar^2}{8m\rho({\bf x},\tau)} 
                            (\mbox{\boldmath $\nabla$} \rho({\bf x},\tau))^2
               + V^{\rm ex}({\bf x}) \rho({\bf x},\tau) \right.  \nonumber \\
     &&\hspace*{2.4in} + \left. \frac{2\pi a\hbar^2}{m} \rho^2({\bf x},\tau)
             \right)~. \nonumber
\end{eqnarray}
Being an action for the density fluctuations of the condensate, 
$S^{\rm eff}[\rho]$ also describes all the collisionless modes of the 
condensate. This is important for our purposes, because the mode which becomes 
unstable first, determines precisely how the condensate collapses. Moreover, it 
determines the probability with which the collapse is going to take place, both 
for quantum and thermal fluctuations, since the energy barrier is smallest in 
that direction of the configuration space. It should be noted that as long as 
we can neglect the interaction between the condensate and the thermal cloud, 
the action $S^{\rm eff}[\rho]$ 
describes also the collective modes of a gas with positive scattering length. 
For various other theoretical approaches that have been applied under these 
conditions see, for example, 
Refs.~\cite{sandro,singh,keith2,yvan2,peter3,keith3,juha,li,stig}. The actual 
measurements of the collective mode frequencies have been performed by 
Jin {\it et al.} \cite{coll1} and Mewes {\it et al.} \cite{coll2} and are at 
sufficiently low temperatures indeed in good agreement with the theoretical 
predictions \cite{L1,L2}. We expect the same to be true for a gas with 
effectively attractive interactions and, therefore, the action 
$S^{\rm eff}[\rho]$ to be a good starting point for the following discussion. 

To obtain the collisionless modes explicitly we consider first the case of an 
ideal Bose gas by putting $a=0$. For the ideal Bose gas we expect the gaussian 
profile 
\begin{equation}
\label{gprof}
\rho({\bf x};q(\tau)) = N_{\bf 0} \left( \frac{1}{\pi q^2(\tau)} \right)^{3/2}
                       \exp \left( - \frac{{\bf x}^2}{q^2(\tau)} \right)
\end{equation}
to describe an exact mode of the condensate. The reason is that in the 
noninteracting case we can make a density fluctuation by taking one particle 
from the condensate and putting that in one of the excited states of the 
external potential. The corresponding density fluctuation obeys
\begin{eqnarray}
\delta\rho({\bf x},t) \propto e^{-i(\epsilon_{\bf n}-\epsilon_{\bf 0})t/\hbar}
    \chi^*_{\bf n}({\bf x}) \chi_{\bf 0}({\bf x})~.                      
\nonumber
\end{eqnarray} 
For the experimentally relevant case of an isotropic harmonic oscillator 
\cite{Rice} it is more convenient to use instead of the cartesian quantum 
numbers ${\bf n}$, the two angular momentum quantum numbers $\ell$ and $m$ and 
the quantum number $n$ that counts the number of nodes in the radial 
wave function $\chi_{n\ell}(x)$. The density fluctuation then becomes
\begin{eqnarray}
\delta\rho({\bf x},t) \propto e^{-i(2n + \ell)\omega t}
    \chi_{n\ell}(x) Y_{\ell m}^*({\bf \hat{x}})
                          \frac{e^{-x^2/2l^2}}{(\pi l^2)^{3/4}}~,    \nonumber
\end{eqnarray} 
with $\epsilon_{n\ell m} - \epsilon_{000} = (2n + \ell) \hbar \omega$ the 
excitation energy and $l=(\hbar/m\omega)^{1/2}$ the size of the condensate 
wave function. Comparing this now with the expansion of the gaussian profile in 
equation~(\ref{gprof}) around the groundstate density profile, which is 
obtained 
by 
substituting $q(\tau)=l + \delta q(\tau)$, we find that 
\begin{equation}
\delta\rho({\bf x},\tau) = 
    - \sqrt{6} N_{\bf 0} \frac{\delta q(\tau)}{l}
        \chi_{10}(x) Y_{00}^*({\bf \hat{x}})
                             \frac{e^{-x^2/2l^2}}{(\pi l^2)^{3/4}}
\end{equation}
has precisely the same form as a density fluctuation in which one particle is 
taken from the condensate and put into the harmonic oscillator state with 
quantum numbers $(n\ell m) = (100)$. The frequency of this so-called 
`breathing' 
mode described by the gaussian density profile must therefore be equal to 
$2\omega$.

To proof that this is indeed correct, we need to evaluate the effective action 
$S^{\rm eff}[\rho]$, and hence the Green's function $G({\bf x},{\bf x}';\rho)$, 
for a gaussian density profile. Substituting such a profile in 
equation~(\ref{geq}) immediately leads to 
$G({\bf x},{\bf x}';\rho) = G({\bf x},{\bf x}';q)/\rho({\bf x}';q)$, with
\begin{equation}
\frac{\hbar}{m} \left( - \frac{2}{q^2} {\bf x} \cdot \mbox{\boldmath $\nabla$}
      + \mbox{\boldmath $\nabla$}^2 \right) G({\bf x},{\bf x}';q) =
                                 \delta({\bf x}-{\bf x}')~.
\end{equation} 
The latter equation can be solved, if we can solve the eigenvalue problem
\begin{equation}
\left( \mbox{\boldmath $\nabla$}^2 
       - \frac{2x}{q^2} \frac{\partial}{\partial x} \right) 
    \xi({\bf x}) = \lambda \xi({\bf x})~.
\end{equation}
This turns out to be an easy task, because substituting 
\begin{equation}
\xi_{n\ell m}({\bf x}) 
    = \xi_{n\ell}(x) \frac{e^{x^2/2q^2}}{x} Y_{\ell m}({\bf \hat{x}})
\end{equation}
gives essentially the radial Schr\"odinger equation for an isotropic harmonic 
oscillator with frequency $\omega_q = \hbar/mq^2$, i.e.,
\begin{eqnarray}
- \frac{2m}{\hbar^2} 
  \left( - \frac{\hbar^2}{2m} \frac{\partial^2}{\partial x^2} 
         + \frac{1}{2} m\omega_q^2 x^2 + \frac{\hbar^2 \ell(\ell+1)}{2mx^2}
         - \frac{3}{2} \hbar\omega_q \right) \xi_{n\ell}(x) \hspace*{0.5in} \\
         = \lambda_{n\ell} \xi_{n\ell}(x)~. \nonumber
\end{eqnarray}
The desired eigenfunctions are therefore 
$\xi_{n\ell m}({\bf x};q) = \varphi_{n\ell m}({\bf x}) e^{x^2/2q^2}$, with 
$\varphi_{n\ell m}({\bf x})$ the properly normalized harmonic 
oscillator states with the energies $(2n + \ell + 3/2)\hbar\omega_q$, and the 
corresponding eigenvalues are $\lambda_{n\ell}(q) = - 2 (2n + \ell)/q^2$. 
Introducing finally the `dual' eigenfunctions 
$\bar{\xi}_{n\ell m}({\bf x};q)
                          \equiv \varphi^*_{n\ell m}({\bf x}) e^{-x^2/2q^2}$,
the Green's function $G({\bf x},{\bf x}';q)$ is given by
\begin{equation}
\label{gq}
G({\bf x},{\bf x}';q) 
  = {\sum_{n\ell m}}' \xi_{n\ell m}({\bf x};q) 
       \frac{m}{\hbar \lambda_{n\ell}(q)} \bar{\xi}_{n\ell m}({\bf x}';q)~.
\end{equation}
Note that prime on the summation sign indicates that the sum is over all 
quantum numbers except $(n\ell m) = (000)$. The latter is excluded because the 
associated eigenfunction $\xi_{000}({\bf x};q)$ is just a constant and thus 
does not contribute to $G({\bf x},{\bf x}';\rho)$, which is defined as the 
Green's function for all phase fluctuations with nonvanishing momenta.

Putting all these results together, we see that the dynamics of the collective 
variable $q(\tau)$ is determined by the action
\begin{eqnarray}
&& \hspace*{-0.4in}
S^{\rm eff}[q] = \int_0^{\hbar\beta} d\tau~
   \left\{ \frac{3mN_{\bf 0}}{4} \left( \frac{dq}{d\tau} \right)^2 
          + N_{\bf 0} \left( \frac{3\hbar^2}{4mq^2} + \frac{3}{4}m\omega^2q^2
            \right) \right\} \\
&&\equiv \int_0^{\hbar\beta} d\tau~
   \left\{ \frac{1}{2} m^* \left( \frac{dq}{d\tau} \right)^2 + V(q) 
   \right\}~, \nonumber
\end{eqnarray}
that is equivalent to the action of a particle with effective mass 
$m^* = 3mN_{\bf 0}/2$ in a potential 
$V(q) = N_{\bf 0}(3\hbar^2/4mq^2 + 3m\omega^2q^2/4)$. As expected from our 
previous remarks, this potential has a minimum for $q=l$ and can be expanded 
near its minimum as
\begin{equation}
V(q) \simeq \frac{3}{2}N_{\bf 0}\hbar\omega 
               + \frac{1}{2} m^* (2\omega)^2 (\delta q)^2~.
\end{equation}
It thus comfirms that the gaussian profile describes a breathing mode with 
frequency $2\omega$ around an equilibrium density profile that is given by 
$\rho({\bf x};l) = N_{\bf 0} |\chi_{000}({\bf x})|^2$. 

Our next task is to investigate how interactions affect this result.
Considering again only gaussian density profiles, the action $S^{\rm eff}[q]$ 
is 
again that of a particle with effective mass $m^* = 3mN_{\bf 0}/2$ but now in 
the potential \cite{BP}
\begin{equation}
V(q) =  N_{\bf 0} \left( \frac{3\hbar^2}{4mq^2} + \frac{3}{4}m\omega^2q^2
              - \frac{N_{\bf 0}}{\sqrt{2\pi}} \frac{\hbar^2 |a|}{m q^3} 
\right)~.
\end{equation} 
The physically most important feature of this potential is that it is unbounded 
from below, since $V(q) \rightarrow -\infty$ if $q \downarrow 0$. Hence, the 
condensate indeed always has the tendency to collapse to the high-density state 
$\lim_{q \downarrow 0} \rho({\bf x};q) = N_{\bf 0} \delta({\bf x})$. However, 
if the number of condensate particles is sufficiently small, or more precisely 
if 
\cite{fetter2}
\begin{equation}
N_{\bf 0} < \frac{2\sqrt{2\pi}}{5^{5/4}} \frac{l}{|a|} \simeq 0.68 
\frac{l}{|a|}~,
\end{equation}
the condensate has to overcome a macroscopic energy barrier before it can 
collapse. Under these conditions the condensate is therefore really metastable 
and can in principle be observed experimentally. The most important question in 
this respect is of course: How metastable is the condensate? Within the 
gaussian approximation this question is easily answered, because then the 
dynamics of the condensate is equivalent to the dynamics of a particle in an 
unstable potential, as we have just seen. We therefore only need to evaluate 
the WKB-expression for the tunnneling rate \cite{shuryak} and compare this to 
the rate of decay due to thermal fluctuations by calculating also the height of 
the energy barrier. The outcome of this comparison for the conditions of the 
experiment with atomic $^7$Li is presented in reference~\cite{cass2} and shows 
that, for the relatively high temperatures $T \gg \hbar\omega/k_B$ that have 
been obtained thusfar \cite{Rice}, the decay by means of thermal fluctuations 
over the energy barrier is the dominant decay mechanism of the condensate.

More important for our purposes, however, is that sufficiently close to the 
maximum number of condensate particles $N_{\rm max}$ the collective decay of 
the condensate discussed above is always much more probable than the decay due 
to two and three-body collisions that lead to a spin-flip or the formation of 
$^7$Li molecules, respectively. As a result the collapse of the condensate 
should be observable within the finite lifetime of the gas. In fact, on the 
basis of this separation of time scales we expect the condensate to go through 
a number of growth and collapse cycles \cite{cass2,cass3}. Physically this 
picture arises as follows. Starting from a gas with a number of atoms 
$N \gg N_{\rm max}$, the condensate will initially grow as a response to 
evaporative cooling. However, if the number of condensate atoms starts to come 
close to $N_{\rm max}$, the condensate fluctuates over the energy barrier and 
collapses in a very short time of ${\cal O}(1/\omega)$ \cite{lev}. During the 
collapse the condensate density increases rapidly and two and three-body 
inelastic processes quickly remove almost all the atoms from the condensate. 
After this has occurred the condensate grows again from the noncondensed part 
of the gas and a new growth and collapse cycle begins. It is only after many of 
these cycles that enough atoms are removed for the gas to relax to an 
equilibrium with a number of condensate particles that is less than $N_{\rm 
max}$. This is shows quantitatively in figure 16 for the experimental 
conditions of interest.

\begin{figure}
\begin{center}
\includegraphics[height=0.7\hsize]{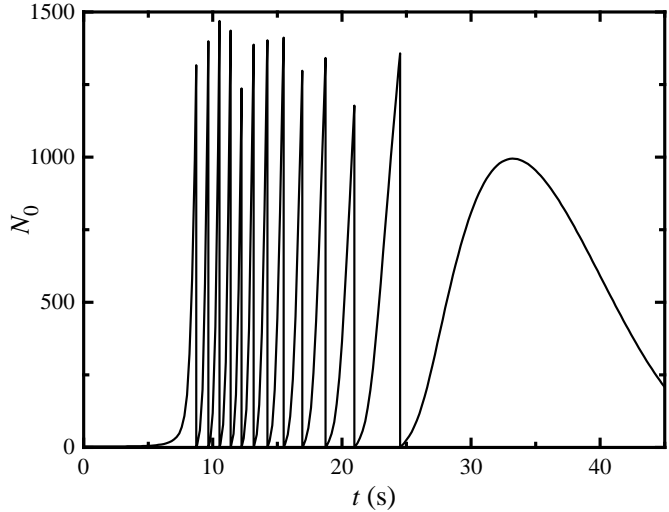}
\end{center}
\caption{Typical evolution of condensate number $N_{\bf 0}$ in response to
         evaporative cooling. The long-time decay of the condensate is due
         to two and three-body inelastic collisions.}
\end{figure} 

A final issue which needs to be addressed at this point is the actual dynamics 
of the collapse and, in particular, how we must include the effect of the 
inelastic growth and decay processes on this dynamics. Unfortunately, the 
inclusion of these effects is rather complicated and we need the results of 
section \ref{CF} to be able to correctly address this problem. It is, however, 
of some interest because Sackett {\it et al.} have recently observed that after 
a single collapse there remains a remnant of the condensate with about 10\% of 
the initial number of atoms \cite{cass4}. At present, it is an important open 
problem to theoretically understand the magnitude of this remnant.  

\subsection{Phase diffusion}
\label{PD}
As we have just seen explicitly, a particularly interesting consequence of the 
finite size of the gas is that quantum fluctuations play a much more important 
role. Although this is especially true for the case of attractive interactions 
that we considered in section \ref{NSL}, it is also true for a Bose gas with 
repulsive interactions. A striking example in this respect is the phenomenon of 
phase `diffusion', which was recently discussed by Lewenstein and You 
\cite{maciek}. We rederive their results for a trapped Bose gas in a moment, 
but 
first consider also the same phenomenon for a neutral and homogeneous 
superconductor. In this manner it is possible to bring out the physics 
involved more clearly.

Using the approach of section \ref{SC}, it can be shown that at zero 
temperature the dynamics of the superconducting order parameter, i.e., 
the BCS gap parameter $\Delta({\bf x},t)$ that is proportional to the 
wave function of the condensate of Cooper pairs, is in a good approximation 
determined by a time-dependent Ginzburg-Landau theory 
\cite{anderson,abrahams,hagen1} with the action 
\begin{eqnarray}
&& \hspace*{-0.7in}
S^{\rm eff}[\Delta^*,\Delta] =
 \frac{N(0)}{4} \int dt \int d{\bf x}~ 
   \left\{ \frac{\hbar^2}{|\Delta_0|^2}
      \left| \frac{\partial \Delta}{\partial t} \right|^2 \right. \\
&&- \left. \frac{\hbar^2 v_F^2}{3|\Delta_0|^2} 
      \left| \mbox{\boldmath $\nabla$} \Delta \right|^2
      + 2 |\Delta|^2 \left( 1 - \frac{|\Delta|^2}{2|\Delta_0|^2} \right)
   \right\}~, \nonumber
\end{eqnarray}
where $N(0)$ is the density of states for one spin projection at the Fermi 
energy $\epsilon_F = mv_F^2/2$ and $\Delta_0$ is the equilibrium value of the 
order parameter \cite{micheal}. Writing the complex order parameter in terms of 
an amplitude and a phase, we immediately observe that the amplitude 
fluctuations are gapped \cite{gap} and can, therefore, be safely neglected at 
large length scales. The long-wavelength dynamics of the superconductor is thus 
dominated by the phase fluctuations, according to the action
\begin{equation}
S^{\rm eff}[\theta] =
 \frac{N(0)\hbar^2}{4} \int dt \int d{\bf x}~ 
   \left\{ \left( \frac{\partial \theta}{\partial t} \right)^2
      - \frac{v_F^2}{3} 
      \left( \mbox{\boldmath $\nabla$} \theta \right)^2
   \right\}~. 
\end{equation}
This also implies that the global phase 
$\theta_{\bf 0}(t) = \int d{\bf x}~ \theta({\bf x},t)/V$ of the superconductor 
has a dynamics that is governed by 
\begin{equation}
\label{phase}
S^{\rm eff}[\theta_{\bf 0}] =
  \frac{N(0)N\hbar^2}{4n} \int dt~ 
             \left( \frac{d \theta_{\bf 0}}{dt} \right)^2~,
\end{equation}
using the fact that the total volume $V$ of the system is given by $N/n$. 

Up to now our discussion has again been semiclassical. To consider also the 
quantum fluctuations, we have to quantize this theory by applying the usual 
rules of quantum mechanics. Doing so, we find that the wave function of the 
overall phase obeys a Schr\"odinger equation
\begin{equation}
i\hbar \frac{\partial}{\partial t} \Psi(\theta_{\bf 0};t) =
   - \frac{n}{N(0)N}~ \frac{\partial^2}{\partial \theta_{\bf 0}^2}
                                                   \Psi(\theta_{\bf 0};t)~,
\end{equation}
with a `diffusion' constant that can easily be shown to be equal to the 
quantity
$(2/\hbar) \partial \epsilon_F/\partial N$ \cite{dan} and is, most importantly 
for our purposes, proportional to $1/N$. In the thermodynamic limit $N 
\rightarrow \infty$ a state with a well defined stationary phase is clearly a 
solution and we are then dealing with a system having a spontaneously broken 
$U(1)$ symmetry. However, for a finite (and fixed) number of particles the 
global phase cannot be well defined at all times and always has to `diffuse' in 
accordance with the above Schr\"odinger equation. Note also that in the 
groundstate the phase is fully undetermined and 
$|\Psi(\theta_{\bf 0};t)|^2 = 1/2\pi$. Maybe surprisingly, the same 
calculation is somewhat more complicated for a Bose gas because the amplitude 
fluctuations of the order parameter cannot be neglected even at the largest 
length scales. However, taking these amplitude fluctuations into account 
properly, we nevertheless arrive at an action that is equivalent to 
equation~(\ref{phase}) and hence again leads to the phenomenon of phase 
`diffusion'. 

We start again from the action $S[\rho,\theta;\mu]$ for the condensate. The 
difference with the previous subsection is, however, that now we are not so 
much 
interested in the dynamics of the density but in the phase dynamics instead. 
Therefore we now want to integrate over the density field $\rho({\bf x},\tau)$. 
This cannot be done exactly and we therefore here consider only the 
strong-coupling limit, which was also treated by Lewenstein and You 
\cite{maciek}. In that limit we are allowed to neglect the gradient of the 
average density profile \cite{tony2} and the action $S[\rho,\theta;\mu]$ is 
for the longest wavelengths well approximated by
\begin{eqnarray}
&& \hspace*{-0.55in}
S[\rho,\theta;\mu] = 
   \int_0^{\hbar\beta} d\tau \int d{\bf x}~ 
   \left( i\hbar \rho({\bf x},\tau) 
                 \frac{\partial \theta({\bf x},\tau)}{\partial \tau} \right. \\
   &&+ \left. V^{\rm ex}({\bf x})\rho({\bf x},\tau) - \mu\rho({\bf x},\tau)  
            + \frac{2\pi a\hbar^2}{m} \rho^2({\bf x},\tau) \right)~. \nonumber
\end{eqnarray}
In equilibrium the average density profile of the condensate thus obeys
\begin{equation}
\langle \rho({\bf x}) \rangle = \frac{m}{4\pi a \hbar^2}
   \left( \mu - V^{\rm ex}({\bf x}) \right) 
                                    \Theta( \mu - V^{\rm ex}({\bf x}) )~.
\end{equation}
Performing now the shift 
$\rho({\bf x},\tau) 
           = \langle \rho({\bf x}) \rangle + \delta\rho({\bf x},\tau)$, we find 
for the zero-momentum part of the action \cite{top}
\begin{equation}
\hspace*{-0.095in}
~~S[\delta N_{\bf 0},\theta_{\bf 0};\mu] =  \hbar\beta E_{\bf 0}(\mu)  
 + \int_0^{\hbar\beta} d\tau  
   \left( i\hbar \delta N_{\bf 0} \frac{d \theta_{\bf 0}}{d\tau}
          + \frac{2\pi a\hbar^2}{mV_{\bf 0}(\mu)} (\delta N_{\bf 0})^2 
\right)~,
\end{equation}
where $E_{\bf 0}(\mu)$ and $V_{\bf 0}(\mu)$ correspond, respectively, to the 
energy and the volume of the condensate in the so-called Thomas-Fermi 
approximation \cite{TF}. Moreover, 
$\delta N_{\bf 0}(\tau) = \int d{\bf x}~ \delta\rho({\bf x},\tau)$ represents 
the fluctuations in the total number of condensate particles in that 
same approximation, implying that the density fluctuations 
$\delta\rho({\bf x},\tau)$ are only nonzero in that region of space where the 
condensate density does not vanish.

Performing now the integration over the number fluctuations 
$\delta N_{\bf 0}(\tau)$ 
and the usual Wick rotation to real times $\tau \rightarrow it$, we immediately 
see that the effective action for the global phase of the condensate has 
precisely the same form as in equation~(\ref{phase}), i.e.,
\begin{equation}
S^{\rm eff}[\theta_{\bf 0};\mu] = \frac{mV_{\bf 0}(\mu)}{8\pi a} \int dt~
                                 \left( \frac{d\theta_{\bf 0}}{dt} \right)^2~.
\end{equation}
The appropriate `diffusion' constant is therefore equal to 
$2\pi a\hbar/mV_{\bf 0}(\mu)$, which can easily be shown to be equal to 
$(1/2\hbar) \partial \mu/\partial N_{\bf 0}$ if we make use of the fact that in 
the Thomas-Fermi approximation the chemical potential obeys
$\mu = m\omega^2R_{TF}^2/2$ and the radius of the condensate is given by
$R_{TF} = (15a\hbar^2 N_{\bf 0}/m^2\omega^2)^{1/5}$ \cite{BP}. Hence, the 
`diffusion' constant is proportional to $1/N_{\bf 0}^{3/5}$. Note that if the 
condensate where contained in a box the `diffusion' constant would be 
proportional to $1/N_{\bf 0}$ instead. It is important to note also that, in 
contrast to the case of a fermionic superfluid, we have to integrate over the 
amplitude fluctuations of the order parameter to arrive at a quadratic action 
for the phase fluctuations. This leads to the important conclusion that for a 
bosonic superfluid it is impossible to be in a state with only phase 
fluctuations and no density fluctuations, even at the largest length scales. 

\subsection{Quantum kinetic theory}
\label{QKT}
For simplicity we restrict ourselves for the rest of this course to the case of 
a doubly-polarized atomic Bose gas, which for most practical purposes can be 
seen as a gas of spin-less bosons \cite{jason1}. Once we have understood the 
quantum kinetic theory for this particular case, however, we can easily 
generalize to multi-component Bose gases that are receiving considerable 
attention at present \cite{jason2}. Furthermore, the same methods can also be 
used for fermionic gases and have recently already been applied to the problem 
of Cooper-pair formation in a two-component fermion gas \cite{marianne3}.

\subsubsection{Ideal Bose gas}
In textbooks an ideal Bose gas is generally discussed in terms of the average 
occupation numbers of the one-particle states $\chi_{\bf n}({\bf x})$ 
\cite{huang2}. Given the density matrix $\hat{\rho}(t_0)$ of the gas at an 
initial time $t_0$, these occupation numbers obey
\begin{equation}
N_{\bf n}(t) = {\rm Tr} \left[ \hat{\rho}(t_0) 
      \hat{\psi}^{\dagger}_{\bf n}(t) \hat{\psi}_{\bf n}(t)
                         \right]~, 
\end{equation}
with $\hat{\psi}^{\dagger}_{\bf n}(t)$ and 
$\hat{\psi}_{\bf n}(t)$ the usual 
(Heisenberg picture) creation and annihilation operators of second quantization, 
respectively. Because the hamiltonian of the gas 
\begin{equation}
\hat{H} = \sum_{\bf n} \epsilon_{\bf n}   
   \hat{\psi}^{\dagger}_{\bf n}(t) \hat{\psi}_{\bf n}(t)
\end{equation}
commutes with the number operators 
$\hat{N}_{\bf n}(t) = \hat{\psi}^{\dagger}_{\bf n}(t) \hat{\psi}_{\bf n}(t)$, 
the nonequilibri\-um dynamics of the system is trivial and the average
occupation numbers are at all times equal to their value at the initial time 
$t_0$. If we are also interested in fluctuations, it is convenient to introduce 
the eigenstates of the number operators, i.e., 
\begin{equation}
|\{N_{\bf n}\};t\rangle = \prod_{\bf n} 
    \frac{ \left( \hat{\psi}^{\dagger}_{\bf n}(t) 
           \right)^{N_{\bf n}} }
         { \sqrt{ N_{\bf n}! } } |0\rangle~,
\end{equation}
and to consider the full probability distribution
\begin{equation}
\label{PN}
P(\{N_{\bf n}\};t) = {\rm Tr} \left[ \hat{\rho}(t_0) 
    |\{N_{\bf n}\};t\rangle \langle\{N_{\bf n}\};t| 
                                       \right]~,
\end{equation}
which is again independent of time for an ideal Bose gas. The average 
occupation numbers are then determined by
\begin{equation}
N_{\bf n}(t) = \sum_{\{N_{\bf n}\}} N_{\bf n} P(\{N_{\bf n}\};t)
\end{equation}
and the fluctuations can be obtained from similar expressions. 

As indicated in section \ref{PD}, we are also interested in 
the phase dynamics of the gas. In analogy with the occupation number 
representation in equation~(\ref{PN}), we can 
now obtain a completely different description of the Bose gas by making use of 
coherent states and considering the probability distribution
$P[\phi^*,\phi;t]$. Although we expect that this probability distribution is 
again independent of time, let us nevertheless proceed to derive its equation 
of motion in a way that can be generalized lateron when we consider an 
interacting Bose gas. First, we use that
\begin{equation}
P[\phi^*,\phi;t] = \int d[\phi^*_0]d[\phi_0]~ 
                     P[\phi_0^*,\phi_0;t_0]
                      \frac{|\langle\phi;t|\phi_0;t_0\rangle|^2}
                           {\langle\phi;t|\phi;t\rangle
                              \langle\phi_0;t_0|\phi_0;t_0\rangle}~.
\end{equation}
This is a particularly useful result, because the time dependence is now 
completely determined by the matrix element $\langle\phi;t|\phi_0;t_0\rangle$ 
for which the functional integral representation is well known from our 
discussion in section \ref{FI}. It is 
given by a `path' integral over all complex field evolutions 
$\psi({\bf x},t_+) = \sum_{\bf n} \psi_{\bf n}(t_+) \chi_{\bf n}({\bf x})$ 
from $t_0$ to $t$. More precisely we have
\begin{equation}
\langle\phi;t|\phi_0;t_0\rangle = 
  \int_{\psi({\bf x},t_0)=\phi_0({\bf x})}^{\psi^*({\bf x},t)=\phi^*({\bf x})}
     d[\psi^*]d[\psi]~ \exp \left\{ \frac{i}{\hbar} S_+[\psi^*,\psi] \right\}~,
\end{equation}                           
with the forward action $S_+[\psi^*,\psi]$ given by
\begin{eqnarray}
&& \hspace*{-0.5in}
S_+[\psi^*,\psi]         
= \sum_{\bf n} 
   \left\{ -i\hbar \psi^*_{\bf n}(t) \psi_{\bf n}(t)                             
    \raisebox{0.2in}{} \right.  \\
&& \hspace*{0.6in} + \left. \int_{t_0}^{t} dt_+~ \psi^*_{\bf n}(t_+) 
        \left( i\hbar \frac{\partial}{\partial t_+} 
                             - \epsilon_{\bf n} \right)
                                      \psi_{\bf n}(t_+) \right\}~. \nonumber
\end{eqnarray}
In the same manner the matrix element 
$\langle\phi;t|\phi_0;t_0\rangle^* = \langle\phi_0;t_0|\phi;t\rangle$ can be 
written as a `path' integral over all possible field configurations 
$\psi({\bf x},t_-) = \sum_{\bf n} \psi_{\bf n}(t_-) \chi_{\bf n}({\bf x})$  
evolving backward in time from $t$ to $t_0$, i.e.,
\begin{equation}
\langle\phi;t|\phi_0;t_0\rangle^* = 
  \int_{\psi({\bf x},t)=\phi({\bf x})}^{\psi^*({\bf x},t_0)=\phi^*_0({\bf x})}
     d[\psi^*]d[\psi]~ \exp \left\{ \frac{i}{\hbar} S_-[\psi^*,\psi] \right\}~,
\end{equation}   
with a backward action
\begin{eqnarray}
&& \hspace*{-0.5in}
S_-[\psi^*,\psi] = \sum_{\bf n} 
   \left\{ -i\hbar \psi^*_{\bf n}(t_0) \psi_{\bf n}(t_0) 
   \raisebox{0.2in}{} \right. \\
&& \hspace*{0.6in} + \left. \int_{t}^{t_0} dt_-~ \psi^*_{\bf n}(t_-) 
        \left( i\hbar \frac{\partial}{\partial t_-} 
                          - \epsilon_{\bf n} \right)
                                 \psi_{\bf n}(t_-) \right\} \nonumber \\
&&\hspace*{0.1in} = \sum_{\bf n} 
   \left\{ -i\hbar \psi^*_{\bf n}(t) \psi_{\bf n}(t)
   \raisebox{0.2in}{} \right. \nonumber \\    
&& \hspace*{0.6in} + \left. \int_{t}^{t_0} dt_-~ \psi_{\bf n}(t_-) 
        \left( -i\hbar \frac{\partial}{\partial t_-} 
                             - \epsilon_{\bf n} \right)
                                     \psi^*_{\bf n}(t_-) \right\}~. \nonumber
\end{eqnarray}

Putting all these results together, we see that the probability distribution 
$P[\phi^*,\phi;t]$ can in fact be represented by a functional integral over all 
fields $\psi({\bf x},t)$ that evolve backwards from $t$ to $t_0$ and then 
forward in time from $t_0$ to $t$. Absorbing the factor 
$P[\phi_0^*,\phi_0;t_0]$ 
into the measure of the functional integral, we thus arrive at the desired 
result that
\begin{equation}
\label{prob2}
P[\phi^*,\phi;t] = 
   \int_{\psi({\bf x},t)=\phi({\bf x})}^{\psi^*({\bf x},t)=\phi^*({\bf x})} 
            d[\psi^*]d[\psi]~ 
                     \exp \left\{ \frac{i}{\hbar} S[\psi^*,\psi] \right\}~,
\end{equation}
where the total (backward-forward) action in first instance obeys
\begin{eqnarray}
\label{action}
&& \hspace*{-0.2in}
S[\psi^*,\psi] = S_-[\psi^*,\psi] + S_+[\psi^*,\psi] 
  = -i\hbar \sum_{\bf n} \left( 
       \psi^*_{\bf n}(t) \psi_{\bf n}(t) -
                                 |\phi_{\bf n}|^2 \right) \hspace{0.4in} \\  
&&+ \sum_{\bf n} \int_{{\cal C}^t} dt'~ \left\{ \frac{1}{2} \left(
      \psi^*_{\bf n}(t') i\hbar \frac{\partial}{\partial t'} 
      \psi_{\bf n}(t')
     - \psi_{\bf n}(t') i\hbar \frac{\partial}{\partial t'} 
       \psi^*_{\bf n}(t') \right) \right.            \nonumber \\
&& \hspace*{2.55in} - \left. \raisebox{0.2in}{} 
        \epsilon_{\bf n} \psi^*_{\bf n}(t') 
                                  \psi_{\bf n}(t')  \right\}~, \nonumber
\end{eqnarray}
and the integration along the Schwinger-Keldysh contour ${{\cal C}^t}$ is 
defined by $\int_{{\cal C}^t} dt' =\int_{t}^{t_0} dt_- + \int_{t_0}^{t} dt_+ $ 
\cite{schwinger,keldysh,D}. Note also that in equation~(\ref{prob2}) we have 
explicitly specified the boundary conditions on the functional integral. It is 
interesting to mention that these boundary conditions are essentially dictated 
by the topological terms in the action $S[\psi^*,\psi]$, which is a general 
feature of the path-integral formulation of quantum mechanics due to the fact 
that the quantum theory should have the correct (semi)classical limit 
\cite{bernard}. Making use of the periodicity of the field $\psi({\bf x},t)$ on 
the Schwinger-Keldysh contour, the variational principle 
$\delta S[\psi^*,\psi]/\delta \psi^*_{\bf n}(t_{\pm}) = 0$ indeed leads not 
only to the Euler-Lagrange equation 
\begin{equation}
i\hbar \frac{\partial}{\partial t_{\pm}}\psi_{\bf n}(t_{\pm}) =
                  \epsilon_{\bf n} \psi_{\bf n}(t_{\pm})~,
\end{equation}
which agrees with the Heisenberg equation of motion 
$i\hbar \partial\hat{\psi}_{\bf n}(t_{\pm})/\partial t_{\pm} = 
                                  [\hat{\psi}_{\bf n}(t_{\pm}),H]$
for the annihilation operators, but also with the appropriate boundary 
condition 
$\delta \psi^*_{\bf n}(t) = 0$. In the same way we find the complex conjugate 
results if we require that 
$\delta S[\psi^*,\psi]/\delta \psi_{\bf n}(t_{\pm}) = 0$. Substituting the 
boundary conditions in equation~(\ref{action}) and performing a partial 
integration, we then finally obtain for the action
\begin{equation}
S[\psi^*,\psi] = 
  \sum_{\bf n} \int_{{\cal C}^t} dt'~
         \psi^*_{\bf n}(t') \left( i\hbar \frac{\partial}{\partial t'} 
               - \epsilon_{\bf n} \right) \psi_{\bf n}(t')~,
\end{equation}
which we for completeness sake also rewrite as
\begin{eqnarray}
&& \hspace*{-0.2in}  
S[\psi^*,\psi]  \\
&&= \int_{{\cal C}^t} dt' \int d{\bf x}~
         \psi^*({\bf x},t') \left( i\hbar \frac{\partial}{\partial t'} 
            + \frac{\hbar^2 \mbox{\boldmath $\nabla$}^2}{2m} 
            - V^{\rm ex}({\bf x}) \right)   
            \psi({\bf x},t')~. \nonumber
\end{eqnarray}
                                
We are now in a position to derive the equation of motion, i.e., the 
Fokker-Planck equation, for the probability distibution $P[\phi^*,\phi;t]$. 
This is most easily achieved by performing the variable transformation 
$\psi({\bf x},t_{\pm}) = \phi({\bf x},t') \pm \xi({\bf x},t')/2$ in 
equation~(\ref{prob2}). In this manner the fields $\psi({\bf x},t_-)$ and 
$\psi({\bf x},t_+)$ that live on the backward and forward branch of the 
Schwinger-Keldysh contour, respectively, are `projected' onto the real time 
axis. Moreover, at the same time we effect a separation between the 
(semi)classical dynamics described by $\phi({\bf x},t')$ and the quantum 
fluctuations determined by $\xi({\bf x},t')$. After the transformation we have
\begin{eqnarray}
&& \hspace*{-0.2in}
P[\phi^*,\phi;t] \\ 
&&= \int_{\phi({\bf x},t)=\phi({\bf x})}^{\phi^*({\bf x},t)=\phi^*({\bf x})}    
     d[\phi^*]d[\phi] \int d[\xi^*]d[\xi]~ 
         \exp \left\{ \frac{i}{\hbar} S[\phi^*,\phi;\xi^*,\xi] \right\}~,
         \nonumber
\end{eqnarray}
with 
\begin{eqnarray}
&& \hspace*{-0.38in}
S[\phi^*,\phi;\xi^*,\xi] =
  \sum_{\bf n} \int_{t_0}^t dt'~ \left\{
      \phi^*_{\bf n}(t') \left( i\hbar \frac{\partial}{\partial t'} 
                - \epsilon_{\bf n} \right) \xi_{\bf n}(t')
                \right. \\
&& \hspace*{1.2in} 
   + \left. \xi^*_{\bf n}(t') \left( i\hbar \frac{\partial}{\partial t'} 
                - \epsilon_{\bf n} \right) \phi_{\bf n}(t')
                         \right\}~. \nonumber
\end{eqnarray}
Because this action is linear in $\xi_{\bf n}(t')$ and 
$\xi^*_{\bf n}(t')$, 
the integration over the quantum fluctuations leads just to a constraint and we 
find that
\begin{eqnarray}
&& \hspace*{-0.65in}
P[\phi^*,\phi;t] = 
   \int_{\phi({\bf x},t)=\phi({\bf x})}^{\phi^*({\bf x},t)=\phi^*({\bf x})} 
   d[\phi^*]d[\phi]~ \\
&& \times \prod_{\bf n} \delta \left[
     \left( -i \frac{\partial}{\partial t'} 
             - \frac{\epsilon_{\bf n}}{\hbar} \right) 
                                              \phi^*_{\bf n}(t') 
\cdot
      \left( i \frac{\partial}{\partial t'} 
             - \frac{\epsilon_{\bf n}}{\hbar} \right) 
                                              \phi_{\bf n}(t')         
                                \right] \nonumber
\end{eqnarray}
or equivalently that
\begin{equation}
\label{prob3}
P[\phi^*,\phi;t] = 
      \int d[\phi^*_0]d[\phi_0]~ P[\phi^*_0,\phi_0;t_0]
      \prod_{\bf n} 
      \delta \left( |\phi_{\bf n} - \phi^{\rm cl}_{\bf n}(t)|^2
                              \right)~,
\end{equation}
where we introduced the quantity $\phi^{\rm cl}_{\bf n}(t)$ obeying the 
semiclassical equation of motion
\begin{equation}
i\hbar \frac{\partial}{\partial t} \phi^{\rm cl}_{\bf n}(t) =
                \epsilon_{\bf n} \phi^{\rm cl}_{\bf n}(t)
\end{equation}
and the initial condition 
$\phi^{\rm cl}_{\bf n}(t_0) = \phi_{0;\bf n}$.

The latter equation is thus solved by 
$\phi^{\rm cl}_{\bf n}(t) 
  = \phi_{0;\bf n} e^{-i\epsilon_{\bf n} (t-t_0)/\hbar}$ 
and we conclude from a simple change of variables in equation~(\ref{prob3}) 
that for an ideal Bose gas $P[\phi^*,\phi;t] = P[\phi^*,\phi;t_0]$, as expected. 
We also see from equation~(\ref{prob3}) that the desired equation of motion for 
$P[\phi^*,\phi;t]$ reads
\begin{eqnarray}
\label{FP1}
&& \hspace*{-0.8in} 
i\hbar \frac{\partial}{\partial t} P[\phi^*,\phi;t] =
  - \left( \sum_{\bf n} \frac{\partial}{\partial \phi_{\bf n}} 
      \epsilon_{\bf n} \phi_{\bf n} \right) P[\phi^*,\phi;t] \\ 
 && \hspace*{0.8in} + \left( \sum_{\bf n} \frac{\partial}{\partial 
            \phi^*_{\bf n}} 
      \epsilon_{\bf n} \phi^*_{\bf n} \right) P[\phi^*,\phi;t]~. \nonumber
\end{eqnarray}
This is indeed the correct Fokker-Planck equation for an ideal Bose gas 
\cite{carruthers}. To see that we first consider  
$\langle \phi_{\bf n} \rangle(t) = 
            \int d[\phi^*]d[\phi]~ \phi_{\bf n} P[\phi^*,\phi;t]$.
Multiplying equation~(\ref{FP1}) with $\phi_{\bf n}$ and integrating over 
$\phi({\bf x})$, we easily find after a partial integration that
\begin{equation}
i\hbar \frac{\partial}{\partial t} \langle \phi_{\bf n} \rangle(t) =
        \epsilon_{\bf n} \langle \phi_{\bf n} \rangle(t)~,
\end{equation}
which precisely corresponds to the equation of motion of the expectation value
$\langle \hat{\psi}_{\bf n}(t) \rangle = 
             {\rm Tr}[\hat{\rho}(t_0) \hat{\psi}_{\bf n}(t)]$
in the operator formalism. Similarly, we find that 
\begin{equation}
i\hbar \frac{\partial}{\partial t} \langle \phi^*_{\bf n} \rangle(t) =
      - \epsilon_{\bf n} \langle \phi^*_{\bf n} \rangle(t)~,
\end{equation}
in agreement with the result for 
$\langle \hat{\psi}^{\dagger}_{\bf n}(t) \rangle = 
       {\rm Tr}[\hat{\rho}(t_0) \hat{\psi}^{\dagger}_{\bf n}(t)]$. 
            
Next we consider the average of $|\phi_{\bf n}|^2$, for which we immediately 
obtain
\begin{equation}
\label{phisquare}
i\hbar \frac{\partial}{\partial t} \langle |\phi_{\bf n}|^2 \rangle(t) = 0.
\end{equation}
We expect this result to be related to the fact that in the operator formalism 
the occupation numbers $N_{\bf n}(t)$ are independent of time. Although this 
turns out to be true, to give the precise relation between 
$\langle |\phi_{\bf n}|^2 \rangle(t)$ and $N_{\bf n}(t)$ is 
complicated by the fact that at equal times the operators 
$\hat{\psi}_{\bf n}(t)$ and $\hat{\psi}^{\dagger}_{\bf n}(t)$ do not commute. 
However, the path-integral formulation of quantum mechanics is only capable of 
calculating time-ordered operator products \cite{hagen2}. In our case this 
implies that $\langle |\phi_{\bf n}|^2 \rangle(t)$ is the value at 
$t' = t$ of
\begin{eqnarray}
&& \hspace*{-0.2in}
{\rm Tr} \left[ \hat{\rho}(t_0)
   T_{{\cal C}^t} \left( \hat{\psi}_{\bf n}(t) 
                  \hat{\psi}^{\dagger}_{\bf n}(t') \right) \right] \\
&&=
\Theta(t,t') {\rm Tr} \left[ \hat{\rho}(t_0)
                             \hat{\psi}_{\bf n}(t)
                             \hat{\psi}^{\dagger}_{\bf n}(t') \right]
+ \Theta(t',t) {\rm Tr} \left[ \hat{\rho}(t_0)
                           \hat{\psi}^{\dagger}_{\bf n}(t')
                           \hat{\psi}_{\bf n}(t) \right]~, \nonumber
\end{eqnarray}
with $T_{{\cal C}^t}$ the time-ordening operator on the Schwinger-Keldysh 
contour and $\Theta(t,t')$ the corresponding Heaviside function. Since the 
Heaviside function is equal to $1/2$ at equal times \cite{half}, we conclude 
that 
\begin{equation}
\langle |\phi_{\bf n}|^2 \rangle(t) = N_{\bf n}(t) + \frac{1}{2}
\end{equation}
and that equation~(\ref{phisquare}) is thus fully consistent with the operator 
formalism. An intuitive understanding of the relation between 
$\langle |\phi_{\bf n}|^2 \rangle(t)$ and the occupation numbers can be 
obtained by noting that, since all our manipulations up to now have been exact, 
we expect $\langle |\phi_{\bf n}|^2 \rangle(t)$ to contain both classical and 
quantum fluctuations. These correspond precisely to the contributions 
$N_{\bf n}(t)$ and $1/2$, respectively.

Finally we need to discuss the stationary solutions of the Fokker-Planck 
equation. It is not difficult to show that any functional that only depends on 
the amplitudes $|\phi_{\bf n}|^2$ is a solution. As it stands the 
Fokker-Planck equation, therefore, does not lead to a unique equilibrium 
distribution. This is not surprising, because for an isolated, ideal Bose gas 
there is no mechanism for redistributing the particles over the various energy 
levels and thus for relaxation towards equilibrium. However, the situation 
changes when we allow the bosons in the trap to tunnel back and forth to a 
reservoir at a temperature $T$. The corrections to the Fokker-Planck equation 
that are required to describe the physics in this case are considered next. 
However, to determine these corrections in the most convenient way, we have to 
slightly generalize the above theory because with the probability distibution 
$P[\phi^*,\phi;t]$ we are only able to study spatial, but not temporal 
correlations in the Bose gas.

To study also those we follow the by now well-known procedure in quantum field 
theory and construct a generating functional $Z[J,J^*]$ for all 
(time-ordered) correlation functions. It is obtained by performing two steps. 
First, we introduce the probability distribution $P_J[\phi^*,\phi;t]$ for a 
Bose gas in the presence of the external currents $J({\bf x},t)$ and 
$J^*({\bf x},t)$ by adding to the hamiltonian the terms
\begin{eqnarray}
- \hbar \int d{\bf x}~ \left( \hat{\psi}({\bf x},t) J^*({\bf x},t) +
                              J({\bf x},t) \hat{\psi}^{\dagger}({\bf x},t)
                       \right) =                    \hspace*{1.3in} \nonumber 
\\
- \hbar \sum_{\bf n} \left( \hat{\psi}_{\bf n}(t) 
             J^*_{\bf n}(t) + J_{\bf n}(t) \hat{\psi}^{\dagger}_{\bf n}(t)
                       \right)~.                                    \nonumber
\end{eqnarray}
As a result we have
\begin{eqnarray}
&& \hspace*{-0.3in} 
P_J[\phi^*,\phi;t] = 
   \int_{\psi({\bf x},t)=\phi({\bf x})}^{\psi^*({\bf x},t)=\phi^*({\bf x})} 
            d[\psi^*]d[\psi]~ \exp \left\{ \frac{i}{\hbar} S[\psi^*,\psi]
                               \right\} \\
&& \hspace*{0.38in} 
 \times \exp \left\{ i \int_{{\cal C}^t} dt' \int d{\bf x}~ \left(
            \psi({\bf x},t') J^*({\bf x},t') + J({\bf x},t') \psi^*({\bf x},t')
                                                 \right) \right\}~. \nonumber
\end{eqnarray}
Second, we integrate this expression over $\phi({\bf x})$ to obtain the desired 
generating functional. Hence
\begin{eqnarray}
\label{z}
&& \hspace*{-0.2in}
Z[J,J^*] = \int d[\phi^*]d[\phi]~ P_J[\phi^*,\phi;t]
         = \int d[\psi^*]d[\psi]~ \exp \left\{ \frac{i}{\hbar} S[\psi^*,\psi]
                               \right\} \hspace{0.4in} \\
&& \hspace*{0.32in}
\times \exp \left\{ i \int_{{\cal C}^{\infty}} dt \int d{\bf x}~ \left(
            \psi({\bf x},t) J^*({\bf x},t) + J({\bf x},t) \psi^*({\bf x},t)
                                                  \right) \right\}~. \nonumber
\end{eqnarray}

It is important to realize that $Z[J,J^*]$ is indeed independent of the time 
$t$ because of the fact that $P_J[\phi^*,\phi;t]$ is a probability distribution 
and thus properly normalized. We are therefore allowed to deform the contour 
${\cal C}^t$ to any closed contour that runs through $t_0$. Since we are in 
principle interested in all times $t \ge t_0$, the most convenient choice is 
the 
countour that runs backward from infinity to $t_0$ and then forwards from $t_0$ 
to infinity. This contour is denoted by ${\cal C}^{\infty}$ and also called the 
Schwinger-Keldysh contour in the following because there is in practice never 
confusion with the more restricted contour ${\cal C}^t$ that is required when 
we consider a probability distribution. With this choice it is also clear that 
$Z[J,J^*]$ is a generating functional. Indeed, equation~(\ref{z}) shows 
explicitly that all time-ordered correlation functions can be obtained by 
functional differentiation with respect to the currents $J({\bf x},t)$ and 
$J^*({\bf x},t)$. We have, for instance, that
\begin{equation}
{\rm Tr}[\hat{\rho}(t_0) \hat{\psi}({\bf x},t)] =
   \left. \frac{1}{i} \frac{\delta}{\delta J^*({\bf x},t)} Z[J,J^*] 
\right|_{J,J^*=0}
\end{equation}
and similarly that
\begin{eqnarray}
\label{corr}
&& \hspace*{-0.5in} 
{\rm Tr} \left[ \hat{\rho}(t_0)
         T_{{\cal C}^{\infty}} \left( \hat{\psi}({\bf x},t) 
                           \hat{\psi}^{\dagger}({\bf x}',t') \right) \right] \\ 
&& = \left. \frac{1}{i^2}
          \frac{\delta^2}{\delta J^*({\bf x},t) \delta J({\bf x}',t')} 
                                                    Z[J,J^*] 
\right|_{J,J^*=0}~. \nonumber
\end{eqnarray} 
Note that the times $t$ and $t'$ always have to be larger or equal to $t_0$ for 
these identities to be valid.

\subsubsection{Ideal Bose gas in contact with a reservoir}
Inspired by the Caldeira-Leggett model for a particle experiencing friction 
\cite{tony1}, we take for the reservoir an ideal gas of $N$ bosons in a box with 
volume $V$. The states in this box are labeled by the momentum $\hbar{\bf k}$ 
and equal to $\chi_{\bf k}({\bf x}) = e^{i {\bf k} \cdot {\bf x} }/\sqrt{V}$. 
They have an energy 
$\epsilon({\bf k}) = \hbar^2 {\bf k}^2/2m + \Delta V^{\rm ex}$, where 
$\Delta V^{\rm ex}$ accounts for a possible bias between the potential energies 
of a particle in the center of the trap and a particle in the reservoir. The 
reservoir is also taken to be sufficiently large that it can be treated in the 
thermodynamic limit and is in an equilibrium with temperature $T$ and chemical 
potential $\mu$ for times $t < t_0$. At $t_0$ it is brought into contact with 
the trap by means of a tunnel hamiltonian
\begin{equation}
\hat{H}^{\rm int} = \frac{1}{\sqrt{V}}
  \sum_{\bf n} \sum_{\bf k}~
  \left( t_{\bf n}({\bf k}) 
           \hat{\psi}_{\bf n}(t) \hat{\psi}^{\dagger}_{\bf k}(t)
       + t^*_{\bf n}({\bf k}) 
           \hat{\psi}_{\bf k}(t) \hat{\psi}^{\dagger}_{\bf n}(t)  
  \right)~,
\end{equation}
with complex tunneling matrix elements $t_{\bf n}({\bf k})$ that for 
simplicity are assumed to be almost constant for momenta $\hbar k$ smaller that 
a cutoff $\hbar k_c$ but to vanish rapidly for momenta larger than this 
ultraviolet cutoff. Moreover, we consider here only the low-temperature regime 
in which the thermal de Broglie wavelength $\Lambda = 
(2\pi\hbar^2/mk_BT)^{1/2}$ 
of the particles obeys $\Lambda \gg 1/k_c$, since this is the most appropriate 
limit for realistic atomic gases.  

To study the evolution of the combined system for times $t \ge t_0$ we thus 
have to deal with the action
\begin{eqnarray}
&& \hspace*{-0.5in}
S[\psi^*,\psi;\psi_R^*,\psi_R] = \sum_{\bf n} \int_{{\cal C}^{\infty}} dt~
         \psi^*_{\bf n}(t) \left( i\hbar \frac{\partial}{\partial t} 
                         - \epsilon_{\bf n} + \mu \right) \psi_{\bf n}(t) \\
&&+ \sum_{\bf k} \int_{{\cal C}^{\infty}} dt~ 
         \psi^*_{\bf k}(t) \left( i\hbar \frac{\partial}{\partial t} 
                         - \epsilon({\bf k}) + \mu \right) \psi_{\bf k}(t)~,
                         \nonumber \\
&&- \frac{1}{\sqrt{V}} \sum_{\bf n} \sum_{\bf k} \int_{{\cal C}^{\infty}} dt~ 
      \left( t_{\bf n}({\bf k}) \psi_{\bf n}(t) \psi^*_{\bf k}(t)
         + t^*_{\bf n}({\bf k}) \psi_{\bf k}(t) \psi^*_{\bf n}(t) \right)
                                                           \nonumber 
\end{eqnarray}
if we measure all energies relative to the chemical potential and also 
introduce the complex field 
$\psi_R({\bf x},t) = \sum_{\bf k} \psi_{\bf k}(t) \chi_{\bf k}({\bf x})$ for 
the degrees of freedom of the reservoir. However, we are only interested in the 
evolution of the Bose gas in the trap and therefore only in the time-ordered 
correlation functions of this part of the system. The corresponding generating 
functional
\begin{eqnarray}
&& \hspace*{-0.5in}
Z[J,J^*] = \int d[\psi^*]d[\psi] \int d[\psi_R^*]d[\psi_R]~ 
   \exp \left\{ \frac{i}{\hbar} S[\psi^*,\psi;\psi_R^*,\psi_R]
                               \right\} \hspace{0.4in} \\
&&\times \exp \left\{ i \int_{{\cal C}^{\infty}} dt \int d{\bf x}~ \left(
            \psi({\bf x},t) J^*({\bf x},t) + J({\bf x},t) \psi^*({\bf x},t)
                                                \right) \right\}~. \nonumber
\end{eqnarray} 
is of the same form as the functional integral in equation~(\ref{z}), but now 
with an effective action that is defined by
\begin{equation}
\exp \left\{ \frac{i}{\hbar} S^{\rm eff}[\psi^*,\psi] \right\} \equiv
  \int d[\psi_R^*]d[\psi_R]~
          \exp \left\{ \frac{i}{\hbar} S[\psi^*,\psi;\psi_R^*,\psi_R] 
\right\}~.
\end{equation}
Hence, our next task is to integrate out the field $\psi_R({\bf x},t)$, which 
can be done exactly because it only requires the integration of a gaussian.

To familiarize ourselves with the Schwinger-Keldysh formalism, we perform the 
gaussian integration here in some detail. In principle, this can of course be 
done explicitly by making use of the fact that the initial density matrix 
$\hat{\rho}_R(t_0)$ of the reservoir can be expanded in the projection operators 
on the coherent states by using the coefficients
\begin{equation}
\rho_R[|\phi_R|^2;t_0] = \prod_{\bf k} 
   \frac{1}{N({\bf k})} e^{ - |\phi_{\bf k}|^2/N({\bf k}) }~,
\end{equation}
with $N({\bf k}) = 1/(e^{\beta(\epsilon({\bf k})-\mu)}-1)$ the appropriate 
Bose distribution function and $\beta = 1/k_BT$. However, in practice it is 
much more easy to use a different procedure. It is based on the observation 
that 
if we introduce the $\delta$ function on the Schwinger-Keldysh contour defined 
by$\int_{{\cal C}^{\infty}} dt'~ \delta(t,t') = 1$ and the Green's function 
$G({\bf k};t,t')$ obeying
\begin{equation}
\label{greena}
\left( i\hbar \frac{\partial}{\partial t} 
                         - \epsilon({\bf k}) + \mu \right) G({\bf k};t,t') =
                                           \hbar \delta(t,t')~,
\end{equation}
the action $S[\psi^*,\psi;\psi_R^*,\psi_R]$ can be written as a complete 
square, or more precisely as the sum of two squares $S_1[\psi^*,\psi]$ and 
$S_2[\psi^*,\psi;\psi_R^*,\psi_R]$ that are given by 
\begin{eqnarray}
&& \hspace*{-0.48in} 
S_1[\psi^*,\psi] = \sum_{\bf n} \int_{{\cal C}^{\infty}} dt~    
         \psi^*_{\bf n}(t) 
            \left( i\hbar \frac{\partial}{\partial t} 
                   - \epsilon_{\bf n} + \mu \right) \psi_{\bf n}(t) \\
&& \hspace*{0.1in} - \frac{1}{\hbar V} \sum_{\bf n,\bf n'} \sum_{\bf k}
   \int_{{\cal C}^{\infty}} dt \int_{{\cal C}^{\infty}} dt'~
      \psi^*_{\bf n}(t) t^*_{\bf n}({\bf k}) G({\bf k};t,t') 
                               t_{\bf n'}({\bf k}) \psi_{\bf n'}(t') \nonumber  
\end{eqnarray}
and 
\begin{eqnarray}
&& \hspace*{-0.3in} 
S_2[\psi^*,\psi;\psi_R^*,\psi_R] \\
&&= \sum_{\bf k} \int_{{\cal C}^{\infty}} dt~
    \left( \psi_{\bf k}^*(t) 
         - \frac{1}{\hbar\sqrt{V}} \sum_{\bf n} \int_{{\cal C}^{\infty}} dt'~
                 t^*_{\bf n}({\bf k}) \psi^*_{\bf n}(t') G({\bf k};t',t)
    \right)                                                    \nonumber \\
&& \hspace*{0.7in} \times \left( i\hbar \frac{\partial}{\partial t} 
                         - \epsilon({\bf k}) + \mu \right)     \nonumber \\
&& \hspace*{0.7in} \times \left( \psi_{\bf k}(t) 
         - \frac{1}{\hbar\sqrt{V}} \sum_{\bf n} \int_{{\cal C}^{\infty}} dt'~
                 G({\bf k};t,t') \psi_{\bf n}(t') t_{\bf n}({\bf k}) 
         \right)~, \nonumber     
\end{eqnarray}
respectively. Since the first term is independent of the field 
$\psi_R({\bf x},t)$, we only need to evaluate
$\int d[\psi_R^*]d[\psi_R]~
                \exp \left( i S_2[\psi^*,\psi;\psi_R^*,\psi_R]/\hbar \right)$. 
Performing a shift in the integration variables, we however see that this 
functional integral is equal to 
\begin{eqnarray}
\int d[\psi_R^*]d[\psi_R]~
   \exp \left\{ \frac{i}{\hbar} 
          \sum_{\bf k} \int_{{\cal C}^{\infty}} dt~ 
             \psi^*_{\bf k}(t) \left( i\hbar \frac{\partial}{\partial t} 
                         - \epsilon({\bf k}) + \mu \right) \psi_{\bf k}(t)
        \right\}  \\
  = {\rm Tr}[\rho_R(t_0)] = 1~.  \nonumber
\end{eqnarray}
As a result the effective action $S^{\rm eff}[\psi^*,\psi]$ is just equal to 
$S_1[\psi^*,\psi]$, which can be slightly rewritten to read
\begin{eqnarray}
\label{Seff}
&& \hspace*{-0.3in}
S^{\rm eff}[\psi^*,\psi] = \sum_{\bf n,\bf n'} 
    \int_{{\cal C}^{\infty}} dt \int_{{\cal C}^{\infty}} dt'~  \\
&& \hspace*{0.37in} \times~ \psi^*_{\bf n}(t) \left\{ 
            \left( i\hbar \frac{\partial}{\partial t} 
                   - \epsilon_{\bf n} + \mu \right) \delta_{\bf n,\bf n'}
                                                     \delta(t,t')
                   - \hbar \Sigma_{{\bf n},{\bf n}'}(t,t') 
                            \right\} \psi_{\bf n'}(t')~, \nonumber
\end{eqnarray}
with a selfenergy $\Sigma_{{\bf n},{\bf n}'}(t,t')$ obeying
\begin{equation}
\hbar \Sigma_{{\bf n},{\bf n'}}(t,t') =
  \frac{1}{\hbar V} \sum_{\bf k} 
                 t^*_{\bf n}({\bf k}) G({\bf k};t,t') t_{\bf n'}({\bf k})~. 
\end{equation}  

This is our first example of an effective action describing the nonequilibrium 
dynamics of a Bose gas. Before we can study its consequences we clearly first 
need to determine the Green's function $G({\bf k};t,t')$ in terms of which the 
selfenergy is expressed. Although we know that this Green's function fulfills 
equation~(\ref{greena}), we cannot directly solve this equation because we do 
not 
know the appropriate boundary condition at $t = t'$. To calculate 
$G({\bf k};t,t')$ 
we therefore have to proceed differently. It is however clear from 
equation~(\ref{greena}) that $G({\bf k};t,t')$ is a property of the reservoir 
and 
we thus expect that it can somehow be related to a time-ordered correlation 
function of this reservior. To see that explicitly we consider again the 
generating functional of these correlation functions, i.e., 
\begin{eqnarray}
&& \hspace*{-0.2in}
Z_R[J,J^*] \\ 
&&= \int d[\psi_R^*]d[\psi_R]~ 
   \exp \left\{ \frac{i}{\hbar} 
          \sum_{\bf k} \int_{{\cal C}^{\infty}} dt~
             \psi^*_{\bf k}(t) \left( i\hbar \frac{\partial}{\partial t} 
                         - \epsilon({\bf k}) + \mu \right) \psi_{\bf k}(t)
                               \right\}                           \nonumber \\
&&\hspace*{0.9in} 
     \times \exp \left\{ i \sum_{\bf k} \int_{{\cal C}^{\infty}} dt~
            \left(
            \psi_{\bf k}(t) J^*_{\bf k}(t) + J_{\bf k}(t) \psi^*_{\bf k}(t)
            \right)
              \right\}~. \nonumber
\end{eqnarray} 
It is again a gaussian integral and can thus be evaluated in the same manner as 
before. The result is now
\begin{equation}
Z_R[J,J^*] = \exp \left\{ -i \sum_{\bf k} \int_{{\cal C}^{\infty}} dt 
                                         \int_{{\cal C}^{\infty}} dt'~ 
               J^*_{\bf k}(t) G({\bf k};t,t') J_{\bf k}(t') \right\}~,
\end{equation}
which shows by means of equation~(\ref{corr}) that
\begin{equation}
i G({\bf k};t,t') = 
  {\rm Tr} \left[ \hat{\rho}_R(t_0)
         T_{{\cal C}^{\infty}} \left( \hat{\psi}_{\bf k}(t) 
                           \hat{\psi}^{\dagger}_{\bf k}(t') \right) \right]~. 
\end{equation}

Note first of all that this is indeed consistent with equation~(\ref{greena}), 
because the right-hand side obeys
\begin{eqnarray}
&& \hspace*{-0.2in}
i\hbar \frac{\partial}{\partial t} 
   {\rm Tr} \left[ \hat{\rho}_R(t_0) 
     T_{{\cal C}^{\infty}} \left( \hat{\psi}_{\bf k}(t) 
                           \hat{\psi}^{\dagger}_{\bf k}(t') \right) \right] \\ 
  &&= i\hbar \delta(t,t') {\rm Tr} \left[ \hat{\rho}_R(t_0) 
                [\hat{\psi}_{\bf k}(t),\hat{\psi}^{\dagger}_{\bf k}(t)] \right]
                                                          \nonumber \\                
  &&\hspace*{0.97in} + {\rm Tr} \left[ \hat{\rho}_R(t_0)
     T_{{\cal C}^{\infty}} \left( 
                      i\hbar \frac{\partial}{\partial t} \hat{\psi}_{\bf k}(t) 
                           \hat{\psi}^{\dagger}_{\bf k}(t') \right) \right]
                                                          \nonumber \\
  &&= i\hbar \delta(t,t') + (\epsilon({\bf k}) - \mu)
                          {\rm Tr} \left[ \hat{\rho}_R(t_0)
                            T_{{\cal C}^{\infty}} \left( \hat{\psi}_{\bf k}(t) 
                            \hat{\psi}^{\dagger}_{\bf k}(t') \right) \right]
                            \nonumber
\end{eqnarray}
in the operator formalism. Moreover, from this identification we see that the 
desired solution fulfilling the appropriate boundary conditions is apparently
\begin{eqnarray}
\label{greens}
&& \hspace*{-0.2in}
G({\bf k};t,t') \\
&&= -i e^{-i(\epsilon({\bf k}) - \mu)(t-t')/\hbar}
 \left\{ \Theta(t,t') \left( 1+ N({\bf k}) \right) +
                           \Theta(t',t) N({\bf k}) \right\}~. \nonumber
\end{eqnarray}
The specific dependence on the backward and forward branches of the 
Schwinger-Keldysh contour is thus solely determined by the Heaviside function 
$\Theta(t,t')$. As a result it is convenient to decompose the Green's function 
into its analytic pieces $G^>({\bf k};t-t')$ and $G^<({\bf k};t-t')$ 
\cite{kadanoff} by means of
\begin{equation}
G({\bf k};t,t') = \Theta(t,t') G^>({\bf k};t-t') 
                                         + \Theta(t',t) G^<({\bf k};t-t')~. 
\end{equation} 
Due to the fact that we are always dealing with time-ordered correlation 
functions, such a decomposition turns out to be a generic feature of all the 
functions on the Schwinger-Keldysh contour that we will encounter in the 
following \cite{langreth}. For a general function $F(t,t')$, it is however also 
possible to have $\delta$-function singularities. If that happens the correct 
decomposition is
\begin{equation}
F(t,t') = F^{\delta}(t) \delta(t,t') + \Theta(t,t') F^>(t,t') 
                                         + \Theta(t',t) F^<(t,t')~. 
\end{equation}
This more general decomposition is not required here, but will be needed in 
section \ref{CF} when we determine the effective interaction between two atoms 
in a gas.

Having obtained the Green's function of the reservoir, we can now return to our 
discussion of the effective action $S^{\rm eff}[\psi^*,\psi]$ for the Bose gas 
in the trap. Although we have chosen to derive in equation~(\ref{Seff}) the 
effective action for the generating functional $Z[J,J^*]$, it is 
straightforward 
to show that the effective action for the probability distribution 
$P[\phi^*,\phi;t]$ is obtained by only replacing the contour ${\cal 
C}^{\infty}$ 
by ${\cal C}^{t}$. In the following we therefore no longer always specify all 
the boundary conditions on the time integration, if the precise details of this 
integration are not important and the discussion applies equally well to both 
cases. Keeping this in mind, we now again perform the transformation 
$\psi_{\bf n}(t_{\pm}) = \phi_{\bf n}(t) \pm \xi_{\bf n}(t)/2$ to explicitly 
separate the (semi)classical dynamics from the effect of fluctuations. It leads 
in first instance to \cite{remark}
\begin{eqnarray}
\label{Sxi}
&& \hspace*{-0.5in}
S^{\rm eff}[\phi^*,\phi;\xi^*,\xi] =   
 \sum_{\bf n} \int_{t_0} dt~ 
      \phi^*_{\bf n}(t) \left( i\hbar \frac{\partial}{\partial t} 
                           - \epsilon_{\bf n} + \mu \right) \xi_{\bf n}(t) \\
&& \hspace*{0.45in} + \sum_{\bf n} \int_{t_0} dt~                                                          
      \xi^*_{\bf n}(t) \left( i\hbar \frac{\partial}{\partial t} 
                           - \epsilon_{\bf n} + \mu \right) \phi_{\bf n}(t)
                                                         \nonumber \\
&& \hspace*{0.45in} - \sum_{\bf n,\bf n'} 
    \int_{t_0} dt \int_{t_0} dt'~
           \phi^*_{\bf n}(t) 
                   \hbar \Sigma^{(-)}_{{\bf n},{\bf n}'}(t-t') \xi_{\bf n'}(t')
                                                         \nonumber \\    
&& \hspace*{0.45in} - \sum_{\bf n,\bf n'} 
    \int_{t_0} dt \int_{t_0} dt'~                                                                         
            \xi^*_{\bf n}(t)
                   \hbar \Sigma^{(+)}_{{\bf n},{\bf n}'}(t-t') \phi_{\bf n'}(t')
                                                \nonumber \\
&& \hspace*{0.45in} - \frac{1}{2} \sum_{{\bf n},{\bf n'}} 
    \int_{t_0} dt \int_{t_0} dt'~ 
                  \xi^*_{\bf n}(t)
                    \hbar \Sigma^{K}_{{\bf n},{\bf n'}}(t-t') \xi_{\bf n'}(t')~,
                    \nonumber
\end{eqnarray}
where we introduced the retarded and advanced components of the selfenergy 
\begin{equation}
\Sigma^{(\pm)}_{{\bf n},{\bf n}'}(t-t') = \pm \Theta(\pm(t-t')) 
  \left( \Sigma^{>}_{{\bf n},{\bf n}'}(t-t') 
                               - \Sigma^{<}_{{\bf n},{\bf n}'}(t-t') \right)
\end{equation} 
that affect the terms in the action that are linear in $\xi^*_{\bf n}(t)$ and 
$\xi_{\bf n}(t)$, respectively, and also the Keldysh component
\begin{equation}
\Sigma^{K}_{{\bf n},{\bf n}'}(t-t') = 
  \left( \Sigma^{>}_{{\bf n},{\bf n}'}(t-t') 
                               + \Sigma^{<}_{{\bf n},{\bf n}'}(t-t') \right)
\end{equation}
that is associated with the part quadratic in the fluctuations. 

The physical content of these various components of the selfenergy is 
understood most clearly if we now apply again the beautiful functional-integral 
procedure that is due to Hubbard and Stratonovich. The basic idea is to write 
the factor 
\begin{eqnarray}
\exp \left\{ - \frac{i}{2} \sum_{{\bf n},{\bf n}'} 
               \int_{t_0} dt \int_{t_0} dt'~ 
                 \xi^*_{\bf n}(t)
                    \Sigma^{K}_{{\bf n},{\bf n}'}(t-t') \xi_{\bf n'}(t')
     \right\}                                               \nonumber
\end{eqnarray}
in the integrant of
$\int d[\phi^*]d[\phi] \int d[\xi^*]d[\xi]~ 
                   \exp\left( iS^{\rm eff}[\phi^*,\phi;\xi^*,\xi]/\hbar 
\right)$
as a gaussian integral over a complex field $\eta({\bf x},t)$. It is 
equivalent, but in practice more convenient, to just multiply the integrant by 
a factor $1$ that is written as a gaussian integral 
$\int d[\eta^*]d[\eta]~ \exp\left( iS^{\rm eff}[\eta^*,\eta]/\hbar \right)$ 
with
\begin{eqnarray}
&& \hspace*{-0.2in} 
S^{\rm eff}[\eta^*,\eta] \\
&&= \frac{1}{2} \sum_{{\bf n},{\bf n}'} \int_{t_0} dt \int_{t_0} dt'~ 
     \left(2 \eta^*_{\bf n}(t) - \sum_{\bf n''} \int_{t_0} dt''~
              \xi^*_{\bf n''}(t'') \hbar \Sigma^K_{{\bf n}'',{\bf n}}(t''-t)
     \right)                                                     \nonumber \\
&&\times~ (\hbar\Sigma^K)^{-1}_{{\bf n},{\bf n}'}(t-t') 
     \left(2 \eta_{\bf n'}(t') - \sum_{\bf n''} \int_{t_0} dt''~
              \hbar \Sigma^K_{{\bf n}',{\bf n}''}(t'-t'') \xi_{\bf n''}(t'') 
     \right) \nonumber
\end{eqnarray}
a complete square. Adding this to $S^{\rm eff}[\phi^*,\phi;\xi^*,\xi]$ the 
total effective action becomes
\begin{eqnarray}
\label{snoise}
&& \hspace*{-0.3in} 
           S^{\rm eff}[\phi^*,\phi;\xi^*,\xi;\eta^*,\eta] \\
&&= \sum_{{\bf n},{\bf n}'} \int_{t_0} dt \int_{t_0} dt'~
      \phi^*_{\bf n}(t) 
       \left\{ \left( i\hbar \frac{\partial}{\partial t} 
                      - \epsilon_{\bf n} + \mu - \eta^*_{\bf n}(t) \right)
                                \delta_{{\bf n},{\bf n}'} \delta(t-t')
                                                \right.   \nonumber \\
&& \hspace*{2.57in} 
   - \left. \raisebox{0.2in}{} \hbar \Sigma^{(-)}_{{\bf n},{\bf n}'}(t-t')
       \right\} \xi_{\bf n'}(t')                          \nonumber \\
&&+ \sum_{{\bf n},{\bf n}'} \int_{t_0} dt \int_{t_0} dt'~
      \xi^*_{\bf n}(t) 
       \left\{ \left( i\hbar \frac{\partial}{\partial t} 
                      - \epsilon_{\bf n} + \mu - \eta_{\bf n}(t) \right)
                                \delta_{{\bf n},{\bf n}'} \delta(t-t')
                                                \right.   \nonumber \\                                
&& \hspace*{2.5in} 
   - \left. \raisebox{0.2in}{} \hbar \Sigma^{(+)}_{{\bf n},{\bf n}'}(t-t')
       \right\} \phi_{\bf n'}(t')                          \nonumber \\
&&+ 2 \sum_{{\bf n},{\bf n}'} \int_{t_0} dt \int_{t_0} dt'~ 
        \eta^*_{\bf n}(t) (\hbar\Sigma^K)^{-1}_{{\bf n},{\bf n}'}(t-t') 
        \eta_{\bf n'}(t') \nonumber
\end{eqnarray}
and is thus linear in $\xi_{\bf n}(t)$ and $\xi^*_{\bf n}(t)$. Integrating 
over these fluctuations we conclude from this action that the field 
$\phi({\bf x},t)$ is constraint to obey the Langevin equations \cite{nico}
\begin{equation}
\label{lan1}
i\hbar \frac{\partial}{\partial t} \phi_{\bf n}(t)
  = (\epsilon_{\bf n} - \mu)\phi_{\bf n}(t) 
    +  \sum_{\bf n'} \int_{t_0}^{\infty} dt'~
           \hbar \Sigma^{(+)}_{{\bf n},{\bf n}'}(t-t') \phi_{\bf n'}(t') 
    + \eta_{\bf n}(t)
\end{equation}
and
\begin{equation}
\label{lan2}
-i\hbar \frac{\partial}{\partial t} \phi^*_{\bf n}(t)
  = (\epsilon_{\bf n} - \mu)\phi^*_{\bf n}(t) 
    +  \sum_{\bf n'} \int_{t_0}^{\infty} dt'~
           \phi^*_{\bf n'}(t') \hbar \Sigma^{(-)}_{{\bf n}',{\bf n}}(t'-t)  
    + \eta^*_{\bf n}(t)
\end{equation}
with gaussian noise terms $\eta_{\bf n}(t)$ and $\eta^*_{\bf n}(t)$ that from 
the last term in the right-hand side of equation~(\ref{snoise}) are seen to 
have 
the time correlations
\begin{eqnarray}
&& \hspace*{-0.4in}
\langle \eta^*_{\bf n}(t) \eta_{\bf n'}(t') \rangle =
                      \frac{i\hbar^2}{2} \Sigma^K_{\bf n,\bf n'}(t-t') \\
&& \hspace*{0.43in} 
  = \frac{1}{2} \int \frac{d{\bf k}}{(2\pi)^3}~ \left( 1+2N({\bf k}) \right)
                 t^*_{\bf n}({\bf k}) 
                    e^{-i(\epsilon({\bf k}) - \mu)(t-t')/\hbar}
                 t_{\bf n'}({\bf k}) \nonumber
\end{eqnarray}
in the thermodynamic limit. Moreover, in that limit the retarded selfenergy is 
equal to
\begin{equation}
\hbar \Sigma^{(+)}_{{\bf n},{\bf n}'}(t-t') =
  - \Theta(t-t') \frac{i}{\hbar} \int \frac{d{\bf k}}{(2\pi)^3}~  
                 t^*_{\bf n}({\bf k}) 
                    e^{-i(\epsilon({\bf k}) - \mu)(t-t')/\hbar}
                 t_{\bf n'}({\bf k})
\end{equation}
and the advanced selfenergy can be found from the relation
$\Sigma^{(-)}_{{\bf n}',{\bf n}}(t'-t) 
                     = \left( \Sigma^{(+)}_{{\bf n},{\bf n}'}(t-t') \right)^*$.
Note that the Heaviside functions in these retarded and advanced selfenergies 
are precisely such that Eqs.~(\ref{lan1}) and (\ref{lan2}) are causal, i.e., 
the evolution of $\phi({\bf x},t)$ and $\phi^*({\bf x},t)$ depends only on 
their 
value at previous times. This is clearly a sensible result and explains why 
these components of the full selfenergy enter into the terms in the action that 
are linear in the fluctuations.

To understand why the Keldysh component enters into the terms that are 
quadratic in the fluctuations is more complicated and is related to the famous 
fluctuation-dissipation theorem \cite{kadanoff,nico,forster}. Physically, the 
fluctuation-dissipation theorem ensures in our case that, due to the 
interactions with the reservoir, the gas in the trap now relaxes towards 
thermal equilibrium in the limit $t \rightarrow \infty$. Let us therefore study 
this limit in somewhat more detail. However, before we do so we would like to 
mention for completeness that if we apply the above Hubbard-Stratonovich 
transformation to the functional integral for the probability distribution 
$P[\phi^*,\phi;t]$,we immediately find that analogous to equation~(\ref{prob3})     
\begin{eqnarray}
&& \hspace*{-0.6in}
P[\phi^*,\phi;t] =
      \int d[\eta^*]d[\eta]~ P[\eta^*,\eta]
      \int d[\phi^*_0]d[\phi_0]~ P[|\phi_0|^2;t_0] \\
&& \hspace*{1.9in} \times
        \prod_{\bf n} \delta \left( |\phi_{\bf n} - \phi^{\rm cl}_{\bf n}(t)|^2
                              \right)~, \nonumber
\end{eqnarray}
where $\phi^{\rm cl}_{\bf n}(t)$ now obeys the Langevin equation in 
equation~(\ref{lan1}) for a particular realization of the noise 
$\eta_{\bf n}(t)$ 
and with the initial condition $\phi^{\rm cl}_{\bf n}(t_0)=\phi_{0;\bf n}$. 
Furthermore, the probability distribution for each realization of the noise is 
given by
\begin{eqnarray}
&& \hspace*{-0.2in} 
P[\eta^*,\eta] \\
&&= \exp \left\{ \frac{2i}{\hbar}
   \sum_{{\bf n},{\bf n}'} \int_{t_0}^t dt' \int_{t_0}^t dt''~ 
        \eta^*_{\bf n}(t') (\hbar\Sigma^K)^{-1}_{{\bf n},{\bf n}'}(t'-t'') 
        \eta_{\bf n'}(t'') \right\}~. \nonumber
\end{eqnarray}
This result shows explicitly that, as expected, the Fokker-Planck equation 
associated with the Langevin equations in Eqs.~(\ref{lan1}) and (\ref{lan2}) is 
in fact the desired Fokker-Planck equation for the probability distribution 
$P[\phi^*,\phi;t]$. 

After this slight digression, we now return to the long-time behaviour of the 
gas. In the long-time limit two important simplifications occur. Looking at the 
expresions for the selfenergies, we see that their width as a function of the 
time difference $t-t'$ is at most of order ${\cal O}(\hbar/k_BT)$ in the 
low-temperature regime of interest where the thermal de Broglie wavelength of 
the particles obeys $\Lambda \gg 1/k_c$. If therefore 
$t \gg t_0 + {\cal O}(\hbar/k_BT)$ we are allowed to take the limit 
$t_0 \rightarrow -\infty$. Taking this limit means physically that we neglect 
the initial transients that are due to the precise way in which the contact 
between the trap and the reservoir is made, and focus on the `universal' 
dynamics that is independent of these details. In addition, at long times the 
dynamics of the gas is expected to be sufficiently slow that we can neglect the 
memory effects altogether. Finally, we consider here first the case of a 
reservoir that is so weakly coupled to the gas in the trap that we can treat 
the coupling with second order perturbation theory. As a result we can also 
neglect the nondiagonal elements of the selfenergies. In total we are then 
allowed to put (but see below) 
\begin{equation}
\Sigma^{(\pm),K}_{{\bf n},{\bf n}'}(t-t') \simeq 
    \Sigma^{(\pm),K}_{{\bf n},{\bf n}}(\epsilon_{\bf n}-\mu)
                \delta_{{\bf n},{\bf n}'} \delta(t-t')
     \equiv \Sigma^{(\pm),K}_{\bf n} \delta_{{\bf n},{\bf n}'} \delta(t-t')~,
\end{equation}
where the Fourier transform of the selfenergies is defined by
\begin{equation}                                                               
\Sigma^{(\pm),K}_{{\bf n},{\bf n}'}(t-t') 
   = \int d\epsilon~ \Sigma^{(\pm),K}_{{\bf n},{\bf n}'}(\epsilon) 
                          \frac{e^{-i\epsilon (t-t')/\hbar}}{2\pi\hbar}~.
\end{equation}
Note that this implies in general that
$\Sigma^{(-)}_{{\bf n}',{\bf n}}(\epsilon) 
                 = \left( \Sigma^{(+)}_{{\bf n},{\bf n}'}(\epsilon) \right)^*$, 
and in particular therefore that 
$\Sigma^{(-)}_{\bf n} = \left( \Sigma^{(+)}_{\bf n} \right)^*$. 

With these simplifications our Langevin equations become 
\begin{equation}
i\hbar \frac{\partial}{\partial t} \phi_{\bf n}(t)
  = (\epsilon_{\bf n} + \hbar\Sigma^{(+)}_{\bf n} - \mu)\phi_{\bf n}(t) 
    + \eta_{\bf n}(t)
\end{equation}
and just the complex conjugate equation for $\phi^*_{\bf n}(t)$. The retarded 
selfenergy in this equation is given by
\begin{equation}
\hbar\Sigma^{(+)}_{\bf n} 
  = \int \frac{d{\bf k}}{(2\pi)^3}~  
                 t^*_{\bf n}({\bf k}) 
                    \frac{1}{\epsilon_{\bf n}^+ - \epsilon({\bf k})} 
                 t_{\bf n}({\bf k})~,
\end{equation}
with $\epsilon^+_{\bf n}$ the usual notation for the limiting procedure
$\epsilon_{\bf n} + i0$. It clearly has real and imaginary parts that we 
denote by $S_{\bf n}$ and $- R_{\bf n}$, respectively. Denoting the Cauchy 
principle value part of an integral by ${\cal P}$, they obey
\begin{equation}
\label{shift}
S_{\bf n}
  = \int \frac{d{\bf k}}{(2\pi)^3}~
                 t^*_{\bf n}({\bf k}) 
                    \frac{{\cal P}}{\epsilon_{\bf n} - \epsilon({\bf k})} 
                 t_{\bf n}({\bf k})
\end{equation}
and
\begin{equation}
\label{decay}
R_{\bf n}
  = \pi \int \frac{d{\bf k}}{(2\pi)^3}~
        \delta( \epsilon_{\bf n} - \epsilon({\bf k}) )         
                                          |t_{\bf n}({\bf k})|^2 ~.
\end{equation} 
The interpretation of these results is quite obvious if we consider the average 
of the Langevin equation, i.e.,
\begin{equation}
i\hbar \frac{\partial}{\partial t} \langle \phi_{\bf n} \rangle(t)
  = (\epsilon_{\bf n} + S_{\bf n} - iR_{\bf n} - \mu)
                                   \langle \phi_{\bf n} \rangle(t)~,
\end{equation}
which is solved by
\begin{equation}
\langle \phi_{\bf n} \rangle(t) = \langle \phi_{\bf n} \rangle(0)
    e^{-i(\epsilon_{\bf n} + S_{\bf n} - \mu)t/\hbar} 
e^{-R_{\bf n}t/\hbar}~.
\end{equation}
Hence the real part of the retarded selfenergy $S_{\bf n}$ represents the 
shift in the energy of state $\chi_{\bf n}({\bf x})$ due to the coupling with 
the reservoir. Indeed, equation~(\ref{shift}) agrees precisely with the value 
of this shift in second order perturbation theory. Moreover, the fact that
$|\langle \phi_{\bf n} \rangle(t)|^2 = 
         |\langle \phi_{\bf n} \rangle(0)|^2 e^{-2R_{\bf n}t/\hbar}$
shows that the average rate of decay $\Gamma_{\bf n}$ of the state 
$\chi_{\bf n}({\bf x})$ is equal to $2R_{\bf n}/\hbar$, which, together with 
equation~(\ref{decay}), exactly reproduces Fermi's Golden Rule.

From the equation of motion for the average $\langle \phi_{\bf n} \rangle(t)$ 
we can also conclude that the right-hand side of the Fokker-Planck equation 
contains the `streaming' terms
\begin{eqnarray}
- \left( \sum_{\bf n} \frac{\partial}{\partial \phi_{\bf n}} 
     (\epsilon_{\bf n} + \hbar\Sigma^{(+)}_{\bf n} - \mu)
                                      \phi_{\bf n} \right) P[\phi^*,\phi;t]
                                               \hspace*{1.5in} \nonumber \\                                     
  + \left( \sum_{\bf n} \frac{\partial}{\partial \phi^*_{\bf n}} 
     (\epsilon_{\bf n} + \hbar\Sigma^{(-)}_{\bf n} - \mu)
                                    \phi^*_{\bf n} \right) P[\phi^*,\phi;t]~.
                                                            \nonumber
\end{eqnarray}
However, there must now also be a `diffusion' term due to the fact that the 
average $\langle |\phi_{\bf n}|^2 \rangle(t)$ is no longer independent of time 
since the interactions with the reservoir can lead to changes in the occupation 
numbers $N_{\bf n}(t)$. To obtain the `diffusion' term we thus need to 
determine $i\hbar \partial \langle |\phi_{\bf n}|^2 \rangle(t)/\partial t$. To 
do so we first formally solve the Langevin equation by
\begin{eqnarray}
&& \hspace*{-0.2in} 
\phi_{\bf n}(t) \\
&&= e^{-i(\epsilon_{\bf n} + \hbar\Sigma^{(+)}_{\bf n} - \mu)t/\hbar}
   \left\{ \phi_{\bf n}(0)
     - \frac{i}{\hbar} \int_0^t dt'~\eta(t') 
          e^{i(\epsilon_{\bf n} + \hbar\Sigma^{(+)}_{\bf n} - \mu)t'/\hbar} 
                                                     \right\}~. \nonumber
\end{eqnarray}
Multiplying this with the complex conjugate expression and taking the average, 
we obtain first of all
\begin{equation}
\langle |\phi_{\bf n}|^2 \rangle(t) = e^{-2R_{\bf n}t/\hbar}
   \left\{ \langle |\phi_{\bf n}|^2 \rangle(0)
           + \frac{i}{2} \Sigma^K_{\bf n} 
                    \int_0^t dt'~ e^{2R_{\bf n}t'/\hbar} \right\}~,  
\end{equation}
which shows that 
\begin{equation}
\label{fluc}
i\hbar \frac{\partial}{\partial t} \langle |\phi_{\bf n}|^2 \rangle(t)
  = -2iR_{\bf n} \langle |\phi_{\bf n}|^2 \rangle(t)
    - \frac{1}{2} \hbar \Sigma^K_{\bf n}~.
\end{equation}
Moreover, the Keldysh component of the selfenergy is given by
\begin{equation}
\hbar \Sigma^K_{\bf n}
= -2\pi i \int \frac{d{\bf k}}{(2\pi)^3}~
                 (1 + 2N({\bf k}))
                    \delta( \epsilon_{\bf n} - \epsilon({\bf k}) )
                                                |t_{\bf n}({\bf k})|^2 
\end{equation} 
and therefore obeys
\begin{equation}
\hbar \Sigma^K_{\bf n} = -2i(1 + 2N_{\bf n}) R_{\bf n}
\end{equation}
with $N_{\bf n} = (e^{\beta(\epsilon_{\bf n} -\mu)} -1 )^{-1}$ the Bose 
distribution function. This is in fact the fluctuation-dissipation theorem, 
because it relates the strength of the fluctuations determined by $\hbar 
\Sigma^K_{\bf n}$, to the amount of dissipation that is given by $R_{\bf n}$.
As mentioned previously, the fluctuation-dissipation theorem ensures that the 
gas relaxes to thermal equilibrium. This we can already see from 
equation~(\ref{fluc}), because substitution of the fluctuation-dissipation 
theorem leads to 
\begin{equation}
\label{fluct}
i\hbar \frac{\partial}{\partial t} \langle |\phi_{\bf n}|^2 \rangle(t)
  = -2iR_{\bf n} \langle |\phi_{\bf n}|^2 \rangle(t)
    + iR_{\bf n}(1 + 2N_{\bf n})~,
\end{equation}
which tells us that in equilibrium 
$\langle |\phi_{\bf n}|^2 \rangle = N_{\bf n} + 1/2$. In fact, we have argued  
that in general 
$\langle |\phi_{\bf n}|^2 \rangle(t) = N_{\bf n}(t) + 1/2$. Substituting this
identity in equation~(\ref{fluct}) and realizing that 
$\Gamma_{\bf n} N_{\bf n} = \Gamma_{\bf n} N({\bf k})$ due to the 
energy-conserving $\delta$ function in $\Gamma_{\bf n}$, we indeed obtain the 
correct rate equation for the average occupation numbers
\begin{eqnarray}
\label{qdBE}
&& \hspace*{-0.55in} 
\frac{\partial}{\partial t} N_{\bf n}(t) 
   = - \Gamma_{\bf n} N_{\bf n}(t)  + \Gamma_{\bf n} N_{\bf n} \\
&&= - \Gamma_{\bf n} N_{\bf n}(t)(1+N({\bf k})) 
                    + \Gamma_{\bf n} (1+N_{\bf n}(t)) N({\bf k})~, \nonumber
\end{eqnarray}
that might justly be called the quantum Boltzmann equation for the gas, because 
the right-hand side contains precisely the rates for scattering into and out of 
the reservoir.  

Furthermore, equation~(\ref{fluc}) shows that the `diffusion' term in the 
Fokker-Planck equation is
\begin{eqnarray}
- \left( \frac{1}{2} \sum_{\bf n} 
    \frac{\partial^2}{\partial \phi^*_{\bf n} \partial \phi_{\bf n}} 
     \hbar \Sigma^K_{\bf n} \right) P[\phi^*,\phi;t]~,  \nonumber
\end{eqnarray}
because the first term in the right-hand side of equation~(\ref{fluc}) is also 
present without the noise and is therefore already accounted for in the 
`streaming' terms. In total we have thus arrived at
\begin{eqnarray}
\label{FP2}
&& \hspace*{-0.7in} i\hbar \frac{\partial}{\partial t} P[\phi^*,\phi;t]
= - \left( \sum_{\bf n} \frac{\partial}{\partial \phi_{\bf n}} 
       (\epsilon_{\bf n} + \hbar\Sigma^{(+)}_{\bf n} - \mu)
                                      \phi_{\bf n} \right) P[\phi^*,\phi;t] \\
&& \hspace*{0.2in}
  + \left( \sum_{\bf n} \frac{\partial}{\partial \phi^*_{\bf n}} 
     (\epsilon_{\bf n} + \hbar\Sigma^{(-)}_{\bf n} - \mu)
                                    \phi^*_{\bf n} \right) P[\phi^*,\phi;t]
                                                         \nonumber \\
&& \hspace*{0.2in}
  - \left( \frac{1}{2} \sum_{\bf n} 
    \frac{\partial^2}{\partial \phi^*_{\bf n} \partial \phi_{\bf n}} 
      \hbar \Sigma^K_{\bf n} \right) P[\phi^*,\phi;t]~. \nonumber                                                                                                       
\end{eqnarray}
Using again the fluctuation-dissipation theorem, it is not difficult to show 
that the stationary solution of this Fokker-Planck equation is 
\begin{equation}
\label{stat}
P[\phi^*,\phi;\infty] = \prod_{\bf n} \frac{1}{N_{\bf n} + 1/2}
     \exp \left\{ - \frac{1}{N_{\bf n} +1/2} |\phi_{\bf n}|^2
          \right\}~.
\end{equation}
Although this appears to be quite reasonable, it is important to note that this 
result is not fully consistent because the `streaming' terms in 
equation~(\ref{FP2}) 
show that the energies of the states $\chi_{\bf n}({\bf x})$ are shifted and 
equal to $\epsilon_{\bf n} + S_{\bf n}$. We therefore expect that the 
equilibrium occupation numbers are equal to the Bose distribution evaluated at 
those energies and not at the unshifted values $\epsilon_{\bf n}$. The reason 
for this inconsistency is, of course, that we have taken 
$\Sigma^{(\pm),K}_{\bf n}$ equal to 
$\Sigma^{(\pm),K}_{{\bf n},{\bf n}}(\epsilon_{\bf n}-\mu)$, which was only 
justified on the basis of second-order perturbation theory. We now, however, 
see that to obtain a fully consistent theory we must take instead
$\Sigma^{(\pm),K}_{\bf n} \equiv  
        \Sigma^{(\pm),K}_{{\bf n},{\bf n}}(\epsilon_{\bf n}+S_{\bf n}-\mu)$~.
This implies that to obtain the retarded selfenergy, we first have to solve
for the real part  
\begin{equation}
S_{\bf n}
  = \int \frac{d{\bf k}}{(2\pi)^3}~
         t^*_{\bf n}({\bf k}) 
            \frac{{\cal P}}{\epsilon_{\bf n} + S_{\bf n} - \epsilon({\bf k})} 
                t_{\bf n}({\bf k})
\end{equation}
and then substitute this into 
\begin{equation}
R_{\bf n}
  = \pi \int \frac{d{\bf k}}{(2\pi)^3}~
        \delta( \epsilon_{\bf n} + S_{\bf n} - \epsilon({\bf k}) )         
                                                   |t_{\bf n}({\bf k})|^2 
\end{equation}      
to obtain the imaginary part. The fluctuation-dissipation theorem then again 
reads $\hbar \Sigma^K_{\bf n} = -2i(1 + 2N_{\bf n}) R_{\bf n}$ but now with 
the equilibrium distribution
$N_{\bf n} = (e^{\beta(\epsilon_{\bf n} + S_{\bf n} -\mu)} -1 )^{-1}$. As a 
consequence the quantum Boltzmann equation and the stationary state of the 
Fokker-Planck equation will now also contain these correct equilibrium 
occupation numbers \cite{PQW}.

Having arrived at the Fokker-Planck equation of our Caldeira-Leggett-like toy 
model, we have now demonstrated the 
use of essentially all the theoretical tools that are required for a discussion 
of a weakly interacting Bose gas. Before we start with this discussion, 
however, we want to make a final remark about the effect of the nondiagonal 
elements of the selfenergies. Physically, including these nondiagonal elements 
accounts for the change in the wave functions $\chi_{\bf n}({\bf x})$ due to 
the interaction with the reservoir. It can clearly be neglected if the coupling 
with the reservoir is sufficiently weak, or more precisely if 
$|\hbar\Sigma_{{\bf n}',{\bf n}}^{(+)}(\epsilon_{\bf n}+S_{\bf n}-\mu)|$
is much smaller than the energy splitting 
$|\epsilon_{{\bf n}'}-\epsilon_{\bf n}|$ between the states in the trap. For a 
quantum dot in solid-state phyisics \cite{kastner} this is often the case and 
for that reason we might call our toy model a bosonic quantum dot. However, 
in magnetically trapped atomic gases the average interaction energy between the 
atoms is already slightly below the Bose-Einstein condensation critical 
temperature larger than the typical energy splitting and the nondiagonal 
elements of the selfenergies are very important. Such a strong-coupling 
situation can, of course, also be treated in our Caldeira-Leggett like model. 
The main difference is that we need to expand our various fields not in the 
eigenstates of the trapping potential $\chi_{\bf n}({\bf x})$ but in the 
eigenstates $\chi'_{\bf n}({\bf x})$ of the nonlocal Schr\"odinger equation
\begin{eqnarray}
\label{nonloc}
&& \hspace*{-0.53in}
\epsilon'_{\bf n} \chi'_{\bf n}({\bf x}) = 
  \left( - \frac{\hbar^2 \mbox{\boldmath $\nabla$}^2}{2m} 
         + V^{\rm ex}({\bf x}) \right) \chi'_{\bf n}({\bf x}) \\
  &&+ \int d{\bf x}'~ 
    {\rm Re}\left[\hbar\Sigma^{(+)}({\bf x},{\bf x}';\epsilon'_{\bf n}-\mu)  
                                      \right] \chi'_{\bf n}({\bf x}')
                                      \nonumber
\end{eqnarray}
where $\epsilon'_{\bf n}$ are the new eigenvalues and the retarded selfenergie 
in coordinate space is
$\hbar\Sigma^{(+)}({\bf x},{\bf x}';\epsilon) 
  = \sum_{{\bf n},{\bf n}'} \chi_{\bf n}({\bf x})
      \hbar\Sigma_{{\bf n},{\bf n}'}^{(+)}(\epsilon) 
      \chi^*_{{\bf n}'}({\bf x}')$.
In this new basis the nondiagonal elements of the selfenergies can now be 
neglected and we find essentially the same results as before. We only need to 
replace $\epsilon_{\bf n}+S_{\bf n}$ by $\epsilon'_{\bf n}$ and 
$t_{\bf n}({\bf k})$ by 
$\sum_{{\bf n}'} \left( \int d{\bf x}~ \chi'_{\bf n}({\bf x}')  
                      \chi^*_{{\bf n}'}({\bf x}) \right) t_{{\bf n}'}({\bf k})$.
Neglecting the nondiagonal elements in this basis only requires that the real 
part of the retarded selfenergy is much larger that its imaginary part, which 
is always fulfilled in our case because $k_c \Lambda \gg 1$.

Summarizing, the coherent and incoherent dynamics of the gas in the trap is for 
our toy model quite generally solved by
\begin{equation}
\langle \phi_{\bf n} \rangle(t) = 
  \langle \phi_{\bf n} \rangle(0) e^{-i(\epsilon'_{\bf n}-\mu)t/\hbar} 
                                                   e^{-\Gamma_{\bf n}t/2}
\end{equation}
and
\begin{equation}
N_{\bf n}(t) = N_{\bf n}(0) e^{-\Gamma_{\bf n}t}
              + N_{\bf n} \left( 1 - e^{-\Gamma_{\bf n}t} \right)~,
\end{equation}
respectively. In the limit $t \rightarrow \infty$, the average of the 
annihilation operators $\langle \phi_{\bf n} \rangle(t)$ thus always vanishes 
but the average occupation numbers $N_{\bf n}(t)$ relax to the equilibrium 
distribution $N_{\bf n} = (e^{\beta(\epsilon'_{\bf n}-\mu)}-1)^{-1}$. 
Although this appears to be an immediately obvious result, its importance stems 
from the fact that it is also true if we tune the potential energy bias 
$\Delta V^{\rm ex}$ such that at low temperatures the groundstate 
$\chi'_{\bf 0}({\bf x})$ acquires a macroscopic occupation, i.e., 
$N_{\bf 0} \gg 1$. The gas therefore never shows a spontaneous breaking of the 
$U(1)$ symmetry, in agreement with the notion that we are essentially dealing 
with an ideal Bose gas in the grand-canonical ensemble \cite{henk2}. The reason 
of the absense of the spontaneous symmetry breaking can also be understood from 
our stationary solution of the Fokker-Planck equation in equation~(\ref{stat}), 
which shows that the probability distribution for $|\phi_{\bf 0}|$ is 
proportional to the Boltzmann factor 
$e^{-\beta(\epsilon'_{\bf 0}-\mu) |\phi_{\bf 0}|^2}$ in the degenerate regime 
of interest. Hence, the corresponding Landau free energy  
$F_L(|\phi_{\bf 0}|) = (\epsilon'_{\bf 0}-\mu) |\phi_{\bf 0}|^2$ never shows an 
instability towards the formation of a nonzero average of $|\phi_{\bf 0}|$ due 
to the fact that $\epsilon'_{\bf 0}-\mu$ can never become less then zero. Once 
we introduce interactions between the atoms in the gas, this picture fully 
changes.

\subsection{Condensate formation}
\label{CF}
We are now in a position to discuss an interacting Bose gas, because in an 
interacting Bose gas, the gas is roughly speaking its own thermal reservoir. The 
Fokker-Planck equation therefore turns out to be quite similar to the one 
presented in equation (\ref{FP2}). To be more precise, however, we should 
mention that we aim in this section to describe the formation of a condensate in 
an interacting Bose gas under the conditions that have recently been realized in 
experiments with atomic rubidium, lithium, sodium and hydrogen gases. In these 
experiments the gas is cooled by means of evaporative cooling. Numerical 
solutions of the Boltzmann equation have shown that during evaporative cooling 
the energy distribution function is well described by an equilibrium 
distribution with time dependent temperature $T(t)$ and chemical potential 
$\mu(t)$, that is truncated at high energies due to the evaporation of the 
highest energetic atoms from the trap \cite{jom}. Moreover, in the experiments 
of interest the densities just above the critical temperature are essentially 
always such that the gas is in the weak-coupling limit, which implies in this 
context that the average interaction energy per atom is always much less than 
the energy splitting of the one-particle states in the harmonic trapping 
potential.

\subsubsection{Weak-coupling limit}
For a trapped atomic gas the appropriate one-particle states are the eigenstates 
$\chi_{\bf n}({\bf x})$ of the external potential $V^{\rm ex}({\bf x})$. As a 
result the Fokker-Planck equation for the interacting gas will now quite 
generally be given by
\begin{eqnarray}
\label{FPinhom}
&& \hspace*{-0.4in}
i\hbar \frac{\partial}{\partial t} P[\phi^*,\phi;t] 
 = - \sum_{{\bf n},{\bf n}'}
   \frac{\partial}{\partial \phi_{{\bf n}}}
       \left\{ \left( (\epsilon_{{\bf n}} - \mu(t)) \delta_{{\bf n},{\bf n}'} 
            + \hbar\Sigma^{(+)}_{{\bf n},{\bf n}'} \right) \phi_{{\bf n}'}
              \raisebox{0.3in}{} \right. \\
&& \hspace*{1.4in} + \left. \sum_{{\bf m},{\bf m}'}
               V^{(+)}_{{\bf n},{\bf m};{\bf n}',{\bf m}'}
                   \phi^*_{{\bf m}}
                   \phi_{{\bf m}'} \phi_{{\bf n}'}
       \right\} P[\phi^*,\phi;t]                         \nonumber \\
&&+ \sum_{{\bf n},{\bf n}'}
  \frac{\partial}{\partial \phi^*_{{\bf n}}}
       \left\{ \left( (\epsilon_{{\bf n}} - \mu(t)) \delta_{{\bf n},{\bf n}'}
            + \hbar\Sigma^{(-)}_{{\bf n}',{\bf n}} \right) \phi^*_{{\bf n}'}
              \raisebox{0.3in}{} \right. \nonumber \\
&& \hspace*{1.4in} + \left. \sum_{{\bf m},{\bf m}'}
               V^{(-)}_{{\bf n}',{\bf m}';{\bf n},{\bf m}}
                   \phi^*_{{\bf n}'} \phi^*_{{\bf m}'}
                   \phi_{{\bf m}}
       \right\} P[\phi^*,\phi;t]                         \nonumber \\
&&- \frac{1}{2} \sum_{{\bf n},{\bf n}'}
  \frac{\partial^2}{\partial \phi_{{\bf n}}
                    \partial \phi^*_{{\bf n}'}}
       \left\{ \hbar\Sigma^K_{{\bf n},{\bf n}'}
            + \sum_{{\bf m},{\bf m}'}
               V^K_{{\bf n},{\bf m};{\bf n}',{\bf m}'}
                   \phi^*_{{\bf m}}\phi_{{\bf m}'}
       \right\} P[\phi^*,\phi;t]~. \nonumber
\end{eqnarray}
For this equation to be useful, however, we need to determine the various 
selfenergies and interactions. To do so, let us for simplicity first consider 
the normal state of the gas which is usually in the weak-coupling limit 
$|\hbar\Sigma^{(+)}_{{\bf n},{\bf n}'}| 
                           \ll |\epsilon_{{\bf n}} - \epsilon_{{\bf n}'}|$ 
for the experiments with atomic alkali gases that are of interest 
here. In this regime we can neglect the nondiagonal elements of the 
selfenergies and also the nonlinear terms in the right-hand side of 
equation~(\ref{FPinhom}). The Fokker-Planck equation then reduces to
\begin{eqnarray}
&& \hspace*{-0.7in} 
i\hbar \frac{\partial}{\partial t} P[\phi^*,\phi;t] 
= - \sum_{{\bf n}}
  \frac{\partial}{\partial \phi_{{\bf n}}}
       (\epsilon_{{\bf n}} + \hbar\Sigma^{(+)}_{{\bf n}} - \mu(t)) 
       \phi_{{\bf n}} P[\phi^*,\phi;t] \\
&& \hspace*{0.23in} + \sum_{{\bf n}}
  \frac{\partial}{\partial \phi^*_{{\bf n}}}
       (\epsilon_{{\bf n}} + \hbar\Sigma^{(-)}_{{\bf n}} - \mu(t)) 
\phi^*_{{\bf n}}
            P[\phi^*,\phi;t]                         \nonumber \\
&& \hspace*{0.23in} - \frac{1}{2} \sum_{{\bf n}}
  \frac{\partial^2}{\partial \phi^*_{{\bf n}}
                    \partial \phi_{{\bf n}}}
        \hbar\Sigma^K_{{\bf n}} P[\phi^*,\phi;t]~,   \nonumber
\end{eqnarray} 
with $\hbar\Sigma^{(\pm),K}_{{\bf n}} 
         = \hbar\Sigma^{(\pm),K}_{{\bf n},{\bf n}}(\epsilon'_{{\bf n}} - \mu)$
and $\epsilon'_{{\bf n}} = \epsilon_{{\bf n}} + S_{{\bf n}}$. This equation 
looks identical to the one we derived for our Caldeira-Leggett toy model but 
describes in fact quite different physics because now the selfenergies are not 
due to an interaction with a reservoir but due to the interactions between the 
atoms themselves. This is, however, completely hidden in the selfenergies as we 
show now explicitly. Performing the same manipulations as in sections \ref{IFD} 
and \ref{AP}, we find in the many-body T-matrix approximation first of all that
\begin{eqnarray}
&& \hspace*{-0.5in}
\hbar\Sigma^{(+)}_{{\bf n}} = \left( \hbar\Sigma^{(-)}_{{\bf n}} \right)^* =
  2 \sum_{{\bf m}} V^{(+)}_{{\bf n},{\bf m};{\bf n},{\bf m}}
                         (\epsilon'_{{\bf n}} + \epsilon'_{{\bf m}} - 2\mu)
                                    N_{{\bf m}} \\
&& \hspace*{0.68in} + 2i \sum_{{\bf m}} \int \frac{d\epsilon}{2\pi\hbar}~
       V^<_{{\bf n},{\bf m};{\bf n},{\bf m}}(\epsilon) 
 \frac{\hbar}{\epsilon^- - (\epsilon'_{{\bf n}} + \epsilon'_{{\bf m}} - 2\mu)}
 \nonumber
\end{eqnarray}  
and similarly that
\begin{eqnarray}
&& \hspace*{-0.33in}
\hbar\Sigma^K_{{\bf n}} = 
  2 \sum_{{\bf m}} V^K_{{\bf n},{\bf m};{\bf n},{\bf m}}
                         (\epsilon'_{{\bf n}} + \epsilon'_{{\bf m}} - 2\mu)
                                    N_{{\bf m}} \\
&&+ 2 \sum_{{\bf m}} V^<_{{\bf n},{\bf m};{\bf n},{\bf m}}
                           (\epsilon'_{{\bf n}} + \epsilon'_{{\bf m}} - 2\mu) 
        (1 + 2N_{{\bf m}})~. \nonumber
\end{eqnarray}

Our next task is therefore the evaluation of the various many-body T-matrix 
elements 
$V^{(+),<,K}_{{\bf n},{\bf m};{\bf n}',{\bf m}'}(\epsilon) \equiv 
    (T^{(+),<,K}_{{\bf n},{\bf m};{\bf n}',{\bf m}'}(\epsilon) +    
             T^{(+),<,K}_{{\bf m},{\bf n};{\bf n}',{\bf m}'}(\epsilon))/2$
that occur in these expressions, which is easily carried out with our 
experience of the homogeneous gas and by noting that the Lippmann-Schwinger 
equation is in fact an operator equation that only needs 
to be expressed in a new basis. In terms of the appropriate matrix elements of 
the interatomic interaction, i.e.,
\begin{equation}
V_{{\bf n},{\bf m};{\bf n}',{\bf m}'} = \int d{\bf x} \int d{\bf x}'~
   \chi^*_{{\bf n}}({\bf x}) \chi^*_{{\bf m}}({\bf x}') V({\bf x}-{\bf x}')
        \chi_{{\bf n}'}({\bf x}) \chi_{{\bf m}'}({\bf x}')~,
\end{equation}
the retarded and advanced components of the many-body T matrix thus obey the 
Bethe-Salpeter equation \cite{nick}
\begin{eqnarray}
&& \hspace*{-0.4in}
T^{(\pm)}_{{\bf n},{\bf m};{\bf n}',{\bf m}'}(\epsilon)
  = V_{{\bf n},{\bf m};{\bf n}',{\bf m}'} \\
&& \hspace*{0.42in}
  + \sum_{{\bf n}'',{\bf m}''} V_{{\bf n},{\bf m};{\bf n}'',{\bf m}''}
      \frac{1 + N_{{\bf n}''} + N_{{\bf m}''} }
        {\epsilon^{\pm} - (\epsilon'_{{\bf n}''} + \epsilon'_{{\bf m}''} - 
                                                                       2\mu)}
    T^{(\pm)}_{{\bf n}'',{\bf m}'';{\bf n}',{\bf m}'}(\epsilon) \nonumber
\end{eqnarray}
and the optical theorem
\begin{eqnarray}
&& \hspace*{-0.4in}
T^{(+)}_{{\bf n},{\bf m};{\bf n}',{\bf m}'}(\epsilon)
  - T^{(-)}_{{\bf n},{\bf m};{\bf n}',{\bf m}'}(\epsilon) = 
- 2\pi i \sum_{{\bf n}'',{\bf m}''} 
  \delta(\epsilon - (\epsilon'_{{\bf n}''} + \epsilon'_{{\bf m}''} - 2\mu)) \\
&& \hspace*{0.9in} \times~        
  T^{(+)}_{{\bf n},{\bf m};{\bf n}'',{\bf m}''}(\epsilon) 
       (1 + N_{{\bf n}''} + N_{{\bf m}''}) 
                   T^{(-)}_{{\bf n}'',{\bf m}'';{\bf n}',{\bf m}'}(\epsilon)~.
                   \nonumber
\end{eqnarray}
Finally, we also need
\begin{eqnarray}
&& \hspace*{-0.6in}
T^<_{{\bf n},{\bf m};{\bf n}',{\bf m}'}(\epsilon) =
  - 2\pi i \sum_{{\bf n}'',{\bf m}''} 
    \delta(\epsilon - (\epsilon'_{{\bf n}''} + \epsilon'_{{\bf m}''} - 2\mu)) 
\\    
&& \hspace*{0.2in} \times~  
       T^{(+)}_{{\bf n},{\bf m};{\bf n}'',{\bf m}''}(\epsilon)
          N_{{\bf n}''} N_{{\bf m}''} 
               T^{(-)}_{{\bf n}'',{\bf m}'';{\bf n}',{\bf m}'}(\epsilon)
               \nonumber
\end{eqnarray}
and the Keldysh component
\begin{eqnarray}
&& \hspace*{-0.2in}
T^K_{{\bf n},{\bf m};{\bf n}',{\bf m}'}(\epsilon) =  
- 2\pi i \sum_{{\bf n}'',{\bf m}''} 
  \delta(\epsilon - (\epsilon'_{{\bf n}''} + \epsilon'_{{\bf m}''} - 2\mu)) \\
&&\times~ T^{(+)}_{{\bf n},{\bf m};{\bf n}'',{\bf m}''}(\epsilon)
    (1 + N_{{\bf n}''} + N_{{\bf m}''} + 2 N_{{\bf n}''} N_{{\bf m}''}) 
                   T^{(-)}_{{\bf n}'',{\bf m}'';{\bf n}',{\bf m}'}(\epsilon)~.
                   \nonumber
\end{eqnarray}
Notice that the appearance of a $\delta$ function in the last three equations 
in practice never leads to any problems, because in the experiments with 
trapped atomic gases the number of particles is always so large that the 
critical temperature of the gas is much larger than the energy splitting of the 
trap and the discrete sum over states is well approximated by an integral over 
a continuous spectrum.

Making use of the above relations we find that for the states with 
thermal energies the renormalized energy is given by the pseudopotential result
\begin{equation}
\epsilon'_{{\bf n}}(t) = \epsilon_{{\bf n}} 
   + \frac{8\pi a \hbar^2}{m} 
        \int d{\bf x}~
             \chi^*_{{\bf n}}({\bf x}) n({\bf x},t) \chi_{{\bf n}}({\bf x})~,
\end{equation}
with $n({\bf x},t)$ the density profile of the gas cloud. For the states near 
the one-particle ground state there is, however, a correction given by
\begin{eqnarray}
-2 \sum_{{\bf m}} \sum_{{\bf n}'',{\bf m}''} 
    |V^{(+)}_{{\bf n},{\bf m};{\bf n}'',{\bf m}''}(t)|^2
          N_{{\bf n}''}(t) N_{{\bf m}''}(t)
           \frac{ {\cal P} } 
                { \epsilon'_{{\bf n}}(t) + \epsilon'_{{\bf m}}(t) 
                 - \epsilon'_{{\bf n}''}(t) - \epsilon'_{{\bf m}''}(t) } 
\nonumber
\end{eqnarray}
and 
$V^{(+)}_{{\bf n},{\bf m};{\bf n}'',{\bf m}''}(t) = 
    V^{(+)}_{{\bf n},{\bf m};{\bf n}'',{\bf m}''}
             (\epsilon'_{{\bf n}''}(t) + \epsilon'_{{\bf m}''}(t) - 2\mu(t))$. 
In the weak-coupling limit, the renormalization of the one-particle energies 
plays by definition not such an important role and we are much more interested 
in the imaginary part of the retarded and advanced selfenergies. They are given 
by
\begin{eqnarray}
&& \hspace*{-0.2in}
R_{{\bf n}}(t) = 2\pi \sum_{{\bf m}} \sum_{{\bf n}'',{\bf m}''}
  \delta(\epsilon'_{{\bf n}''}(t) + \epsilon'_{{\bf m}''}(t) 
         - \epsilon'_{{\bf n}}(t) - \epsilon'_{{\bf m}}(t)) \\
&&\times |V^{(+)}_{{\bf n},{\bf m};{\bf n}'',{\bf m}''}(t)|^2
            [(1 + N_{{\bf n}''}(t) + N_{{\bf m}''}(t)) N_{{\bf m}}(t) 
                                   - N_{{\bf n}''}(t) N_{{\bf m}''}(t)]~.
                                   \nonumber  
\end{eqnarray}
Furthermore, the Keldysh component of the selfenergies equals
\begin{eqnarray}
&& \hspace*{-0.48in}
\hbar\Sigma^K_{{\bf n}}(t) = - 4\pi \sum_{{\bf m}} \sum_{{\bf n}'',{\bf m}''}
  \delta(\epsilon'_{{\bf n}''}(t) + \epsilon'_{{\bf m}''}(t) 
         - \epsilon'_{{\bf n}}(t) - \epsilon'_{{\bf m}}(t))  \\
&& \times |V^{(+)}_{{\bf n},{\bf m};{\bf n}'',{\bf m}''}(t)|^2
            [(1 + N_{{\bf n}''}(t) + N_{{\bf m}''}(t)) N_{{\bf m}}(t)
            \nonumber \\ 
&& \hspace*{1.27in} + N_{{\bf n}''}(t) N_{{\bf m}''}(t)(1 + 2N_{{\bf m}}(t))]~.
                \nonumber  
\end{eqnarray}
As a result the quantum Boltzmann equation for the gas becomes
\begin{equation}
\label{BEinhom}
\frac{\partial}{\partial t} N_{{\bf n}}(t) =
  - \Gamma^{\rm out}_{{\bf n}}(t) N_{{\bf n}}(t) 
     + \Gamma^{\rm in}_{{\bf n}}(t) (1 + N_{{\bf n}}(t))~,
\end{equation}
with the rate to scatter out of the state $\chi_{{\bf n}}({\bf x})$ given by 
\begin{eqnarray}
&& \hspace*{-0.7in} 
\Gamma^{\rm out}_{{\bf n}}(t) =  \frac{4\pi}{\hbar} 
 \sum_{{\bf m}} \sum_{{\bf n}'',{\bf m}''}
  \delta(\epsilon'_{{\bf n}''}(t) + \epsilon'_{{\bf m}''}(t) 
         - \epsilon'_{{\bf n}}(t) - \epsilon'_{{\bf m}}(t)) \\
&&\times |V^{(+)}_{{\bf n},{\bf m};{\bf n}'',{\bf m}''}(t)|^2
        (1 + N_{{\bf n}''}(t))(1 + N_{{\bf m}''}(t)) N_{{\bf m}}(t) \nonumber 
\end{eqnarray}
and the rate to scatter into this state by
\begin{eqnarray}
&& \hspace*{-0.7in} 
\Gamma^{\rm in}_{{\bf n}}(t) =  \frac{4\pi}{\hbar} 
 \sum_{{\bf m}} \sum_{{\bf n}'',{\bf m}''}
  \delta(\epsilon'_{{\bf n}''}(t) + \epsilon'_{{\bf m}''}(t) 
         - \epsilon'_{{\bf n}}(t) - \epsilon'_{{\bf m}}(t))  \\
&&\times |V^{(+)}_{{\bf n},{\bf m};{\bf n}'',{\bf m}''}(t)|^2
            N_{{\bf n}''}(t) N_{{\bf m}''}(t) (1 + N_{{\bf m}}(t))~, \nonumber 
\end{eqnarray}
as expected from Fermi's Golden Rule.

The quantum Boltzmann equation found above, in combination with the expressions 
for the shift in the energy levels, fully describes the dynamics of the gas in 
the normal state. However, close to the critical temperature when the 
occupation numbers of the one-particle groundstate $\chi_{\bf 0}({\bf x})$ start 
to become large, we are no longer allowed to neglect the nonlinear terms in the 
Fokker-Planck equation in equation~(\ref{FPinhom}). Applying again the 
Hartree-Fock approximation, which is essentially exact in the weak-coupling 
limit as we have seen in the equilibrium case, and substituting the ansatz 
$P[\phi^*,\phi;t] = P_0[\phi^*_{\bf 0},\phi_{\bf 0};t] P_1[\phi'^*,\phi';t]$ 
into equation~(\ref{FPinhom}), we obtain for the probability distribution of 
the condensate
\begin{eqnarray}
\label{FPCinhom}
&& \hspace*{-0.4in}
i\hbar \frac{\partial}{\partial t} P_0[\phi^*_{\bf 0},\phi_{\bf 0};t] \\
&&= - \frac{\partial}{\partial \phi_{\bf 0}} 
       \left( \epsilon_{\bf 0} + \hbar\Sigma^{(+)}_{\bf 0} - \mu(t) 
         + T^{(+)}_{{\bf 0},{\bf 0};{\bf 0},{\bf 0}}(t)
              |\phi_{\bf 0}|^2 \right) \phi_{\bf 0} 
                                P_0[\phi^*_{\bf 0},\phi_{\bf 0};t] \nonumber \\ 
&&+ \frac{\partial}{\partial \phi^*_{\bf 0}} 
      \left( \epsilon_{\bf 0} + \hbar\Sigma^{(-)}_{\bf 0} - \mu(t) 
        + T^{(+)}_{{\bf 0},{\bf 0};{\bf 0},{\bf 0}}(t)
             |\phi_{\bf 0}|^2 \right) \phi^*_{\bf 0} 
                                P_0[\phi^*_{\bf 0},\phi_{\bf 0};t] \nonumber \\
&&- \frac{1}{2} 
     \frac{\partial^2}{\partial \phi^*_{\bf 0} \partial \phi_{\bf 0}} 
       \hbar \Sigma^K_{\bf 0} P_0[\phi^*_{\bf 0},\phi_{\bf 0};t]~, \nonumber                                                                                                      
\end{eqnarray}     
where it should be remembered that for consistency reasons the selfenergies 
need to be evaluated at the energies 
$\epsilon''_{\bf 0}(t) = \epsilon'_{\bf 0}(t) 
        + T^{(+)}_{{\bf 0},{\bf 0};{\bf 0},{\bf 0}}(t) |\phi_{\bf 0}|^2$. 
In equilibrium the fluctuation-dissipation theorem then reads
\begin{equation}
iR_{{\bf n}} = - \frac{\beta}{4} \hbar \Sigma^K_{\bf 0} 
   \left( \epsilon'_{\bf 0} - \mu + T^{(+)}_{{\bf 0},{\bf 0};{\bf 0},{\bf 0}} 
|\phi_{\bf 0}|^2 \right)~,
\end{equation}
which guarantees that the probability distribution for the condensate relaxes 
to the correct equilibrium distribution
\begin{eqnarray}
P_0[\phi^*_{\bf 0},\phi_{\bf 0};\infty] \propto \exp \left\{ -\beta
   \left( \epsilon'_{\bf 0} - \mu + 
          \frac{T^{(+)}_{{\bf 0},{\bf 0};{\bf 0},{\bf 0}}}{2} |\phi_{\bf 0}|^2
   \right) |\phi_{\bf 0}|^2 \right\}~.          \nonumber
\end{eqnarray}
The Fokker-Planck equation, and hence the Boltzmann equation, for the 
noncondensed part of the gas is the same as in the normal state because for a 
discrete system the contributions from the condenstate to the selfenergies is 
already included in our expressions for $\hbar \Sigma^{(\pm),K}_{{\bf n}}$ and 
needs not be separated out explicitly. This implies of course, that if we 
evaluate the rates $\Gamma^{\rm out}_{{\bf n}}(t)$ and 
$\Gamma^{\rm in}_{{\bf n}}(t)$ by 
replacing the sum over states by an integral over a continuum, we first need to 
separate the contribution from the condensate to obtain the correct result. 

In summary we have thus found that in the weak-coupling limit the dynamics of 
Bose-Einstein condensation in an external trapping potential is described by 
the coupled equations in equation~(\ref{BEinhom}) and 
equation~(\ref{FPCinhom}), 
i.e., by a Boltzmann equation for the noncondensed part of the gas and a 
Fokker-Planck equation for the condensate, respectively. If we apply these 
equations to a Bose gas with a positive scattering length, we essentially 
arrive at the theory recently put forward by Gardiner and coworkers 
\cite{peter1}. Indeed, neglecting the energy shifts and assuming that the 
noncondensed cloud is in equilibrium with a chemical potential 
$\mu > \epsilon_{\bf 0}$, our Fokker-Planck equation for a $U(1)$ invariant 
probability distribution of the condensate becomes
\begin{eqnarray}
&& \hspace*{-0.7in}
i\hbar \frac{\partial}{\partial t} P_0[|\phi_{\bf 0}|;t] 
= - \frac{\beta}{4} \hbar \Sigma^K_{\bf 0} 
      \frac{\partial}{\partial \phi_{\bf 0}} 
       \left( \epsilon_{\bf 0} - \mu 
         + T^{(+)}_{{\bf 0},{\bf 0};{\bf 0},{\bf 0}}
              |\phi_{\bf 0}|^2 \right) \phi_{\bf 0} P_0[|\phi_{\bf 0}|;t] \\ 
&& \hspace*{0.2in} - \frac{\beta}{4} \hbar \Sigma^K_{\bf 0} 
    \frac{\partial}{\partial \phi^*_{\bf 0}} 
      \left( \epsilon_{\bf 0} - \mu 
        + T^{(+)}_{{\bf 0},{\bf 0};{\bf 0},{\bf 0}}
             |\phi_{\bf 0}|^2 \right) \phi^*_{\bf 0} 
                                P_0[|\phi_{\bf 0}|;t] \nonumber \\ 
&& \hspace*{0.2in} - \frac{1}{2} \hbar \Sigma^K_{\bf 0}
     \frac{\partial^2}{\partial \phi^*_{\bf 0} \partial \phi_{\bf 0}} 
        P_0[|\phi_{\bf 0}|;t]~. \nonumber                                                                                                      
\end{eqnarray}     

\begin{figure}
\begin{center}
\includegraphics[height=0.7\hsize]{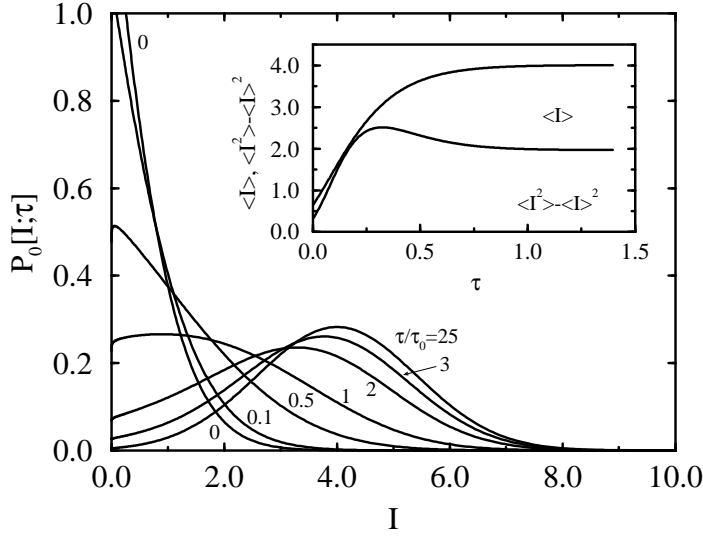}
\end{center}
\caption{Typical evolution of the condensate probability distribution during
         Bose-Einstein condensation. The slowest relaxation rate is 
         $1/\tau_0 \simeq 5.7$. The inset shows the evolution of the
         average condensate number and its fluctuations.}
\end{figure} 

\noindent
It is equivalent to the Fokker-Planck equation for a single-mode laser, since
introduction of the dimensionless time 
$\tau = t (i\Sigma^K_{\bf 0}/8)
          (2\beta T^{(+)}_{{\bf 0},{\bf 0};{\bf 0},{\bf 0}})^{1/2}$, the 
dimensionless condensate number 
$I = |\phi_{\bf 0}|^2 (2\beta T^{(+)}_{{\bf 0},{\bf 0};{\bf 0},{\bf 0}})^{1/2}$,
and the so-called `pump parameter'
$a = 2\beta(\mu-\epsilon_{\bf 0})/
           (2\beta T^{(+)}_{{\bf 0},{\bf 0};{\bf 0},{\bf 0}})^{1/2}$
leads to the Fokker-Planck equation
\begin{equation}
\label{sml}
\frac{\partial}{\partial \tau} P[I;\tau] =
  \frac{\partial}{\partial I} \left\{
    2(I-a)I + 4I \frac{\partial}{\partial I} \right\} P[I;\tau]
\end{equation}  
well-known from laser theory \cite{risken}. The typical behaviour of $P[I;\tau]$ 
during Bose-Einstein condensation in shown in figure 17.

Calculating $\langle I \rangle (\tau)$ from equation (\ref{sml}) by making use 
of the mean-field assumption 
$\langle I^2 \rangle (\tau) = \left( \langle I \rangle (\tau) \right)^2$, 
precisely reproduces the theory of Gardiner {\it et al.} A comparison with the 
experiments of H.-J. Miesner {\it et al.} shows that the theory of Gardiner and 
coworkers is unfortunately not fully satisfactory \cite{wolfgang2, peter1a}. The 
reason for this is presumably that these experiments are really in the 
strong-coupling limit, where the above theory does not apply. As mentioned 
previously, the theory is however applicable to gases with a negative scattering 
length. Since we know from our experience with homogeneous gases that the 
chemical potential will never be larger than $\epsilon'_{\bf 0}$ in this case, 
we clearly see from the equilibrium probability distribution 
$P_0[\phi^*_{\bf 0},\phi_{\bf 0};\infty]$ 
that a condensate in a gas with effectively attractive interactions is only 
metastable and will ultimately collapse to a dense state after it has overcome 
an energy barrier, either by thermal fluctuations or by quantum mechanical 
tunneling. This is clearly in full agreement with the discussion in section 
\ref{NSL}. Furthermore, we now also understand how the dynamics of the collapse 
must be treated. It requires the solution of a stochastic, dissipative nonlinear 
Schr\"odinger equation, in which the dissipative terms and the noise obey an 
appropriate fluctuation-dissipation theorem. Having arrived at this important 
result, we are finally left with a discussion of the strong-coupling limit.

\subsubsection{Strong-coupling limit}
The difficulty with the strong-coupling limit is that the appropriate 
one-particle states of the gas are no longer the states 
$\chi_{{\bf n}}({\bf x})$ of the trapping potential because these states are now 
strongly coupled by the mean-field interaction. For our Caldeira-Leggett model 
this problem could be easily resolved by introducing the states 
$\chi'_{{\bf n}}({\bf x})$ that 
incorporate the average interaction with the reservoir. However, for a trapped 
Bose gas the same procedure would lead to an expansion into states 
$\chi'_{{\bf n}}({\bf x};t)$ that parametrically depend on time since the 
mean-field interaction depends on time also. In principle, we thus have to deal 
with the situation that not only the occupation numbers vary with time but also 
the states to which these occupation numbers belong. This results in highly 
nontrivial dynamics since it is {\it a priori} clear that an adiabatic 
approximation, in which we neglect the nondiagonal matrix elements 
$\int d{\bf x}~\chi'^*_{{\bf n}}({\bf x};t) (\partial/\partial t) 
                                                 \chi'_{{\bf n}'}({\bf x};t)$,
is not valid, because of the lack of a separation of time scales. Notice that 
we were able to circumvent the above problems for a Bose gas with effectively 
attractive interactions due to the fact that on the one hand the growth of the 
condensate essentially takes place in the weak-coupling regime and that on the 
other hand the collapse, which certainly occurs in the strong-coupling regime, 
requires in first instance only knowledge of the collisionless dynamics of the 
condensate.                                                 

Notwithstanding the above remarks, we can make progress in the strong-coupling 
limit if we realize that in this limit the typical length scale on which the 
(noncondensate) density of the gas varies is always much larger than the trap 
size. We are therefore justified in performing a gradient expansion. To do so 
in a systematic fashion, it is convenient to express the various selfenergies 
and interactions in the coordinate representation, i.e., by introducing
\begin{equation}
\hbar\Sigma^{(\pm),K}({\bf x},{\bf x}';\epsilon) =
 \sum_{{\bf n},{\bf n}'} \chi_{{\bf n}}({\bf x})
 \hbar\Sigma^{(\pm),K}_{{\bf n},{\bf n}'}(\epsilon) \chi^*_{{\bf n}'}({\bf x}')
\end{equation}
and similarly
\begin{eqnarray}
&& \hspace*{-0.2in}
V^{(\pm),K}({\bf x},{\bf x}';{\bf y},{\bf y}';\epsilon) \\
&&= \sum_{{\bf n},{\bf n}'} \sum_{{\bf m},{\bf m}'}
   \chi_{{\bf n}}({\bf x}) \chi_{{\bf m}}({\bf y})
         V^{(\pm),K}_{{\bf n},{\bf m};{\bf n}',{\bf m}'}(\epsilon)
   \chi^*_{{\bf n}'}({\bf x}') \chi^*_{{\bf m}'}({\bf y}')~. \nonumber
\end{eqnarray}
In this coordinate representation, our Fokker-Planck equation acquires in all 
generality the form
\begin{eqnarray}
\label{FPgen}
&& \hspace*{-0.2in}
i\hbar \frac{\partial}{\partial t} P[\phi^*,\phi;t] 
= - \int d{\bf x} d{\bf x}'~
   \frac{\delta}{\delta \phi({\bf x})} \\
&& \hspace*{0.2in}   
       \left\{ \left( \left(
              - \frac{\hbar^2\mbox{\boldmath $\nabla$}^2}{2m} 
              + V^{\rm ex}({\bf x}) - \mu(t)
                      \right) \delta({\bf x}-{\bf x}')  
          + \hbar\Sigma^{(+)}({\bf x},{\bf x}') \right) \phi({\bf x}')
            \right. \nonumber \\
&& \hspace*{0.4in} + \left. \int d{\bf y} d{\bf y}'~
               V^{(+)}({\bf x},{\bf x}';{\bf y},{\bf y}') 
                   \phi^*({\bf y})
                   \phi({\bf y}') \phi({\bf x}')
     \raisebox{0.2in}{} \right\} P[\phi^*,\phi;t] \nonumber \\
&& + \int d{\bf x} d{\bf x}'~
  \frac{\delta}{\delta \phi^*({\bf x})} \nonumber \\
&& \hspace*{0.2in}   
       \left\{ \left( \left(
                    - \frac{\hbar^2\mbox{\boldmath $\nabla$}^2}{2m} 
                    + V^{\rm ex}({\bf x}) - \mu(t)
                      \right) \delta({\bf x}-{\bf x}')  
          + \hbar\Sigma^{(-)}({\bf x}',{\bf x}) \right) \phi^*({\bf x}')
            \right. \nonumber \\
&& \hspace*{0.4in} +  \left. \int d{\bf y} d{\bf y}'~
               V^{(-)}({\bf x}',{\bf x};{\bf y}',{\bf y})
                   \phi^*({\bf x}') \phi^*({\bf y}')
                   \phi({\bf y})
     \raisebox{0.2in}{} \right\} P[\phi^*,\phi;t] \nonumber \\
&&- \frac{1}{2} \int d{\bf x} d{\bf x}'~
  \frac{\delta^2}{\delta \phi({\bf x})
                    \delta \phi^*({\bf x}')}
       \left\{ \hbar\Sigma^K({\bf x},{\bf x}')
                    \raisebox{0.2in}{} \right. \nonumber \\
&& \hspace{0.4in} +  \left. \int d{\bf y} d{\bf y}'~
               V^K({\bf x},{\bf x}';{\bf y},{\bf y}') 
                   \phi^*({\bf y}) \phi({\bf y}') 
       \right\} P[\phi^*,\phi;t]~. \nonumber
\end{eqnarray}
To study its physical content and to familiarize ourselves with how to perform 
the desired gradient expansion, we first analyze again the normal state of the 
gas in which we can neglect the nonlinear terms representing the effect of the 
condensate.

In the normal state the condensate is absent and we are only interested in the 
behavior of the quantity $\langle \phi({\bf x}) \phi^*({\bf x}') \rangle(t)$. 
From the above Fokker-Planck equation we see that it obeys the equation of 
motion
\begin{eqnarray}
&& \hspace*{-0.2in} 
i\hbar \frac{\partial}{\partial t}
           \langle \phi({\bf x}) \phi^*({\bf x}') \rangle(t) \\
&&= \int d{\bf x}''~
      \left( \left( - \frac{\hbar^2}{2m} \frac{\partial^2}{\partial {\bf x}^2}
                   + V^{\rm ex}({\bf x}) - \mu(t)
            \right) \delta({\bf x}-{\bf x}'') \right. \nonumber \\
&& \hspace*{2.0in} + \left. \raisebox{0.2in}{}           
              \hbar\Sigma^{(+)}({\bf x},{\bf x}'') \right)
                       \langle \phi({\bf x}'') \phi^*({\bf x}') \rangle(t)
                                                                  \nonumber \\
&&- \int d{\bf x}''~
     \left( \left( - \frac{\hbar^2}{2m} \frac{\partial^2}{\partial {\bf x}'^2}
                   + V^{\rm ex}({\bf x}') - \mu(t)
            \right) \delta({\bf x}''-{\bf x}') \right. \nonumber \\
&& \hspace*{2.0in} + \left. \raisebox{0.2in}{}               
              \hbar\Sigma^{(-)}({\bf x}'',{\bf x}') \right)
                       \langle \phi({\bf x}) \phi^*({\bf x}'') \rangle(t)                                                                  
                                                                  \nonumber \\
&&- \frac{1}{2} \hbar\Sigma^K({\bf x},{\bf x}')~.                                                                  
\end{eqnarray}
Introducing now the Wigner distribution \cite{wigner} by means of
\begin{equation}
N({\bf x},{\bf k};t) + \frac{1}{2} =
  \int d{\bf x}'~ e^{-i {\bf k} \cdot {\bf x}' }
                 \langle \phi({\bf x} + {\bf x}'/2) 
                         \phi^*({\bf x} - {\bf x}'/2) \rangle(t)
\end{equation}
and similarly the selfenergies $\hbar\Sigma^{(\pm),K}({\bf x},{\bf k})$ by
\begin{equation}
\hbar\Sigma^{(\pm),K}({\bf x},{\bf k}) =
  \int d{\bf x}'~ e^{-i {\bf k} \cdot {\bf x}' }
     \hbar\Sigma^{(\pm),K}({\bf x} + {\bf x}'/2,{\bf x} - {\bf x}'/2)~,
\end{equation}
we find after substituting these relations in lowest nonvanishing order in the 
gradient $\partial/\partial{\bf x}$ the expected quantum Boltzmann equation 
\cite{D,kadanoff}
\begin{eqnarray}
\label{qBe}
&& \hspace*{-0.4in}
\frac{\partial}{\partial t} N({\bf x},{\bf k};t) \\
&&+ \frac{1}{\hbar} 
    \frac{\partial \epsilon'({\bf x},{\bf k};t)}{\partial {\bf k}} \cdot 
      \frac{\partial}{\partial {\bf x}} N({\bf x},{\bf k};t) 
  - \frac{1}{\hbar}
    \frac{\partial \epsilon'({\bf x},{\bf k};t)}{\partial {\bf x}} \cdot
      \frac{\partial}{\partial {\bf k}} N({\bf x},{\bf k};t) \nonumber \\
&&= - \Gamma^{\rm out}({\bf x},{\bf k};t) N({\bf x},{\bf k};t) 
      + \Gamma^{\rm in}({\bf x},{\bf k};t) (1 + N({\bf x},{\bf k};t))~,
      \nonumber
\end{eqnarray}
where the energies of the atoms have to be determined selfconsistently from
\begin{equation}
\epsilon'({\bf x},{\bf k};t) = \epsilon({\bf k}) + V^{\rm ex}({\bf x}) 
     + {\rm Re}[\hbar\Sigma^{(+)}({\bf x},{\bf k};
                                      \epsilon'({\bf x},{\bf k};t)-\mu(t))]~.    
\end{equation}
Moreover, the expressions for the retarded energy 
$\hbar\Sigma^{(+)}({\bf x},{\bf k};\epsilon'({\bf x},{\bf k};t)-\mu(t))$,
and the collision rates $\Gamma^{\rm out}({\bf x},{\bf k};t)$ and 
$\Gamma^{\rm in}({\bf x},{\bf k};t)$ turn out to be identical to the ones 
for a homogeneous gas \cite{jom,ST,stoof}. We only need to replace the 
one-particle energies $\epsilon'({\bf k};t)$ and occupation numbers 
$N({\bf k};t)$ by 
their local counterparts $\epsilon'({\bf x},{\bf k};t)$ and 
$N({\bf x},{\bf k};t)$, respectively. Notice also that the same substitution has 
to be made in the Bethe-Salpeter equation for the many-body T matrix, which 
implies that the effective interaction now 
implicitly depends both on the position ${\bf x}$ and the time $t$. 

As an example, let us consider the gas not too close to the critical 
temperature. Then the effective interaction is well approximated by the 
two-body T matrix and we simply have 
\begin{equation}
\epsilon'({\bf x},{\bf k};t) = \epsilon({\bf k}) + V^{\rm ex}({\bf x}) 
        + \frac{8\pi n({\bf x},t) a \hbar^2}{m}~.
\end{equation}
As a result the equilibrium solution $N({\bf x},{\bf k};\infty)$ of the quantum 
Boltzmann equation equals a Bose-Einstein distribution evaluated at the energy
$\epsilon({\bf k}) + V^{\rm ex}({\bf x}) + 8\pi n({\bf x}) a \hbar^2/m - \mu$. 
The equilibrium density profile 
$n({\bf x}) = \int d{\bf k}~N({\bf x},{\bf k};\infty)/(2\pi)^3$ thus 
corresponds to a local-density approximation but is indeed very accurate for the 
atomic gases of interest. Apart from describing the relaxation of the gas 
towards equilibrium, the quantum Boltzmann equation can also be used to study 
the collective modes of the trapped gas above the critical temperature. Using 
the same methods that have recently been applied to inhomogeneous Bose gases by 
Zaremba, Griffin, and Nikuni \cite{eugene2}, we can for example derive the 
hydrodynamic equations of motion the gas. Moreover, the collisionless modes are 
obtained from
\begin{eqnarray}
\label{clqBe}
&& \hspace*{-0.37in}
\frac{\partial}{\partial t} N({\bf x},{\bf k};t)
 + \frac{\hbar {\bf k}}{m} \cdot
      \frac{\partial}{\partial {\bf x}} N({\bf x},{\bf k};t) \\
&& \hspace*{0.4in} - \frac{1}{\hbar}
    \frac{\partial}{\partial {\bf x}} 
    \left( V^{\rm ex}({\bf x}) + \frac{8\pi n({\bf x},t) a \hbar^2}{m} \right) 
\cdot
      \frac{\partial}{\partial {\bf k}} N({\bf x},{\bf k};t) = 0~. \nonumber
\end{eqnarray}                    
Solving this equation is equivalent to the Generalized Random-Phase 
Approximation \cite{kadanoff}. More important is that the above collisionless 
Boltzmann equation can be shown to satisfy the Kohn theorem \cite{kohn} by 
substituting the shifted equilibrium profile 
$N({\bf x}-{\bf x}_{\rm cm}(t),{\bf k}
                        -(m/\hbar)d{\bf x}_{\rm cm}(t)/dt;\infty)$. For 
an harmonic external potential the center-of-mass of the gas cloud then 
precisely follows Newton's law 
$d^2 {\bf x}_{\rm cm}(t)/dt^2
                = - \mbox{\boldmath $\nabla$} V^{\rm ex}({\bf x}_{\rm cm}(t))$, 
which shows that the gas has three collisionless modes with frequencies that are 
equal to the three frequencies of the external trapping potential.

The full quantum Boltzmann equation in equation~(\ref{qBe}) also describes the 
relaxation of the gas towards equilibrium. It can thus be used to study forced  
evaporative cooling by adding an appropriate loss term to account for the 
escaping high-energetic atoms. In agreement with our previous picture we then 
expect that, if the evaporative cooling is performed slowly, the distribution 
function $N({\bf x},{\bf k};t)$ can be well approximated by a truncated 
equilibrium distribution with a time-dependent temperature and chemical 
potential. As a result, we can use the results from the homogeneous gas to show 
that in the critical regime the gas again becomes unstable towards 
Bose-Einstein condensation. This is the case when 
$\epsilon'({\bf x},{\bf 0};t) < \mu(t)$ for 
positions near the center of the trap where the density of the gas is highest. 
If that happens we are no longer allowed to neglect the nonlinear terms in the 
Fokker-Planck equation and some corrections are neccesary. Applying again the 
Hartee-Fock approximation, which is correct sufficiently close to the critical 
temperature where $k_BT \gg 4\pi n_{\bf 0}({\bf x},t) a \hbar^2/m$, 
we find first of all that the quantum Boltzmann equation for the noncondensed 
part of the gas has the same form as in equation~(\ref{qBe}). We only have to 
add the by now familiar mean-field interaction 
$2 n_{\bf 0}({\bf x},t) T^{(+)}({\bf 0},0)$ due 
to the condensate to the renormalized energy $\epsilon'({\bf x},{\bf k};t)$. 
Finally, the scattering rates $\Gamma^{\rm out}({\bf x},{\bf k};t)$ and 
$\Gamma^{\rm in}({\bf x},{\bf k};t)$ have to be adjusted in precisely  
the same way as in the homogeneous case to account for the macroscopic 
occupation of the condensate \cite{ST,bob,eugene3}. 

In addition, the resulting quantum Boltzmann equation for the noncondensed 
atoms is coupled to a Fokker-Planck equation for the condensate. After a 
gradient expansion the latter equals
\begin{eqnarray}
\label{FPsc}
&& \hspace*{-0.5in}
i\hbar \frac{\partial}{\partial t} P_0[\phi_{\bf 0}^*,\phi_{\bf 0};t] \\
&& = - \int d{\bf x}~
   \frac{\delta}{\delta \phi_{\bf 0}({\bf x})}
        \left( - \frac{\hbar^2\mbox{\boldmath $\nabla$}^2}{2m} 
               + V^{\rm ex}({\bf x}) - \mu(t)
               + \hbar\Sigma^{(+)}({\bf x},{\bf 0})
               \right. \nonumber \\
&& \hspace{1.22in} +  \left. {}^{{}^{{}^{{}^{{}^{{}^{}}}}}} 
                T^{(+)}({\bf 0},0) |\phi_{\bf 0}({\bf x})|^2
        \right) \phi_{\bf 0}({\bf x}) P_0[\phi_{\bf 0}^*,\phi_{\bf 0};t]         
  \nonumber \\
&& + \int d{\bf x}~
  \frac{\delta}{\delta \phi_{\bf 0}^*({\bf x})}
        \left( - \frac{\hbar^2\mbox{\boldmath $\nabla$}^2}{2m} 
               + V^{\rm ex}({\bf x}) - \mu(t)
               + \hbar\Sigma^{(-)}({\bf x},{\bf 0})
               \right. \nonumber \\
&& \hspace{1.22in} +  \left. {}^{{}^{{}^{{}^{{}^{{}^{}}}}}} 
                T^{(+)}({\bf 0},0) |\phi_{\bf 0}({\bf x})|^2
        \right) \phi_{\bf 0}^*({\bf x}) P_0[\phi_{\bf 0}^*,\phi_{\bf 0};t]       
    \nonumber \\
&&- \frac{1}{2} \int d{\bf x}~
  \frac{\delta^2}{\delta \phi_{\bf 0}({\bf x}) \delta \phi_{\bf 0}^*({\bf x})}
     \hbar\Sigma^K({\bf x},{\bf 0}) P_0[\phi_{\bf 0}^*,\phi_{\bf 0};t]~,
     \nonumber
\end{eqnarray}
where $\hbar\Sigma^{(\pm),K}({\bf x},{\bf 0})$ denotes  
$\hbar\Sigma^{(\pm),K}({\bf x},{\bf k}={\bf 0})$. In more detail we have for 
the real part of the retarded and advanced selfenergies that 
$S({\bf x},{\bf 0}) = {\rm Re}[\hbar\Sigma^{(\pm)}({\bf x},{\bf 0};
                                 \epsilon''({\bf x},{\bf 0};t) - \mu(t))]$
and $\epsilon''({\bf x},{\bf 0};t)$ the local energy of a condensate atom. 
Moreover, using our picture of the evaporative cooling of the gas, the 
imaginary parts follow most conveniently from the fluctuation-dissipation 
theorem
\begin{eqnarray}
&& \hspace*{-0.3in} iR({\bf x},{\bf 0}) 
  = - \frac{\beta(t)}{4} \hbar\Sigma^K({\bf x},{\bf 0}) \\
&& \hspace*{0.23in} \times~
   \left( - \frac{\hbar^2\mbox{\boldmath $\nabla$}^2}{2m} 
          + \epsilon'({\bf x},{\bf 0};t) - \mu(t)
          + T^{(+)}({\bf 0},0) |\phi_{\bf 0}({\bf x})|^2
        \right) \nonumber
\end{eqnarray} 
where $\hbar\Sigma^K({\bf x},{\bf 0}) 
               = \hbar\Sigma^{(\pm),K}({\bf x},{\bf 0};
                                       \epsilon''({\bf x},{\bf 0};t) - \mu(t))$ 
can in a good approximation be considered as independent of 
$|\phi_{\bf 0}({\bf x})|$. 
Near equilibrium the Fokker-Planck equation thus simplifies to
\begin{eqnarray}
&&\hspace*{-0.5in} 
i\hbar \frac{\partial}{\partial t} P_0[|\phi_{\bf 0}|;t] \\
&&=
 - \frac{\beta}{4} \int d{\bf x}~ \hbar\Sigma^K({\bf x},{\bf 0})
   \frac{\delta}{\delta \phi_{\bf 0}({\bf x})}
        \left( - \frac{\hbar^2\mbox{\boldmath $\nabla$}^2}{2m} 
               + \epsilon'({\bf x},{\bf 0}) - \mu
        \right. \nonumber \\
&& \hspace{1.2in} +  \left. {}^{{}^{{}^{{}^{{}^{{}^{}}}}}} 
                T^{(+)}({\bf 0},0) |\phi_{\bf 0}({\bf x})|^2
        \right) \phi_{\bf 0}({\bf x}) P_0[|\phi_{\bf 0}|;t]    \nonumber \\
&&- \frac{\beta}{4} \int d{\bf x}~ \hbar\Sigma^K({\bf x},{\bf 0})
  \frac{\delta}{\delta \phi_{\bf 0}^*({\bf x})}
        \left( - \frac{\hbar^2\mbox{\boldmath $\nabla$}^2}{2m} 
               + \epsilon'({\bf x},{\bf 0}) - \mu
        \right. \nonumber \\
&& \hspace{1.2in} +   \left. {}^{{}^{{}^{{}^{{}^{{}^{}}}}}} 
                T^{(+)}({\bf 0},0) |\phi_{\bf 0}({\bf x})|^2
        \right) \phi_{\bf 0}^*({\bf x}) P_0[|\phi_{\bf 0}|;t] \nonumber \\
&&- \frac{1}{2} \int d{\bf x}~ \hbar\Sigma^K({\bf x},{\bf 0}) 
  \frac{\delta^2}{\delta \phi_{\bf 0}({\bf x}) \delta \phi_{\bf 0}^*({\bf x})}
     P_0[|\phi_{\bf 0}|;t] \nonumber
\end{eqnarray}
and the probability distribution of the condensate relaxes to the expected 
stationary state
\begin{eqnarray}
&& \hspace{-0.2in}
P_0[|\phi_{\bf 0}|;\infty] \propto 
\exp \left\{ - \beta \int d{\bf x}~ \phi_{\bf 0}^*({\bf x})
  \left( - \frac{\hbar^2\mbox{\boldmath $\nabla$}^2}{2m} 
         + \epsilon'({\bf x},{\bf 0}) - \mu \right. \right. \\
&& \hspace*{2.0in} \left. \left.         
      + \frac{T^{(+)}({\bf 0},0)}{2} |\phi_{\bf 0}({\bf x})|^2
  \right) \phi_{\bf 0}({\bf x}) \right\}~. \nonumber
\end{eqnarray}

From this result we conclude that Eqs.~(\ref{qBe}) and (\ref{FPsc}) give an 
accurate discription of the dynamics of Bose-Einstein condensation in the 
strong-coupling regime. Unfortunately, the solution of these coupled equations 
is not possible analytically and we therefore have to refer to future 
work for a numerical study and a comparison of this theory with 
experiment. We here only remark that the semiclassical approximation in this 
case amounts to the Thomas-Fermi result for the condensate density 
\begin{equation}
n_{\bf 0}({\bf x};t) = \frac{1}{T^{(+)}({\bf 0},0)}
   \left( \mu(t) - \epsilon'({\bf x},{\bf 0};t) \right) 
       \Theta( \mu(t) - \epsilon'({\bf x},{\bf 0};t) )~.
\end{equation}
Such an approximation implicitly assumes, however, that the dominant mechanism 
for the formation of the condensate is by means of a coherent evolution and not 
by incoherent collisions. This might be expected not to be very accurate on the 
basis of the homogeneous result that the second mechanism only diverges 
logarithmically with the system size \cite{henk2}, whereas the first diverges 
linearly \cite{subir}. Nevertheless it seems that only a full solution of 
equations~(\ref{qBe}) and (\ref{FPsc}) can settle this important issue.

\subsection{Collective modes}
\label{modes}
Although we have up to now primarily been focussed on the dynamics of 
Bose-Einstein condensation itself, it is interesting that we have in fact, 
similarly as for temperatures above the critical temperature, also 
arrived at an accurate discription of the collective modes of a condensed gas 
at nonzero temperatures. Since these collective modes are receiving much 
attention at present, we now briefly want to discuss how such a description 
emerges from our Fokker-Planck equation. Since most experiments at present 
are performed in the collisionless limit, we consider only this limit here. It 
is in principle possible, however, to derive the hydrodynamic equations of 
motions as well \cite{eugene2}. Let us for simplicitly first assume that the 
temperature of the gas is sufficiently far below the critical temperature that 
the many-body T matrix is well approximated by the two-body T matrix, but not 
so low that the Hartree-Fock approximation is no longer valid. In this regime we 
have that 
$\hbar\Sigma^{(\pm)}({\bf x},{\bf k}) = 8\pi n'({\bf x},t) a\hbar^2/m$ 
and the dynamics of the noncondensed part of the gas is again determined by the 
collisionless quantum Boltzmann equation given in equation~(\ref{clqBe}), where 
the total density now consists of two contributions
\begin{equation}
n({\bf x},t) =  n_{\bf 0}({\bf x},t) + n'({\bf x},t) \equiv
         |\langle \phi({\bf x}) \rangle(t)|^2
                     + \int \frac{d{\bf k}}{(2\pi)^3}~N({\bf x},{\bf k};t)~.
\end{equation}
The dynamics of the condensate follows again from the nonlinear Schr\"oding\-er 
equation
\begin{eqnarray}
\label{nlse}
&&\hspace*{-0.2in}
i\hbar \frac{\partial}{\partial t} \langle \phi({\bf x}) \rangle(t) \\
&&= \left\{ - \frac{\hbar^2 \mbox{\boldmath $\nabla$}^2}{2m} 
            + V^{\rm ex}({\bf x}) - \mu 
            + \frac{4\pi a\hbar^2}{m} (2n'({\bf x},t) + n_{\bf 0}({\bf x},t)) 
    \right\} \langle \phi({\bf x}) \rangle(t)~, \nonumber
\end{eqnarray}
which can be rewritten as the continuity equation
\begin{equation}
\label{nlse1}
\frac{\partial}{\partial t} n_{\bf 0}({\bf x},t) 
                = - \mbox{\boldmath $\nabla$} \cdot 
                    ( {\bf v}_s({\bf x},t) n_{\bf 0}({\bf x},t) )
\end{equation}
together with the Josephson relation for the superfluid velocity 
\begin{eqnarray}
\label{nlse2}
&& \hspace*{-0.2in}
\frac{\partial}{\partial t} {\bf v}_s({\bf x},t) \\
&&=- \mbox{\boldmath $\nabla$} \left( V^{\rm ex}({\bf x}) - \mu 
            + \frac{4\pi a\hbar^2}{m} (2n'({\bf x},t) + n_{\bf 0}({\bf x},t)) 
            + \frac{1}{2} m {\bf v}_s^2({\bf x},t) \right)~. \nonumber
\end{eqnarray}
Here, we used that $\langle \phi({\bf x}) \rangle(t) 
                     = \sqrt{n_{\bf 0}({\bf x},t)} e^{i\theta({\bf x},t)}$,
${\bf v}_s({\bf x},t) = \hbar \mbox{\boldmath $\nabla$} \theta({\bf x},t)/m$, 
and have applied the Thomas-Fermi approximation, which is appropriate in the 
strong-coupling limit. It will presumably not come as a surprise that these 
coupled equations for 
$N({\bf x},{\bf k};t)$, $n_{\bf 0}({\bf x},t)$, and ${\bf v}_s({\bf x},t)$ can 
be shown to fulfill the Kohn theorem exactly. This is an important constraint 
on any theory for the collective modes of a trapped gas, as stressed recently by 
Zaremba, Griffin, and Nikuni \cite{eugene2}.

As a first step towards this goal we here consider the time-dependent 
Hartree-Fock theory for the collective modes of the Bose condensed gas, in 
which the collisionless Boltzmann equation in equation~(\ref{clqBe}) is coupled 
to the nonlinear Schr\"odinger equation in equation~(\ref{nlse}). It turns out 
that a numerical determination of the eigenmodes of the collisionless Boltzmann 
equation is rather difficult. Therefore, we have recently proposed a 
variational approach to solve these coupled equations \cite{michiel2}. Denoting 
the anisotropic harmonic trapping potential used in the experiments 
\cite{coll1,coll2} by $V^{\rm ex}({\bf x}) = \sum_j m\omega_j^2 x_j^2/2$, we 
first of all take for the condensate wave function the gaussian {\it ansatz}
\begin{eqnarray}
&& \hspace*{-0.2in}
\langle \phi({\bf x}) \rangle(t) \\
&&= \sqrt{N_{\bf 0}} 
  \prod_j \left( \frac{1}{\pi q_j^2(t)} \right)^{1/4} 
    \exp \left\{ - \frac{x_j^2}{2q_j^2(t)} 
                      \left( 1 - i \frac{m q_j(t)}{\hbar}
                                       \frac{d q_j(t)}{dt} \right)
         \right\}~, \nonumber
\end{eqnarray}  
with the functions $q_j(t)$ describing the `breathing' of the condensate in 
three independent directions. Clearly, this is the anisotropic generalization 
of the gaussian {\it ansatz} that we used in section \ref{NSL} to study the 
stability of a condensate with negative scattering length. Although it is not 
immediately obvious that such an {\it ansatz} is also appropriate in the 
strong-coupling regime of interest here, it has been shown by Perez-Garcia {\it 
et al.} \cite{peter3} that at zero temperature the assumption of a gaussian 
density profile nevertheless leads to rather accurate results for the 
frequencies of the collective modes. We therefore anticipate that at nonzero 
temperatures the same will be the case.

Next we also need an {\it ansatz} for the distribution function 
$N({\bf x},{\bf k};t)$ of the thermal cloud. Here, we consider the Bose 
distribution
\begin{eqnarray}
&&\hspace*{-0.3in}
N({\bf x},{\bf k};t) = N' \frac{1}{\zeta(3)} \left( \prod_j 
    \frac{\hbar\omega_j}{k_BT \langle \alpha_j \rangle^2} \right) \\
&&\times \left(
        \exp \left\{ \sum_j \frac{1}{k_BT \langle \alpha_j \rangle^2}
               \left( \frac{\hbar^2 \alpha_j^2(t)}{2m}
                        \left( k_j - \frac{m x_j}{\hbar\alpha_j(t)}
                                     \frac{d \alpha_j(t)}{dt} \right)^2
                                     \right. \right. \right. \nonumber \\
&& \hspace*{2.35in} + \left. \left. \left.
        \frac{m\omega_j^2}{2} \frac{x_j^2}{\alpha_j^2(t)}
               \right) \raisebox{0.3in}{}
             \right\} - 1 \right)^{-1}~, \nonumber
\end{eqnarray}
where $\alpha_j(t)$ are again three scaling parameters, $N'=N-N_{\bf 0}$ is the 
number of noncondensed particles, $\langle \alpha_j \rangle$ denotes the 
equilibrium value of $\alpha_j(t)$ that in general is greater than $1$ due to 
the repulsive interactions between the atoms, and $\zeta(3) \simeq 1.202$. The 
functional form of the distribution function is physically motivated by the 
following reasons. First, by integrating over the momentum $\hbar{\bf k}$ we see 
that the time-dependent density profile of the noncondensed cloud 
$n'({\bf x},t)$ is a time-independent function of $x_j/\alpha_j(t)$, as desired. 
Second, for such a density profile we 
must have that the local velocity $\hbar\langle k_j \rangle({\bf x},t)/m$ is 
given by $(x_j/\alpha_j(t)) d\alpha_j(t)/dt$. Third, we want the equilibrium 
distribution to be only anisotropic in coordinate space, and not in momentum 
space. This explains the appearance of the factor $1/\langle \alpha_j 
\rangle^2$ in the exponent.

\begin{figure}
\begin{center}
\includegraphics[height=0.7\hsize]{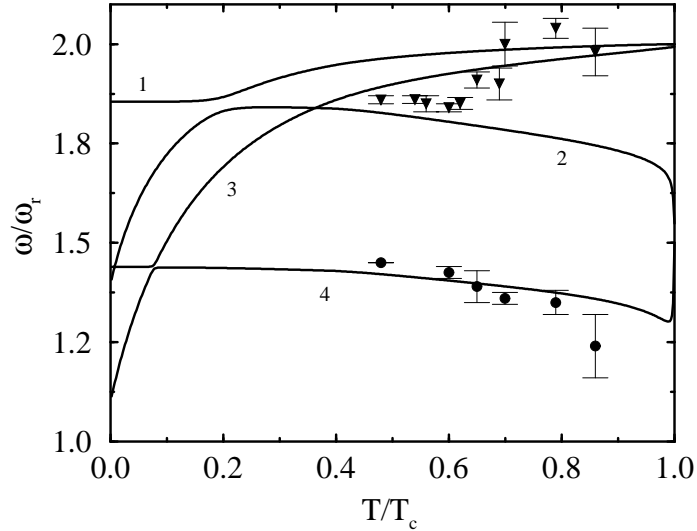}
\end{center}
\caption{Collsionless modes in a Bose condensed $^87$Rb gas. The full lines
         give the results with a Bose distribution {\it ansatz} for the 
         distribution function of the thermal cloud. Curves 1 and 2 correspond 
         to the $m=0$ in  and out-of-phase modes, respectively. Similarly, 
         curves 3 and 4 give  the frequencies of the $m=2$ in and out-of-phase 
         modes. The experimental data is also shown. Triangles are for a $m=0$ 
         mode and circles for a $m=2$ mode. }
\end{figure} 

Using the ideal gas result for the number of condensate particles and 
substituting the above {\it ansatz} for the condensate wave function and the 
Wigner distribution function into the nonlinear Schr\"odinger and collisionless 
Boltzmann equations, it is straightforward but somewhat tedious to derive the 
six coupled equations of motion for the `breathing' parameters $q_j(t)$ and 
$\alpha_j(t)$. Linearizing these equations around the equilibrium solutions 
$\langle q_j \rangle$ and $\langle \alpha_j \rangle$, respectively, we can then 
also obtain the desired eigenmodes and eigenfrequencies. The results of this 
calculation for the trap parameters of Jin {\it et al.} \cite{coll1} are shown 
in figure 18 together with the experimental data. Note that the trap has a 
disk geometry with $\omega_x = \omega_y \equiv \omega_r$ and 
$\omega_z = \sqrt{8} \omega_r$. Due to the axial symmetry, the modes of the gas 
can be labeled by the usual quantum number $m$ that describes the angular 
dependence around the $z$-axis. On the whole there appears to be a quite 
reasonable agreement between this simple theory and experiment. The only 
exceptions are the two data point almost halfway in between the in and 
out-of-phase $m=0$ modes. A possible explanation for this discrepancy is that 
experimentally both modes are excited simultaneously \cite{eric}. To make sure 
of this, however, we need to determine the oscillator strength of the various 
modes and it appears that the in-phase mode indeed dominates the response for 
temperatures larger than about $0.6 T_c$ \cite{michiel2}.

\section{Outlook}
With the latter remark we end this course on the use of field-theoretical 
methods for the study of the equilibrium and nonequilibrium properties of 
trapped atomic gases. Although we have in principle illustrated all the 
necessary tools for an {\it ab initio} treatment of these new quantum systems, 
many interesting topics still need to be considered in detail. We already 
mentioned such topics as Fermi-Bose mixtures, the dynamics of condensate 
collapse, the strong-coupling theory for the formation of a condensate in either 
atomic Bose of Fermi gases, and the hydrodynamics of single or multicomponent 
atomic gases. Several other important areas of research, which can also be 
easily addressed within the context of quantum field theory, are atom lasers, 
the damping of collective modes, the behaviour of spinor condensates, the 
dynamics of topological excitations such as kinks, vortices and skyrmions, 
quantum critical phenomena, the optical properties of superfluid gases, and 
two-dimensional phase transitions. In view of these and many other possibilities 
that are now experimentally feasible, it appears certain that the physics of 
degenerate atomic gases will remain very exciting for years to come.


\begin{thebibliography}{106}
\bibitem{gap} The magnitude of the frequency gap is actually not accurately
              described by this time-dependent Ginzburg-Landau theory, because 
              the action $S[\Delta^*,\Delta]$ is implicitly only valid for
              long wavelengths $\hbar v_F k \ll |\Delta_0|$ and low 
              frequencies $\hbar \omega \ll |\Delta_0|$.
\bibitem{top} For simplicity we here again neglect a topological term in the
              action, which essentially plays no role in the following. This is
              also in agreement with our discussion of the superconductor, 
              where exactly the same topological term was not included in the
              effective action for the global phase \cite{micheal}.
\bibitem{half} This follows from the requirement that we want to be able to
               Fourier transform the Heaviside function.
\bibitem{remark} In agreement with our previous remarks, the notation
                 $\int_{t_0}$ implies that the time integration
                 can be either from $t_0$ to $\infty$ or from $t_0$ to $t$.
\end{thebibliography}
\end{document}